\documentclass[11pt]{article}

\usepackage{amsmath}
\usepackage{amsfonts}
\usepackage{amssymb}
\usepackage{bbm}
\usepackage{graphicx}
\usepackage{tikz}
\usetikzlibrary{arrows,shapes,positioning}
\usetikzlibrary{calc}
\usepackage{pgfplots}
\usepackage{cleveref}
\usepackage{graphicx}
\usepackage{subcaption}
\usetikzlibrary{trees,snakes}
\usepackage{verbatim}
\usepackage{mathtools}
    
\usepackage{mathrsfs} 
\usepackage{xfrac} 
\usepackage{tabu}
\usepackage{color}
\usepackage{tikz}

\definecolor{myGOODNESS}{rgb}{0.82, 0.1, 0.26}
\definecolor{myAMBER}{rgb}{1, 0.75, 0}
\definecolor{myred}{RGB}{255, 0, 0}
\definecolor{myblue}{RGB}{0, 0, 255}
\newtheorem{theorem}{Theorem}
\newtheorem{lemma}{Lemma}
\newtheorem{proposition}{Proposition}
\newtheorem{fact}{Fact}
\newtheorem{corollary}{Corollary}
\newcommand{\LambdaBOOM}{\Phi}
\newcommand{\RE}{\mathsf{E}(\calC_{n})}
\newcommand{\expVAR}{E_{0}}
\newcommand{\E}{\mathbb{E}}
\newcommand{\nn}{\nonumber}
\newcommand {\prob} {\mathbb{P}}
\newcommand{\IND}{\calI}
\newcommand{\DEF}{\overset{\Delta}{=}}

\newcommand {\lexe} {\stackrel{\cdot} {\leq}}
\newcommand {\gexe} {\stackrel{\cdot} {\geq}}

\newcommand {\DEXE} {\stackrel{\circ}{=}}
\newcommand {\DLEXE} {\stackrel{\circ}{\leq}}
\newcommand {\DGEXE} {\stackrel{\circ}{\geq}}

\newcommand {\hQ} {\hat{Q}}

\newcommand {\bi} {\boldsymbol{i}}
\newcommand {\bj} {\boldsymbol{j}}
\newcommand {\bk} {\boldsymbol{k}}

\newcommand {\bx} {\boldsymbol{x}}
\newcommand {\by} {\boldsymbol{y}}

\newcommand {\bX} {\boldsymbol{X}}
\newcommand{\calA}{{\cal A}}
\newcommand{\calB}{{\cal B}}
\newcommand{\calC}{{\cal C}}
\newcommand{\calD}{{\cal D}}
\newcommand{\calE}{{\cal E}}
\newcommand{\calF}{{\cal F}}
\newcommand{\calG}{{\cal G}}

\newcommand{\calI}{{\cal I}}
\newcommand{\calJ}{{\cal J}}
\newcommand{\calK}{{\cal K}}
\newcommand{\calL}{{\cal L}}
\newcommand{\calM}{{\cal M}}

\newcommand{\calQ}{{\cal Q}}

\newcommand{\calS}{{\cal S}}
\newcommand{\calT}{{\cal T}}
\newcommand{\calU}{{\cal U}}
\newcommand{\calV}{{\cal V}}

\newcommand{\calX}{{\cal X}}
\newcommand{\calY}{{\cal Y}}

\newcommand{\styleA}{\calA}

\begin{document}
\thispagestyle{empty}
\title{Large Deviations Behavior of the Logarithmic Error Probability of Random Codes\\}
\author{\\ Ran Tamir (Averbuch), Neri Merhav, Nir Weinberger, and Albert Guill\'en i F\`abregas \\}
\maketitle
\vspace{1.5\baselineskip}
\setlength{\baselineskip}{1.5\baselineskip}

\begin{abstract}
\let\thefootnote\relax\footnote{
R.\ Tamir and N.\ Merhav are with the Andrew \& Erna Viterbi Faculty of Electrical Engineering, Technion -- Israel Institute of Technology,
Technion City, Haifa 32000, Israel (e--mails:
rans@campus.technion.ac.il and merhav@ee.technion.ac.il).

N.\ Weinberger is with IDSS and LIDS, Massachusetts Institute of Technology, Cambridge, MA, 02139, USA.\ (e--mail: nirw@mit.edu).

A.\ Guill\'en i F\`abregas is with the Department of Information and Communication Technologies, Universitat Pompeu Fabra, Barcelona 08018, Spain, also with the Instituci\'o Catalana de Recerca i Estudis Avan\c{c}ats (ICREA), Barcelona 08010, Spain, and also with the Department of Engineering, University of Cambridge, Cambridge CB2 1PZ, U.K.\ (e--mail: guillen@ieee.org).

The research of R.\ Tamir and N.\ Merhav was supported by Israel Science Foundation (ISF) grant no. 137/18.
The research of N. Weinberger was partially supported by the MIT--Technion fellowship, and the Viterbi scholarship, Technion.
The research of A. Guill\'en i F\`abregas was funded in part by the European Research Council under ERC grant 725411 and by the Spanish Ministry of Economy and Competitiveness under grant TEC2016-78434-C3-1-R.}
This work studies the deviations of the error exponent of the constant composition code ensemble around its expectation, known as the error exponent of the typical random code (TRC). In particular, it is shown that the probability of randomly drawing a codebook whose error exponent is smaller than the TRC exponent is exponentially small; 
upper and lower bounds for this exponent are given, which coincide in some cases. In addition, the probability of randomly drawing a codebook whose error exponent is larger than the TRC exponent is shown to be double--exponentially small; upper and lower bounds to the double--exponential exponent are given. The results suggest that codebooks whose error exponent is larger than the error exponent of the TRC are extremely rare.  The key ingredient in the proofs is a new large deviations result of type class enumerators
with dependent variables.  \\

\noindent
\end{abstract}

\clearpage
\section{Introduction}

Random coding is the 
most common method 
to show that the probability of error vanishes for rates below the channel capacity. 
In 1955, Feinstein \cite{Feinstein} proved that, for a sequence of codes of fixed rate and increasing length, the probability of error decays to zero exponentially with the length of the codes, provided that the rate of the code is below the mutual information of the channel. In the same year, Elias \cite{Elias55} derived the random coding and sphere--packing bounds and observed that they exponentially coincide at high rates,
for the cases of the binary symmetric channel (BSC) and the binary erasure channel (BEC). 
Fano \cite{FANO61} derived the random coding exponent, 
namely,
\begin{align}
\label{RCE}
E_{\mbox{\tiny r}}(R) = \lim_{n \to \infty} \left\{ - \tfrac{1}{n} \log \mathbb{E} \left[P_{\mbox{\tiny e}}(\calC_{n}) \right] \right\},
\end{align}
where the expectation is with respect to (w.r.t.) a given ensemble of codes, and heuristically also the sphere--packing bound for the general discrete memoryless channel (DMC). 
In 1965, Gallager \cite{Gal65} derived $E_{\mbox{\tiny r}}(R)$ in a much simpler way and improved on $E_{\mbox{\tiny r}}(R)$ at low rates by the idea of expurgation. 

In random coding analysis, the code is selected at random and remains fixed, and thus, it seems reasonable to study the performance in terms of error exponent of the very chosen code, rather than considering the exponent of the averaged probability of error, as in $E_{\mbox{\tiny r}}(R)$. Therefore, it is natural to ask what would be the error exponent associated with the typical randomly selected code. The error exponent of the typical random code (TRC) is defined as 
\begin{align} \label{TRC_DEF}
E_{\mbox{\tiny trc}}(R) = \lim_{n \to \infty} \left\{- \tfrac{1}{n} \mathbb{E} \left[\log P_{\mbox{\tiny e}}(\calC_{n}) \right] \right\}.
\end{align}
We find the exponent of the TRC to be the more relevant performance metric as 
it captures the true exponential behavior of the probability of error, as opposed to the random coding error exponent, which is dominated by the relatively poor codes of the ensemble, rather than the channel noise, at relatively low coding rates.

To the best of our knowledge, not much is known on typical random codes. 
In \cite{BargForney}, Barg and Forney considered typical random codes with independently and identically distributed codewords for the BSC with maximum--likelihood (ML) decoding. They also considered typical linear codes.
It was shown that at a certain range of low rates, $E_{\mbox{\tiny trc}}(R)$ lies between $E_{\mbox{\tiny r}}(R)$ and the expurgated exponent, $E_{\mbox{\tiny ex}}(R)$. 
In \cite{PRAD2014} Nazari \textit{et al}.\ provided bounds on the error exponent of the TRC for both DMCs and multiple--access channels. 
In a recent article \cite{MERHAV_TYPICAL}, an exact single--letter expression has been derived for the error exponent of typical, random, constant composition codes, over DMCs, and a wide class of (stochastic) decoders, collectively referred to as the generalized likelihood decoder (GLD), 
which includes the ML decoder as a special case.
For such decoders, the probability of deciding on a given message is proportional to a general exponential function of the joint empirical distribution of the codeword and the received channel output vector.
Recently, Merhav has studied error exponents of TRCs for the colored Gaussian channel \cite{MERHAV_GAUSS}, as well as typical random trellis codes \cite{MERHAV_TRELLIS}.

Note that the TRC exponent can be 
viewed as the limit of the expectation of the random variable
\begin{align}
\RE = - \tfrac{1}{n} \log P_{\mbox{\tiny e}}(\calC_{n}),
\end{align} 
where $P_{\mbox{\tiny e}}(\calC_{n})$ is the error probability of a given code $\calC_{n}$, governed by the randomness of the codebook $\calC_{n}$. 
Having defined this random variable, it is interesting to study, not only its expectation, but also other, more refined, quantities associated with its probability distribution. 
One of them is the tail behavior, i.e., the large deviations (LD) rate functions. 
In particular, it is partially implied\footnote{More specifically, for every $\epsilon > 0$, $\prob \{\RE \leq E_{\mbox{\tiny trc}}(R) +\epsilon\}$ converges to one exponentially fast as $n \to \infty$.} 
from \cite{MERHAV_TYPICAL}, that $\RE$ concentrates around its expectation, i.e., the error exponent $E_{\mbox{\tiny trc}}(R)$. In this work we prove that $\RE$ indeed concentrates around $E_{\mbox{\tiny trc}}(R)$.


In this paper we are interested in probabilities of large fluctuations around $E_{\mbox{\tiny trc}}(R)$.
More specifically, we investigate the probability of randomly choosing a {\it bad} codebook, i.e., a codebook with a relatively small value of $\RE$. On the other hand, the probability of randomly drawing a {\it good} codebook, i.e., a codebook with a relatively large value of $\RE$ is of interest as well, 
since obtaining tight LD bounds is an alternative method to prove upper or lower bounds on the channel reliability function, a long--standing problem.

To the best of our knowledge, the only known bounds on the probability of drawing codebooks with relatively low error exponents are given in \cite[Appendix III]{GoodCodes}. It is proved in \cite{GoodCodes} that $\prob \left\{ \RE < E_{\mbox{\tiny r}}(R) \right\}$ is upper bounded by $\exp\{-\exp\{n (R - E_{\mbox{\tiny r}}(R))\}\}$, as long as $R > E_{\mbox{\tiny r}}(R)$, while the entire range of relatively low rates, namely $R \leq E_{\mbox{\tiny r}}(R)$, was hardly considered in \cite{GoodCodes}, and is one of the main topics in the current work. 
Furthermore, in this paper, we study the deviations of $\RE$ w.r.t.\ its actual expected value $E_{\mbox{\tiny trc}}(R)$, and not as in \cite{GoodCodes}, in which considered deviations w.r.t.\ $E_{\mbox{\tiny r}}(R)$.

Accordingly, the main purpose of this paper is to study the probabilistic behavior of the tails of $\RE$, i.e., to characterize its large deviations properties. 
For a given $\expVAR < E_{\mbox{\tiny trc}}(R)$, we assess the probability $\prob \left\{ \RE \leq \expVAR \right\}$ and provide exponentially small lower and upper bounds on it, which proves that bad codebooks are rare.
More refined questions concerning the lower tail are as follows. Does the probability $\prob \left\{ \RE \leq \expVAR \right\}$ tend to zero with a finite exponent in the entire range $[0, E_{\mbox{\tiny trc}}(R))$? 
If not, what is the range of $\expVAR$ for which $\prob \left\{ \RE \leq \expVAR \right\}$ decays faster than exponentially? 
Indeed, we prove that a {\it phase transition} occurs in the behavior of this probability, i.e., at some point below $E_{\mbox{\tiny trc}}(R)$, we observe an abrupt change between an ordinary exponential decay to a super--exponential decay. 
In addition, we consider the probability $\prob \left\{ \RE \geq \expVAR \right\}$, for $\expVAR > E_{\mbox{\tiny trc}}(R)$, and derive double--exponentially small lower and upper bounds on it. We find the largest value $\expVAR$, for which $\prob \left\{ \RE \geq \expVAR \right\}$ is strictly positive, thereby proving the existence of exceptionally good codebooks.

The remaining part of the paper is organized as follows. 
In Section 2, we establish notation conventions.
In Section 3, we formalize the model, the decoder, LD quantities, and provide some preliminaries. 
In Section 4, we summarize and discuss the main results, and provide numerical example for the binary $z$--channel. 
Sections 5, 6 and 7 include the proofs of our main theorems.

\section{Notation Conventions}

Throughout the paper, random variables will be denoted by capital letters, realizations will be denoted by the corresponding lower case letters, 
and their alphabets in calligraphic font.
Random vectors and their realizations will be denoted, 
respectively, by boldfaced capital and lower case letters. 
Their alphabets will be superscripted by their dimensions. 
For a generic joint distribution $Q_{XY} = \{Q_{XY}(x,y), x \in \mathcal{X}, y \in \mathcal{Y} \}$, which will often be abbreviated by $Q$, information measures will be denoted in 
the conventional manner, but with a subscript $Q$, that is, $I_{Q}(X;Y)$ is the mutual information between $X$ and $Y$, and similarly for other quantities. 
Logarithms are taken to the natural base.
The probability of an event $\mathcal{E}$ will be denoted by 
$\prob \{\cal{E}\}$, and the expectation operator will be denoted by $\mathbb{E}[\cdot]$. 
The indicator function of an event $\calE$ 
will be denoted by $\IND\{\calE\}$. 
The notation $[t]_{+}$ will stand for $\max \{0,t\}$.

For two positive sequences, $\{a_{n}\}$ and $\{b_{n}\}$, the notation $a_{n} \doteq b_{n}$ will stand for equality in the exponential scale, that is, $\lim_{n \to \infty} (1/n) \log \left(a_{n}/b_{n}\right) = 0$. Similarly, $a_{n} \lexe b_{n}$ means that $\limsup_{n \to \infty} (1/n) \log \left(a_{n}/b_{n}\right) \leq 0$, and so on.
Accordingly, the notation $a_{n} \doteq e^{-n \infty}$ means that $a_{n}$ decays at a super--exponential rate (e.g.\ double--exponentially).

By the same token, for two positive sequences, $\{a_{n}\}$ and $\{b_{n}\}$, the notation $a_{n} \DEXE b_{n}$ will stand for equality in the double--exponential scale, that is, 
\begin{align}
\lim_{n \to \infty} \frac{1}{n} \log \left(\frac{\log a_{n}}{\log b_{n}} \right) = 0.
\end{align}
Similarly, $a_{n} \DLEXE b_{n}$ means that 
\begin{align}
\liminf_{n \to \infty} \frac{1}{n} \log \left(\frac{\log a_{n}}{\log b_{n}} \right) \geq 0,
\end{align}
and $a_{n} \DGEXE b_{n}$ stands for 
\begin{align}
\limsup_{n \to \infty} \frac{1}{n} \log \left(\frac{\log a_{n}}{\log b_{n}} \right) \leq 0.
\end{align}

The empirical distribution of a sequence $\boldsymbol{x} \in \mathcal{X}^{n}$, which will 
be denoted by $\hat{P}_{\boldsymbol{x}}$, is the vector of 
relative frequencies, $\hat{P}_{\bx}(x)$, 
of each symbol $x \in \mathcal{X}$ in $\bx$.
The joint empirical distribution of a pair of sequences, denoted by $\hat{P}_{\bx \by}$, is similarly defined.
The type class of $Q_{X}$, denoted $\mathcal{T}(Q_{X})$, is the set of all vectors $\bx \in \calX^{n}$ with $\hat{P}_{\bx} = Q_{X}$. 
In the same spirit, the joint type class of $Q_{XY}$, denoted $\calT(Q_{XY})$, is the set of all pairs of sequences $(\bx,\by) \in \calX^{n} \times \calY^{n}$ with $\hat{P}_{\bx\by} = Q_{XY}$.

Throughout the paper, we will make a frequent use of the fact that 
\begin{align}
\label{SME}
\sum_{i=1}^{k_{n}} a_{n}(i) \doteq \max_{1 \leq i \leq k_{n}} a_{n}(i)
\end{align}
as long as $\{a_{n}(i)\}$ are positive and $k_{n} \doteq 1$. This exponential equivalence will be termed henceforth the {\it summation--maximization equivalence} (SME). The sequence $k_{n}$ will represent the number of joint types possible for a given block length $n$, which is polynomial in $n$.

\section{Problem Formulation}

Consider a DMC $W=\{W(y|x),~x \in \calX,~y \in \calY\}$, where $\calX$ and $\calY$ are the finite input and output alphabets, respectively. When the channel is fed with a sequence $\bx = (x_{1}, \dotsc , x_{n}) \in \calX^{n}$, it produces $\by = (y_{1}, \dotsc , y_{n}) \in \calY^{n}$ according to
\begin{align}
W(\by|\bx) = \prod_{t=1}^{n} W(y_{t}|x_{t}).
\end{align}
Let $\calC_{n}$ be a codebook, i.e., a collection $\{\bx_{0},\bx_{1}, \dotsc, \bx_{M-1}\}$ of $M=e^{nR}$ codewords, $n$ being the block--length and $R$ the coding rate in nats per channel use. When the transmitter wishes to convey a message $m \in \{0,1, \dotsc, M-1\}$, it feeds the channel with $\bx_{m}$.
We assume that messages are chosen with equal probability.
We consider the ensemble of constant composition codes: for a given distribution $Q_{X}$ over $\calX$, all vectors in $\calC_{n}$ are uniformly and independently drawn from the type class $\calT(Q_{X})$.
As in \cite{MERHAV_TYPICAL}, \cite{MERHAV2017}, we consider here the GLD, 
which is a stochastic decoder, that chooses the estimated message $\hat{m}$ according to the following posterior probability mass function, induced by the channel output $\by$:
\begin{align}
\label{StrongGLD}
\prob \left\{ \hat{M}=m \middle| \by \right\}  =
\frac{\exp \{n g( \hat{P}_{\bx_{m} \by } ) \}} {\sum_{m'=0}^{M-1}  
	\exp \{n g( \hat{P}_{\bx_{m'} \by } ) \} } ,
\end{align}
where $\hat{P}_{\bx_{m} \by }$ is the empirical distribution of $(\bx_{m}, \by)$, and $g(\cdot)$ is a given continuous, real--valued functional of this empirical distribution.
The GLD provides a unified framework which covers several important special cases, e.g., matched likelihood decoding, mismatched decoding, ML decoding, and universal decoding (similarly to the $\alpha$--decoders described in \cite{CKgraph}). 
In particular, we recover the ML decoder by choosing the decoding metric
\begin{align}
g(Q_{XY}) = \beta \sum_{x \in \calX} \sum_{y \in \calY} Q_{XY}(x,y) \log W(y|x),
\end{align}
and letting $\beta \to \infty$.
A more detailed discussion is given in \cite{MERHAV2017}.

The probability of error, associated with a given code $\calC_{n}$ and the GLD, is given by
\begin{align}
\label{PROBABILITYofErrorDEF}
P_{\mbox{\tiny e}}(\calC_{n})
=\frac{1}{M} \sum_{m=0}^{M-1} \sum_{m' \neq m} \sum_{\by \in \calY^{n}} W(\by|\bx_{m}) \cdot \frac{\exp\{n g(\hat{P}_{\bx_{m'}\by}) \}}{\sum_{\tilde{m}=0}^{M-1} \exp\{n g(\hat{P}_{\bx_{\tilde{m}}\by}) \}}.
\end{align}

For the constant composition ensemble, Merhav \cite{MERHAV_TYPICAL} has derived a single--letter expression for
\begin{align}
E_{\mbox{\tiny trc}}(R)
= \lim_{n \to \infty} \left\{- \tfrac{1}{n} \mathbb{E} \left[\log P_{\mbox{\tiny e}}(\calC_{n}) \right] \right\}.
\end{align}
In order to present this expression, we define first a few quantities. 
Define the set $\calQ(Q_{X}) = \{Q_{XX'}: Q_{X'}=Q_{X}\}$ and 
\begin{align}
\label{ALPHA_DEF}
\alpha(R,Q_{Y}) = \max_{Q_{\tilde{X}|Y} \in \calS(Q_{X},Q_{Y})} \{g(Q_{\tilde{X}Y}) - I_{Q}(\tilde{X};Y)\} + R,
\end{align}
where $\calS(Q_{X},Q_{Y})=\{Q_{\tilde{X}|Y}:~I_{Q}(\tilde{X};Y) \leq R,~ Q_{\tilde{X}}=Q_{X}\}$, as well as 
\begin{align}
\label{Gamma_DEF}
\Gamma(Q_{XX'},R) &= \min_{Q_{Y|XX'}} \{ D(Q_{Y|X} \| W |Q_{X}) + I_{Q}(X';Y|X) \nn \\
&~~+ [\max\{g(Q_{XY}), \alpha(R,Q_{Y})\} - g(Q_{X'Y})]_{+} \},
\end{align}
where $D(Q_{Y|X} \| W |Q_{X})$ is the conditional divergence between $Q_{Y|X}$ and $W$, averaged by $Q_{X}$:
\begin{equation}
D(Q_{Y|X} || W | Q_{X}) = \sum_{x \in \calX} Q_{X}(x) 
\sum_{y \in \calY} Q_{Y|X}(y|x) \log \frac{Q_{Y|X}(y|x)}{W(y|x)}. 
\end{equation}
The TRC error exponent is given by \cite{MERHAV_TYPICAL}
\begin{align} \label{TRCexponent}
E_{\mbox{\tiny trc}}(R)
= \min_{\{\calQ(Q_{X}):~I_{Q}(X;X') \leq 2R\}} \{\Gamma(Q_{XX'},R) + I_{Q}(X;X') - R\} . 
\end{align}
In the sequel, we prove that the exponent $E_{\mbox{\tiny trc}}(R)$ is the exact value around which the random variable $\RE$ concentrates, as was partially implied from the proof in \cite[Subsection 5.2]{MERHAV_TYPICAL}. 
The expurgated exponent $E_{\mbox{\tiny ex}}(R)$, proved in \cite{MERHAV2017}, has exactly the same expression, 
but with the minimization constraint in (\ref{TRCexponent}) $I_{Q}(X;X') \leq 2R$ replaced by $I_{Q}(X;X') \leq R$.
In case of ML decoding, define
\begin{align}
a(R,Q_{Y}) = \max_{Q_{\tilde{X}|Y} \in \calS(Q_{X},Q_{Y})} \mathbb{E}_{Q} [\log W(Y|\tilde{X})]
\end{align}
and the set
\begin{align}
\calA(R) &= \{Q_{X'Y|X}:~ I_{Q}(X;X') \leq 2R,~Q_{X'}=Q_{X}, \nn \\
&~~~~~ \mathbb{E}_{Q}[\log W(Y|X')] \geq \max \left\{\mathbb{E}_{Q}[\log W(Y|X)], a(R,Q_{Y}) \right\} \}.
\end{align}
Then, \eqref{TRCexponent} particularizes to \cite[Sec.\ 4]{MERHAV_TYPICAL}
\begin{align} \label{TRCexponentML}
E_{\mbox{\tiny trc}}^{\mbox{\tiny ML}}(R)
= \min_{Q_{X'Y|X} \in \calA(R)} \{D(Q_{Y|X} \| W |Q_{X}) + I_{Q}(X,Y;X') - R\} .
\end{align}

We are interested in the lower and the upper tails of the distribution of $\RE$. 
The first is
\begin{align}
\label{BAD_probability}
\prob \left\{ \RE \leq \expVAR \right\},~~\expVAR < E_{\mbox{\tiny trc}}(R),
\end{align}
which is the probability of drawing a {\it bad} codebook. The second one is
\begin{align}
\label{GOOD_probability}
\prob \left\{ \RE \geq \expVAR \right\},~~\expVAR > E_{\mbox{\tiny trc}}(R),
\end{align}
which is the probability of drawing a {\it good} codebook. 
Finding exact expressions for (\ref{BAD_probability}) and (\ref{GOOD_probability}) appears to be difficult. We derive lower and upper bounds on both (\ref{BAD_probability}) and (\ref{GOOD_probability}).

\section {Main Results}

\subsection {The Lower Tail}

In order to present the error exponents of the lower tail, we define the quantities:
\begin{align}
\label{Beta_DEF}
&\beta(R,Q_{Y}) = \max_{\{Q_{\tilde{X}|Y}:~ Q_{\tilde{X}}=Q_{X}\}} \{g(Q_{\tilde{X}Y}) + [R - I_{Q}(\tilde{X};Y)]_{+}\} ,  \\
\label{LAMBDA_DEF}
&\Lambda(Q_{XX'},R) = \min_{Q_{Y|XX'}} \{ D(Q_{Y|X} \| W |Q_{X}) + I_{Q}(X';Y|X) + \beta(R,Q_{Y}) - g(Q_{X'Y}) \}, 
\end{align}
and,
\begin{align}
&\Psi(R,\expVAR,Q_{XX'}) = \Gamma(Q_{XX'},R) + R - \expVAR, \\
\label{XI_DEF}
&\Xi(R,\expVAR,Q_{XX'}) = \Lambda(Q_{XX'},R) + R - \expVAR.
\end{align}
Also, define the sets
\begin{align}
\label{L_DEF}
\calL(R,\expVAR) &= \{Q_{XX'} \in \calQ(Q_{X}): ~  [2R-I_{Q}(X;X')]_{+}  \geq \Psi(R,\expVAR,Q_{XX'}) \}, \\
\label{M_DEF}
\calM(R,\expVAR) &= \{Q_{XX'} \in \calQ(Q_{X}): ~  [2R-I_{Q}(X;X')]_{+}  \geq \Xi(R,\expVAR,Q_{XX'}) \},
\end{align}
and the error exponent functions
\begin{alignat}{2}
\label{LT_UB_EXPONENT}
E_{\mbox{\tiny lt}}^{\mbox{\tiny ub}}(R,\expVAR) 
&= \min_{Q_{XX'} \in \calL(R,\expVAR)}  &&[I_{Q}(X;X')-2R]_{+}, \\
\label{LT_LB_EXPONENT} 
E_{\mbox{\tiny lt}}^{\mbox{\tiny lb}}(R,\expVAR) 
&= \min_{Q_{XX'} \in \calM(R,\expVAR)}  &&[I_{Q}(X;X')-2R]_{+}.
\end{alignat}
Our first result in this section is the following theorem, which is proved in Section \ref{SEC_V}.

\begin{theorem} \label{THEOREM_LOWER_TAIL}
	Consider the ensemble of random constant composition codes $\calC_{n}$ of rate $R$ and composition $Q_{X}$. Then,
	\begin{align}
	\prob \left\{ \RE \leq \expVAR \right\} \lexe \exp \{-n \cdot E_{\mbox{\tiny lt}}^{\mbox{\tiny ub}}(R,\expVAR)\}.
	\end{align}
	Also,
	\begin{align}
	\prob \left\{ \RE \leq \expVAR \right\} \gexe \exp \{-n \cdot E_{\mbox{\tiny lt}}^{\mbox{\tiny lb}}(R,\expVAR)\}.
	\end{align}
\end{theorem}

An expression for the special case of ML decoding can be derived, but turns out to be relatively cumbersome, since it consists of a nested optimization problem. Instead, let us recall the result of \cite{LCV2017} (see also \cite{Mondelli2016}), which asserts that the probability of error for ordinary likelihood decoding (\cite[Eq.\ (3)]{MERHAV2017}) is at most twice the error probability of ML decoding. Hence, it is enough to use the decoding metric $g(Q) = \mathbb{E}_{Q} [\log W(Y|X)]$ (here and in all of the results later on) in order to study the LD rate functions under the ML decoder. For example, \eqref{ALPHA_DEF} particularizes to
\begin{align}
\alpha(R,Q_{Y}) = \max_{Q_{\tilde{X}|Y} \in \calS(Q_{X},Q_{Y})} \{\mathbb{E}_{Q} [\log W(Y|\tilde{X})] - I_{Q}(\tilde{X};Y)\} + R,
\end{align}      
and similarly for $\Gamma(Q_{XX'},R)$, $\beta(R,Q_{Y})$, and $\Lambda(Q_{XX'},R)$.
 
In order to characterize the behavior of the error exponent functions (\ref{LT_UB_EXPONENT}) and (\ref{LT_LB_EXPONENT}), let us first define
\begin{equation} 
\tilde{E}(R) = \min_{\{\calQ(Q_{X}):~I_{Q}(X;X') \leq 2R\}} \{\Lambda(Q_{XX'},R) + I_{Q}(X;X') - R\} .
\end{equation}
The following proposition is proved in Appendix D.

\begin{proposition} \label{LT_Properties}
	$E_{\mbox{\tiny lt}}^{\mbox{\tiny ub}}(R,\expVAR)$ and $E_{\mbox{\tiny lt}}^{\mbox{\tiny lb}}(R,\expVAR)$ have the following properties:
	\begin{enumerate}
		\item For fixed $R$, 
		$E_{\mbox{\tiny lt}}^{\mbox{\tiny ub}}(R,\expVAR)$ and $E_{\mbox{\tiny lt}}^{\mbox{\tiny lb}}(R,\expVAR)$ 
		are decreasing in $\expVAR$.
		\item $E_{\mbox{\tiny lt}}^{\mbox{\tiny ub}}(R,\expVAR) > 0$ if and only if $\expVAR < E_{\mbox{\tiny trc}}(R)$.
		\item $E_{\mbox{\tiny lt}}^{\mbox{\tiny lb}}(R,\expVAR) > 0$ if and only if $\expVAR < \tilde{E}(R)$.
		\item $E_{\mbox{\tiny lt}}^{\mbox{\tiny ub}}(R,\expVAR) = \infty$ for any $\expVAR < \expVAR^{\mbox{\tiny min}}(R)$, where
		\begin{align} \label{E_0}
		\expVAR^{\mbox{\tiny min}}(R) = 
		 \min_{\calQ(Q_{X})} \{\Gamma(Q_{XX'},R) - [2R - I_{Q}(X;X')]_{+}\} +R. 
		\end{align}
	\end{enumerate}
\end{proposition}
Note that $\tilde{E}(R)$ is defined similarly as $E_{\mbox{\tiny trc}}(R)$, with $\Lambda(Q_{XX'},R)$ replacing $\Gamma(Q_{XX'},R)$. Generally, $\tilde{E}(R) \geq E_{\mbox{\tiny trc}}(R)$, but in some special cases, e.g.\ the $z$--channel and the BEC, it can be easily proved that $\tilde{E}(R) = E_{\mbox{\tiny trc}}(R)$, as can be seen in Figure \ref{Z-Channel-Numeric-UT} below.
Moreover, since $E_{\mbox{\tiny lt}}^{\mbox{\tiny ub}}(R,\expVAR)$ is defined similarly as $E_{\mbox{\tiny lt}}^{\mbox{\tiny lb}}(R,\expVAR)$, also with $\Lambda(Q_{XX'},R)$ replacing $\Gamma(Q_{XX'},R)$, it turns out that for the same special cases, $E_{\mbox{\tiny lt}}^{\mbox{\tiny ub}}(R,\expVAR) = E_{\mbox{\tiny lt}}^{\mbox{\tiny lb}}(R,\expVAR)$. Hence, we conclude that there exist channels for which $\prob \left\{ \RE \leq \expVAR \right\}$ has an exponentially tight expression.    

Proposition \ref{LT_Properties} answers the questions we raised in the Introduction. First, it asserts that drawing a codebook for which $\RE$ is strictly below the TRC exponent has an exponentially vanishing probability. 
This implies that only for a small fraction of constant composition codes, $\RE$ is significantly
lower than the TRC error exponent. Second, the probability that $\RE$ falls in the range $(\expVAR^{\mbox{\tiny min}}(R), E_{\mbox{\tiny trc}}(R))$ tends to zero with a finite exponent, but for $\expVAR \in [0,\expVAR^{\mbox{\tiny min}}(R))$, 
the probability of $\RE \leq \expVAR$ converges to zero faster than exponentially; these codebooks are extremely rare. 

\begin{figure}[ht!]
	\centering
	\begin{tikzpicture}[scale=1.2]
	\begin{axis}[
	disabledatascaling,
	scaled y ticks=false,
	yticklabel style={/pgf/number format/fixed,
		/pgf/number format/precision=3},
	xlabel={$R$},
	xmin=0, xmax=0.7,
	ymin=0, ymax=1.75,
	legend pos=north east,
	]
	
	\addplot[smooth,color=blue,thick]
	table[row sep=crcr] {
		0	0.662014394	\\
		0.01	0.652014394	\\
		0.02	0.642014394	\\
		0.03	0.632014394	\\
		0.04	0.622014394	\\
		0.05	0.612014394	\\
		0.06	0.602014394	\\
		0.07	0.592014394	\\
		0.08	0.582014394	\\
		0.09	0.572014394	\\
		0.1	0.562014394	\\
		0.11	0.552014394	\\
		0.12	0.542014394	\\
		0.13	0.532014394	\\
		0.14	0.522014394	\\
		0.15	0.512014394	\\
		0.16	0.502014394	\\
		0.17	0.492014394	\\
		0.18	0.482014394	\\
		0.19	0.472014394	\\
		0.2	0.462014394	\\
		0.21	0.452014394	\\
		0.22	0.442014394	\\
		0.23	0.432014394	\\
		0.24	0.422014394	\\
		0.25	0.412014394	\\
		0.26	0.402014394	\\
		0.27	0.392014394	\\
		0.28	0.382014394	\\
		0.29	0.372014394	\\
		0.3	0.362014394	\\
		0.31	0.352014394	\\
		0.32	0.342014394	\\
		0.33	0.332014394	\\
		0.34	0.322014394	\\
		0.35	0.312014394	\\
		0.36	0.302014394	\\
		0.37	0.292014394	\\
		0.38	0.282014394	\\
		0.39	0.272014394	\\
		0.4	0.262014394	\\
		0.41	0.252014394	\\
		0.42	0.242014394	\\
		0.43	0.232014394	\\
		0.44	0.222014394	\\
		0.45	0.212014394	\\
		0.46	0.202014394	\\
		0.47	0.192014394	\\
		0.48	0.182014394	\\
		0.49	0.172014394	\\
		0.5	0.162014394	\\
		0.51	0.152014394	\\
		0.52	1.42E-01	\\
		0.53	1.32E-01	\\
		0.54	1.22E-01	\\
		0.55	1.12E-01	\\
		0.56	1.02E-01	\\
		0.57	9.20E-02	\\
		0.58	8.20E-02	\\
		0.59	7.20E-02	\\
		0.6	6.20E-02	\\
		0.61	5.20E-02	\\
		0.62	4.20E-02	\\
		0.63	3.23E-02	\\
		0.64	2.37E-02	\\
		0.65	1.62E-02	\\
		0.66	9.83E-03	\\
		0.67	4.86E-03	\\
		0.68	1.38E-03	\\
		0.69	1.93E-05	\\
	};
	\legend{}
	\addlegendentry{$E_{\mbox{\tiny r}}(R)$}	
	
	\addplot[smooth,color=black!40!green,thick,dash pattern={on 4pt off 2pt}]
	table[row sep=crcr]{
		0	1.726943521	\\
		0.01	1.392906268	\\
		0.02	1.26197453	\\
		0.03	1.164910603	\\
		0.04	1.085766788	\\
		0.05	1.018285934	\\
		0.06	0.958581329	\\
		0.07	0.905611798	\\
		0.08	0.857106008	\\
		0.09	0.813228557	\\
		0.1	0.773260996	\\
		0.11	0.735781684	\\
		0.12	0.701440364	\\
		0.13	0.669328903	\\
		0.14	0.639352068	\\
		0.15	0.611413234	\\
		0.16	0.58541399	\\
		0.17	0.56097448	\\
		0.18	0.538309271	\\
		0.19	0.517072377	\\
		0.2	0.496998353	\\
		0.21	0.47849741	\\
		0.22	0.461280955	\\
		0.23	0.445149394	\\
		0.24	0.430251435	\\
		0.25	0.416443931	\\
		0.26	0.403896665	\\
		0.27	0.392406303	\\
		0.28	0.382014394	\\
		0.29	0.372014394	\\
		0.3	0.362014394	\\
		0.31	0.352014394	\\
		0.32	0.342014394	\\
		0.33	0.332014394	\\
		0.34	0.322014394	\\
		0.35	0.312014394	\\
		0.36	0.302014394	\\
		0.37	0.292014394	\\
		0.38	0.282014394	\\
		0.39	0.272014394	\\
		0.4	0.262014394	\\
		0.41	0.252014394	\\
		0.42	0.242014394	\\
		0.43	0.232014394	\\
		0.44	0.222014394	\\
		0.45	0.212014394	\\
		0.46	0.202014394	\\
		0.47	0.192014394	\\
		0.48	0.182014394	\\
		0.49	0.172014394	\\
		0.5	0.162014394	\\
		0.51	0.152014394	\\
		0.52	1.42E-01	\\
		0.53	1.32E-01	\\
		0.54	1.22E-01	\\
		0.55	1.12E-01	\\
		0.56	1.02E-01	\\
		0.57	9.20E-02	\\
		0.58	8.20E-02	\\
		0.59	7.20E-02	\\
		0.6	6.20E-02	\\
		0.61	5.20E-02	\\
		0.62	4.20E-02	\\
		0.63	3.23E-02	\\
		0.64	2.37E-02	\\
		0.65	1.62E-02	\\
		0.66	9.83E-03	\\
		0.67	4.86E-03	\\
		0.68	1.38E-03	\\
		0.69	1.93E-05	\\
	};
	\addlegendentry{$E_{\mbox{\tiny trc}}(R)$} 
	
	\addplot[smooth,color=black!40!red,thick,dash pattern={on 6pt off 3pt}]
	table[row sep=crcr]
	{
		0	1.72693882	\\
		0.01	1.483120752	\\
		0.02	1.382707411	\\
		0.03	1.306056164	\\
		0.04	1.241772054	\\
		0.05	1.185434583	\\
		0.06	1.134772327	\\
		0.07	1.08843505	\\
		0.08	1.04554066	\\
		0.09	1.005476263	\\
		0.1	0.967795241	\\
		0.11	0.932159505	\\
		0.12	0.89830632	\\
		0.13	0.866026934	\\
		0.14	0.835151682	\\
		0.15	0.805540345	\\
		0.16	0.777076011	\\
		0.17	0.74965991	\\
		0.18	0.723207106	\\
		0.19	0.697644791	\\
		0.2	0.672909257	\\
		0.21	0.648944924	\\
		0.22	0.625702976	\\
		0.23	0.603139506	\\
		0.24	0.581215951	\\
		0.25	0.559897542	\\
		0.26	0.539153206	\\
		0.27	0.518954485	\\
		0.28	0.499276195	\\
		0.29	0.480095008	\\
		0.3	0.461389872	\\
		0.31	0.443141489	\\
		0.32	0.42533221	\\
		0.33	0.407945941	\\
		0.34	0.390967816	\\
		0.35	0.374384248	\\
		0.36	0.358182739	\\
		0.37	0.342351627	\\
		0.38	0.326880439	\\
		0.39	0.311759337	\\
		0.4	0.296979293	\\
		0.41	0.282532128	\\
		0.42	0.268410221	\\
		0.43	0.254606728	\\
		0.44	0.24111518	\\
		0.45	0.22793005	\\
		0.46	0.215046029	\\
		0.47	0.202458626	\\
		0.48	0.190163716	\\
		0.49	0.178157836	\\
		0.5	0.166438057	\\
		0.51	0.155001927	\\
		0.52	1.44E-01	\\
		0.53	1.33E-01	\\
		0.54	1.22E-01	\\
		0.55	1.12E-01	\\
		0.56	1.02E-01	\\
		0.57	9.20E-02	\\
		0.58	8.20E-02	\\
		0.59	7.20E-02	\\
		0.6	6.20E-02	\\
		0.61	5.20E-02	\\
		0.62	4.20E-02	\\
		0.63	3.23E-02	\\
		0.64	2.36E-02	\\
		0.65	1.61E-02	\\
		0.66	9.77E-03	\\
		0.67	4.75E-03	\\
		0.68	1.30E-03	\\
		0.69	3.85E-10	\\
	};
	\addlegendentry{$E_{\mbox{\tiny ex}}(R)$}

	\addplot[smooth,color=black!20!orange,thick,dash pattern={on 4pt off 2pt on 1pt off 2pt}]
	table[row sep=crcr]
	{
		0	0.00E+00	\\
		0.01	0.01	\\
		0.02	0.02	\\
		0.03	0.03	\\
		0.04	0.04	\\
		0.05	0.05	\\
		0.06	0.06	\\
		0.07	0.07	\\
		0.08	0.08	\\
		0.09	0.09	\\
		0.1	0.1	\\
		0.11	0.11	\\
		0.12	0.12	\\
		0.13	0.13	\\
		0.14	0.14	\\
		0.15	0.15	\\
		0.16	0.16	\\
		0.17	0.17	\\
		0.18	0.18	\\
		0.19	0.19	\\
		0.2	0.2	\\
		0.21	0.21	\\
		0.22	0.22	\\
		0.23	0.23	\\
		0.24	0.24	\\
		0.25	0.25	\\
		0.26	0.26	\\
		0.27	0.27	\\
		0.28	0.28	\\
		0.29	0.29	\\
		0.3	0.3	\\
		0.31	0.31	\\
		0.32	0.32	\\
		0.33	0.33	\\
		0.34	0.322021571	\\
		0.35	0.312021571	\\
		0.36	0.302021571	\\
		0.37	0.292021571	\\
		0.38	0.282021571	\\
		0.39	0.272021571	\\
		0.4	0.262021571	\\
		0.41	0.252021571	\\
		0.42	0.242021571	\\
		0.43	0.232021571	\\
		0.44	0.222021571	\\
		0.45	0.212021571	\\
		0.46	0.202021571	\\
		0.47	0.192021571	\\
		0.48	0.182021571	\\
		0.49	0.172021571	\\
		0.5	0.162021571	\\
		0.51	0.152021571	\\
		0.52	0.142021571	\\
		0.53	0.132021571	\\
		0.54	0.122021571	\\
		0.55	0.112021571	\\
		0.56	0.102021571	\\
		0.57	0.092021571	\\
		0.58	0.082021571	\\
		0.59	0.072021571	\\
		0.6	0.062021571	\\
		0.61	0.052021571	\\
		0.62	0.042021571	\\
		0.63	0.032587431	\\
		0.64	0.024185721	\\
		0.65	0.01705461	\\
		0.66	0.010202132	\\
		0.67	0.005049971	\\
		0.68	0.002040104	\\
		0.69	0.000532714	\\
	};
	\addlegendentry{$\expVAR^{\mbox{\tiny min}}(R)$}
	\end{axis}
	
	\end{tikzpicture}
	\caption{Various exponents for the $z$--channel with crossover probability 0.001.}\label{Z-Channel-Numeric}
\end{figure}
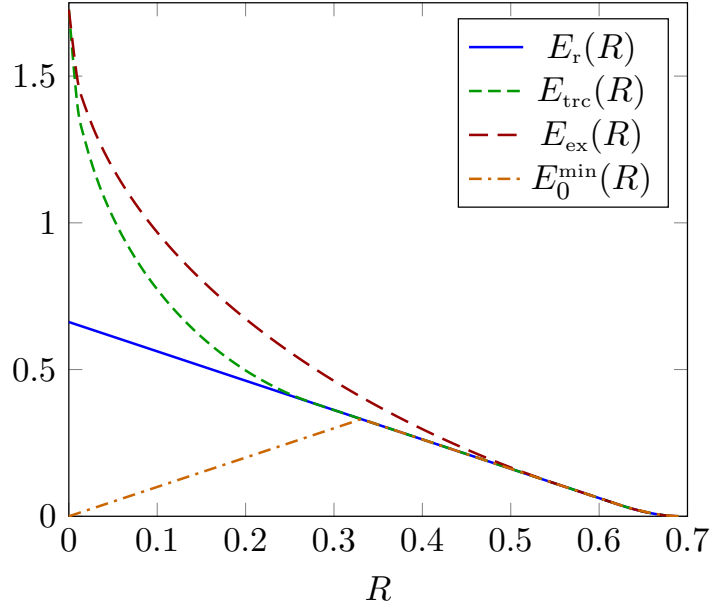

We next describe the behavior of $\expVAR^{\mbox{\tiny min}}(R)$. Denote by $Q_{XX'}^{*}(R)$ the minimizer of (\ref{E_0}) at rate $R$, and let $R^{*}$ be the maximal rate for which $2R \leq I_{Q^{*}(R)}(X;X')$ holds. 
On the one hand, for any $R \in [0,R^{*}]$, the operator $[\cdot]_{+}$ in (\ref{E_0}) is active and $\expVAR^{\mbox{\tiny min}}(R)$ is given by  
\begin{align} 
&\expVAR^{\mbox{\tiny min}}(R) =  \min_{\{\calQ(Q_{X}):~ 2R \leq I_{Q}(X;X')\}} \Gamma(Q_{XX'},R) +R ,
\end{align}
which is a monotonically increasing function. On the other hand, if $R \geq R^{*}$, the operator $[\cdot]_{+}$ in (\ref{E_0}) is neutral and $\expVAR^{\mbox{\tiny min}}(R)$ coincides with the TRC error exponent $E_{\mbox{\tiny trc}}(R)$. 
Figure \ref{Z-Channel-Numeric} illustrates the error exponents, as well as $\expVAR^{\mbox{\tiny min}}(R)$, for the binary $z$--channel with crossover parameter 0.001, the symmetric input distribution, $Q_{X} = (\tfrac{1}{2},\tfrac{1}{2})$, and the ML decoder.
The highest transmission rate is $R \cong 0.685$ [nats/channel use]. As can be seen in Figure \ref{Z-Channel-Numeric}, the exponent $E_{\mbox{\tiny trc}}(R)$ lies between $E_{\mbox{\tiny r}}(R)$ and  $E_{\mbox{\tiny ex}}(R)$, a fact that was already asserted for a general DMC in \cite{MERHAV_TYPICAL}. Moreover, $E_{\mbox{\tiny trc}}(R)$ is strictly higher than $E_{\mbox{\tiny r}}(R)$ for relatively low coding rates, and above $R \cong 0.279$ [nats/channel use], they coincide, i.e., the random coding error exponent provides the true exponential behavior of the typical codes in the ensemble. 
As for $\expVAR^{\mbox{\tiny min}}(R)$, we observe the following phenomena: First, note that $\expVAR^{\mbox{\tiny min}}(0)=0$, which means that all codebooks that have a sub--exponential number of codewords are drawn with a finite exponent. Second, in the range $(0,R^{*})$, $\expVAR^{\mbox{\tiny min}}(R)$ is linear and divides the range $[0,E_{\mbox{\tiny trc}}(R))$ into two intervals; in $(\expVAR^{\mbox{\tiny min}}(R), E_{\mbox{\tiny trc}}(R))$ -- an exponential decay with a finite exponent, and in $[0,\expVAR^{\mbox{\tiny min}}(R))$ -- a super--exponential decay. Third, for rates above $R^{*}$, the curves $\expVAR^{\mbox{\tiny min}}(R)$, $E_{\mbox{\tiny trc}}(R)$, and $E_{\mbox{\tiny r}}(R)$ are all equal. We conclude that for relatively high rates, $\prob \left\{ \RE < E_{\mbox{\tiny trc}}(R) \right\}$ converges to zero super--exponentially fast, a fact that was already proved in \cite[Theorem 5]{GoodCodes}.           

In order to gain some intuitive insight behind the various types of behavior of $E_{\mbox{\tiny lt}}^{\mbox{\tiny ub}}(R,\expVAR)$, it is instructive to examine the properties of the type class enumerators,
\begin{align} \label{NQ_def}
N(Q_{XX'}) \DEF \sum_{m=0}^{M-1} \sum_{m' \neq m} \IND \left\{(\bX_{m},\bX_{m'}) \in \calT(Q_{XX'}) \right\},
\end{align}
which play a pivotal role in the proofs of the main results of the paper. The summation (\ref{NQ_def}) contains $M(M-1) \doteq e^{n2R}$ terms. 
Borrowing from the terminology of binomial random variables, we refer to it as the {\it number of trials} associated with $N(Q_{XX'})$. The expectation of each binary random variable in (\ref{NQ_def}) is given by $\prob \left\{(\bX_{m},\bX_{m'}) \in \calT(Q_{XX'}) \right\} \doteq e^{-n I_{Q}(X;X')}$, which is referred to as the {\it success probability}. 
Unlike its one--dimensional counterpart \cite{SBM}--\cite{WM2017}, $N(Q_{XX'})$ is not a binomial random variable, since its terms are not mutually independent.  

We distinguish between two kinds of joint compositions.
On the one hand, we have the joint types $Q_{XX'}$ for which $I_{Q}(X;X') \leq 2R$, i.e., the exponential rate of the number of trials is higher than the negative exponential rate of the success probability. Thus, with overwhelmingly high probability, the respective $N(Q_{XX'})$ will concentrate around its mean, $\exp\{n(2R-I_{Q}(X;X'))\}$. 
Such compositions are referred to as {\it typically populated} (TP) type classes. 
On the other hand, for $Q_{XX'}$ with $I_{Q}(X;X') > 2R$, $N(Q_{XX'})=0$ with high probability. 
These compositions are referred to as the {\it typically empty} (TE) type classes.

\begin{figure}[h!]
	\definecolor{CODEWORD}{rgb}{0,0,0}
	\definecolor{FULL}{rgb}{0.13,0.65,0.13}
	\definecolor{EMPTY}{rgb}{1.0,0,0.22}	
	\begin{subfigure}[b]{0.5\columnwidth}
		\centering 
		\begin{tikzpicture}[scale=0.3]
		\def\radius{0.35} 
		\def\radus{0.2}
		\foreach \i in {0,6,...,360}{
			\fill[FULL] (5,5)+(\i:8) circle [radius=\radius];
		}
		\foreach \i in {0,7.5,...,360}{
			\fill[FULL] (5,5)+(\i:7) circle [radius=\radius];
		}
		\foreach \i in {0,9,...,360}{
			\fill[FULL] (5,5)+(\i:6) circle [radius=\radius];
		}
		\foreach \i in {0,12,...,360}{
			\fill[FULL] (5,5)+(\i:5) circle [radius=\radius];
		}
		\foreach \i in {0,15,...,360}{
			\draw[line width = 0.4mm,FULL] (5,5)+(\i:4) circle [radius=\radius];
			\fill[FULL] (5,5)+(\i:4) circle [radius=\radus];
		}
		\foreach \i in {0,20,...,360}{
			\draw[line width = 0.4mm,FULL] (5,5)+(\i:3) circle [radius=\radius];
			\fill[FULL] (5,5)+(\i:3) circle [radius=\radus];
		}
		\foreach \i in {0,30,...,360}{
			\draw[line width = 0.4mm,FULL] (5,5)+(\i:2) circle [radius=\radius];
			\fill[FULL] (5,5)+(\i:2) circle [radius=\radus];
		}
		\foreach \i in {0,60,...,360}{
			\draw[line width = 0.4mm,FULL] (5,5)+(\i:1) circle [radius=\radius];
			\fill[FULL] (5,5)+(\i:1) circle [radius=\radus];
		}
		\fill[CODEWORD] (5,5) circle [radius=0.3];
		\end{tikzpicture}
		\caption{For $\expVAR \leq \expVAR^{\mbox{\tiny min}}(R)$} 
		\label{fig:POPULATION1}
	\end{subfigure}%
	\begin{subfigure}[b]{0.5\columnwidth}
		\centering 
		\begin{tikzpicture}[scale=0.3]
		\def\radius{0.35} 
		\def\radus{0.2}
		\foreach \i in {0,6,...,360}{
			\fill[FULL] (5,5)+(\i:8) circle [radius=\radius];
		}
		\foreach \i in {0,7.5,...,360}{
			\fill[FULL] (5,5)+(\i:7) circle [radius=\radius];
		}
		\foreach \i in {0,9,...,360}{
			\fill[FULL] (5,5)+(\i:6) circle [radius=\radius];
		}
		\foreach \i in {0,12,...,360}{
			\fill[FULL] (5,5)+(\i:5) circle [radius=\radius];
		}
		\foreach \i in {0,15,...,360}{
			\draw[line width = 0.4mm,FULL] (5,5)+(\i:4) circle [radius=\radius];
			\fill[FULL] (5,5)+(\i:4) circle [radius=\radus];
		}
		\foreach \i in {0,20,...,360}{
			\draw[line width = 0.4mm,FULL] (5,5)+(\i:3) circle [radius=\radius];
			\fill[FULL] (5,5)+(\i:3) circle [radius=\radus];
		}
		\foreach \i in {0,30,...,360}{
			\draw[line width = 0.4mm,EMPTY] (5,5)+(\i:2) circle [radius=\radius];
			\fill[EMPTY] (5,5)+(\i:2) circle [radius=\radus];
		}
		\foreach \i in {0,60,...,360}{
			\draw[line width = 0.4mm,EMPTY] (5,5)+(\i:1) circle [radius=\radius];
			\fill[EMPTY] (5,5)+(\i:1) circle [radius=\radus];
		}
		\fill[CODEWORD] (5,5) circle [radius=0.3];
		\end{tikzpicture}
		\caption{$\expVAR \in (\expVAR^{\mbox{\tiny min}}(R), E_{\mbox{\tiny trc}}(R))$} 
		\label{fig:POPULATION2}
	\end{subfigure}%
	\\
	\centering
	\begin{subfigure}[b]{0.5\columnwidth}
		\centering 
		\begin{tikzpicture}[scale=0.3]
		\def\radius{0.35} 
		\def\radus{0.2}
		\foreach \i in {0,6,...,360}{
			\fill[FULL] (5,5)+(\i:8) circle [radius=\radius];
		}
		\foreach \i in {0,7.5,...,360}{
			\fill[FULL] (5,5)+(\i:7) circle [radius=\radius];
		}
		\foreach \i in {0,9,...,360}{
			\fill[FULL] (5,5)+(\i:6) circle [radius=\radius];
		}
		\foreach \i in {0,12,...,360}{
			\fill[FULL] (5,5)+(\i:5) circle [radius=\radius];
		}
		\foreach \i in {0,15,...,360}{
			\draw[line width = 0.4mm,EMPTY] (5,5)+(\i:4) circle [radius=\radius];
			\fill[EMPTY] (5,5)+(\i:4) circle [radius=\radus];
		}
		\foreach \i in {0,20,...,360}{
			\draw[line width = 0.4mm,EMPTY] (5,5)+(\i:3) circle [radius=\radius];
			\fill[EMPTY] (5,5)+(\i:3) circle [radius=\radus];
		}
		\foreach \i in {0,30,...,360}{
			\draw[line width = 0.4mm,EMPTY] (5,5)+(\i:2) circle [radius=\radius];
			\fill[EMPTY] (5,5)+(\i:2) circle [radius=\radus];
		}
		\foreach \i in {0,60,...,360}{
			\draw[line width = 0.4mm,EMPTY] (5,5)+(\i:1) circle [radius=\radius];
			\fill[EMPTY] (5,5)+(\i:1) circle [radius=\radus];
		}
		\fill[CODEWORD] (5,5) circle [radius=0.3];
		\end{tikzpicture}
		\caption{Around the $E_{\mbox{\tiny trc}}(R)$} 
		\label{fig:POPULATION3}
	\end{subfigure}%
	\begin{subfigure}[b]{0.5\columnwidth}
		\centering 
		\begin{tikzpicture}[scale=0.3]
		\def\radius{0.35}
		\def\radus{0.2} 
		\foreach \i in {0,6,...,360}{
			\fill[FULL] (5,5)+(\i:8) circle [radius=\radius];
		}
		\foreach \i in {0,7.5,...,360}{
			\fill[FULL] (5,5)+(\i:7) circle [radius=\radius];
		}
		\foreach \i in {0,9,...,360}{
			\fill[EMPTY] (5,5)+(\i:6) circle [radius=\radius];
		}
		\foreach \i in {0,12,...,360}{
			\fill[EMPTY] (5,5)+(\i:5) circle [radius=\radius];
		}
		\foreach \i in {0,15,...,360}{
			\draw[line width = 0.4mm,EMPTY] (5,5)+(\i:4) circle [radius=\radius];
			\fill[EMPTY] (5,5)+(\i:4) circle [radius=\radus];
		}
		\foreach \i in {0,20,...,360}{
			\draw[line width = 0.4mm,EMPTY] (5,5)+(\i:3) circle [radius=\radius];
			\fill[EMPTY] (5,5)+(\i:3) circle [radius=\radus];
		}
		\foreach \i in {0,30,...,360}{
			\draw[line width = 0.4mm,EMPTY] (5,5)+(\i:2) circle [radius=\radius];
			\fill[EMPTY] (5,5)+(\i:2) circle [radius=\radus];
		}
		\foreach \i in {0,60,...,360}{
			\draw[line width = 0.4mm,EMPTY] (5,5)+(\i:1) circle [radius=\radius];
			\fill[EMPTY] (5,5)+(\i:1) circle [radius=\radus];
		}
		\fill[CODEWORD] (5,5) circle [radius=0.3];
		\end{tikzpicture}
		\caption{$\expVAR \in (E_{\mbox{\tiny trc}}(R),E_{\mbox{\tiny ex}}(R))$} 
		\label{fig:POPULATION4}
	\end{subfigure}%
	\caption{Typical populations for different $\expVAR$ values. The center is the true codeword and each concentric circle around it represents a conditional type class.
		The radii of the concentric circles represent distances between codewords, which are measured by the empirical conditional entropy (also proportional to the negative empirical mutual information), induced by the joint composition of the codewords.		   
		Dots denote the TP type classes and circle--dots represent the TE type classes.
		TP type classes are the sets of relatively distant codewords; they include all joint compositions $Q_{XX'}$ with $I_{Q}(X;X') \leq 2R$.  
		Red dots/circle--dots mean empty type classes. For larger $\expVAR$ values, the minimum distance between codewords increases.} 
	\label{fig:POPULATIONS}
\end{figure}
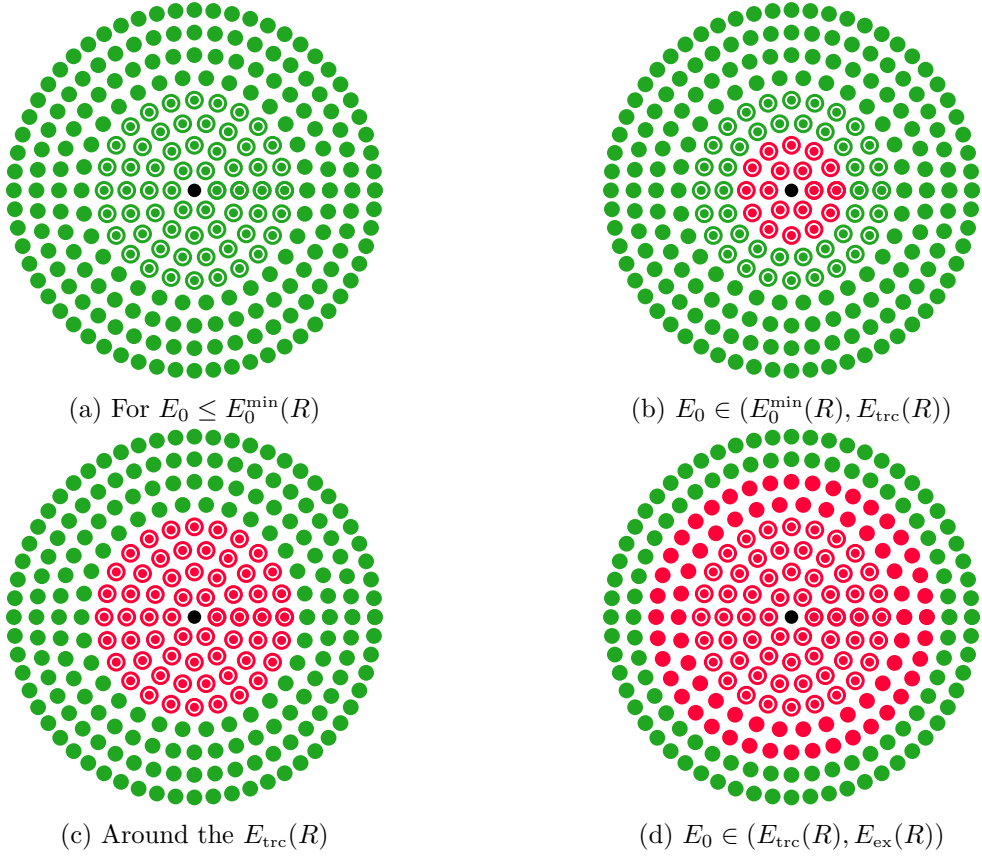

For $\expVAR \in (\expVAR^{\mbox{\tiny min}}(R), E_{\mbox{\tiny trc}}(R))$, let us denote the minimizer of $E_{\mbox{\tiny lt}}^{\mbox{\tiny ub}}(R,\expVAR)$ by $Q_{XX'}^{*}$.
Then, the dominant error event is due to pairs of codewords with joint empirical composition $Q_{XX'}^{*}$. In this range of exponents, all TP type classes are populated, as well as all TE type classes with $I_{Q}(X;X') \leq I_{Q^{*}}(X;X')$. The rest of the TE type classes, those with higher value of $I_{Q}(X;X')$, are still empty (see Figure \ref{fig:POPULATION2}). These are the joint type classes of the ``closest'' pairs of sequences in $\calX^{n}$, in the sense of high empirical mutual information. 

When $\expVAR = \expVAR^{\mbox{\tiny min}}(R)$, 
the constraint set $\calL(R,\expVAR)$ becomes empty, all TE type classes become populated (see Figure \ref{fig:POPULATION1}) and $E_{\mbox{\tiny lt}}^{\mbox{\tiny ub}}(R,\expVAR)$ jumps to infinity. In some sense, the curve $\expVAR^{\mbox{\tiny min}}(R)$ exhibits a {\it phase transition}. 
When $\expVAR > \expVAR^{\mbox{\tiny min}}(R)$, the minimum ``distance'' between pairs of codewords is still positive, but when $\expVAR \leq \expVAR^{\mbox{\tiny min}}(R)$, this minimum distance vanishes.

For $\expVAR < \expVAR^{\mbox{\tiny min}}(R)$, 
the super--exponential behavior of $\prob \left\{ \RE \leq \expVAR \right\}$ follows from the result of Lemma \ref{The_Upper_Tail_Thin} in Appendix B, which states that $\prob  \left\{ N(Q_{XX'}) \geq e^{n\epsilon } \right\}$ tends to zero faster than exponentially for any TE type class.
Now, if all TE type classes are populated by exponentially many pairs, then codebooks with exponentially many identical codewords also exist in the range of these low exponents.
Consider the set $\calD_{n}=\{\calC_{n}\}$ of codebooks, 
such that in each one of them, every TE type class is populated by exponentially many pairs of codewords. 
Obviously, $\RE \leq \expVAR^{\mbox{\tiny min}}(R)$ for every $\calC_{n} \in \calD_{n}$, and it turns out that this set has, in fact, a double--exponentially small probability.
To see why this is true, consider the following upper bound, which only requires from some $e^{n\epsilon}$ codewords to be identical:        
\begin{align}
\prob \left\{\calC_{n} \in \calD_{n}\right\} 
&\leq \binom{e^{nR}}{e^{n\epsilon}} \cdot \left(\frac{1}{|\calT(Q_{X})|}\right)^{e^{n \epsilon}} \\
&\DEXE \binom{e^{nR}}{e^{n\epsilon}} \cdot \exp\left\{-n H_{Q}(X) e^{n \epsilon} \right\} .
\end{align}
The binomial coefficient is upper--bounded as
\begin{align}
\binom{e^{nR}}{e^{n\epsilon}}
\leq \exp\left\{n R e^{n \epsilon} \right\} ,
\end{align}
hence,
\begin{align}
\prob \left\{\calC_{n} \in \calD_{n}\right\} 
&\DLEXE \exp\left\{-n (H_{Q}(X) - R) e^{n \epsilon} \right\},
\end{align}
which decays double--exponentially fast, since $R < I_{Q}(X;Y) \leq H_{Q}(X)$.

At last, we prove that a concentration property holds:
\begin{proposition}
	$\RE$ concentrates around $E_{\mbox{\tiny trc}}(R)$ as $n \to \infty$.
\end{proposition}
\textbf{Proof:} On the one hand, it follows by Theorem \ref{THEOREM_LOWER_TAIL} and Proposition \ref{LT_Properties} that for every $\epsilon > 0$, $\prob \{\RE \leq E_{\mbox{\tiny trc}}(R) -\epsilon\} \to 0$, exponentially fast, as $n \to \infty$. On the other hand, the proof in \cite[Subsection 5.2]{MERHAV_TYPICAL} implies that for every $\epsilon > 0$, $\prob \{\RE \leq E_{\mbox{\tiny trc}}(R) +\epsilon\} \to 1$, also exponentially fast, as $n \to \infty$. Combining these two facts, it follows that $\RE$ concentrates at $E_{\mbox{\tiny trc}}(R)$.

\subsection {The Upper Tail} 
In this subsection, we study the probability $\prob \left\{ \RE \geq \expVAR \right\}$.
On the one hand, we are interested in lower--bounding the probability $\prob \left\{ \RE \geq \expVAR \right\}$, such that we can assure the existence of good codebooks. 
On the other hand, we would also like to provide a tight upper bound on this probability, in order to prove that above some critical exponent value, codebooks cease to exist. We begin with a few definitions. 
Let us define the sets
\begin{align}
\label{V_DEF}
\calV(R,\expVAR) 
&= \{Q_{XX'} \in \calQ(Q_{X}):~ I_{Q}(X;X') \leq 2R, \nonumber \\
&~~~~~  \Lambda(Q_{XX'},R) + I_{Q}(X;X') - R \leq \expVAR \},\\
\calU(R,\expVAR) 
&= \{Q_{XX'} \in \calQ(Q_{X}):~ I_{Q}(X;X') \leq 2R, \nonumber \\ 
\label{U_DEF}
&~~~~~  \Gamma(Q_{XX'},R) + I_{Q}(X;X') - R \leq \expVAR \},
\end{align}
and the error exponent functions
\begin{align} 
\label{UT_UB_EXPONENT}
&E_{\mbox{\tiny ut}}^{\mbox{\tiny ub}}(R,\expVAR) 
= \max_{Q_{XX'} \in \calV(R,\expVAR)}
\min\{2R - I_{Q}(X;X'), \expVAR - \Lambda(Q_{XX'},R) - I_{Q}(X;X') + R,R\}, \\
\label{UT_LB_EXPONENT}
&E_{\mbox{\tiny ut}}^{\mbox{\tiny lb}}(R,\expVAR) = \max_{Q_{XX'} \in \calU(R,\expVAR)} \{2R - I_{Q}(X;X')\}.
\end{align}

The main result in this subsection is the following theorem.
\begin{theorem} \label{THEOREM_UPPER_TAIL}
	Consider the ensemble of random constant composition codes $\calC_{n}$ of rate $R$ and composition $Q_{X}$. Then, 
	\begin{align} \label{THM2_UB}
	\prob \left\{ \RE \geq \expVAR \right\} \DLEXE \exp \left\{- \exp\left\{n \cdot  E_{\mbox{\tiny ut}}^{\mbox{\tiny ub}}(R,\expVAR) \right\} \right\} .
	\end{align}
	If $\expVAR \in (E_{\mbox{\tiny trc}}(R), E_{\mbox{\tiny ex}}(R))$, then 
	\begin{align} \label{THM2_LB}
	\prob \left\{ \RE \geq \expVAR \right\} \DGEXE \exp \left\{- \exp\left\{n \cdot E_{\mbox{\tiny ut}}^{\mbox{\tiny lb}}(R,\expVAR)\right\} \right\}.
	\end{align}
\end{theorem}
The proofs of \eqref{THM2_UB} and \eqref{THM2_LB} appear in Sections \ref{SEC_VI} and \ref{SEC_VII}, respectively. The double--exponential behavior indicates that the relative number of very good codebooks is extremely small. 

The restriction to $(E_{\mbox{\tiny trc}}(R), E_{\mbox{\tiny ex}}(R))$ in the lower bound of Theorem \ref{THEOREM_UPPER_TAIL} stems from the technical condition of \cite[Theorem 9]{Suen}, which is equivalent to the one found in the Lov\'asz local lemma \cite{ALON}.
If a large number of events are all independent and each has probability less than 1, then there is a positive probability that none of the events will occur. 
The Lov\'asz local lemma allows one to slightly relax the independence condition, as long as the events are only ``weakly'' dependent in some sense. 
More specifically, referring to the type class enumerator $N(Q_{XX'})$, it turns out that if $I_{Q}(X;X') > R$, then the binary random variables composing $N(Q_{XX'})$ are only weakly dependent, and the probability $\prob \{N(Q_{XX'})=0\}$, which appears in the derivation of the lower bound of Theorem \ref{THEOREM_UPPER_TAIL}, can be lower--bounded using the Lov\'asz local lemma by $\exp\{-\exp\{n(2R-I_{Q}(X;X'))\}\}$. 
Otherwise, when $I_{Q}(X;X') < R$, 
this probability is very small, but it cannot be lower--bounded by the Lov\'asz local lemma, since its condition is not met.   
In our setting, the condition of the local lemma is met, as long as the number of codewords is not too high, which results in an upper bound on $\expVAR$, given by $E_{\mbox{\tiny ex}}(R)$.

In order to characterize the behavior of the error exponent functions (\ref{UT_UB_EXPONENT}) and (\ref{UT_LB_EXPONENT}), we provide the following proposition, which is proved in Appendix E.
\begin{proposition} \label{UT_Properties}
	$E_{\mbox{\tiny ut}}^{\mbox{\tiny ub}}(R,\expVAR)$ and $E_{\mbox{\tiny ut}}^{\mbox{\tiny lb}}(R,\expVAR)$ have the following properties:
	\begin{enumerate}
		\item For fixed $R$, 
		$E_{\mbox{\tiny ut}}^{\mbox{\tiny ub}}(R,\expVAR)$ and $E_{\mbox{\tiny ut}}^{\mbox{\tiny lb}}(R,\expVAR)$ 
		are increasing in $\expVAR$.
		\item $E_{\mbox{\tiny ut}}^{\mbox{\tiny lb}}(R,\expVAR) > 0$ if and only if $\expVAR > E_{\mbox{\tiny trc}}(R)$.
		\item $E_{\mbox{\tiny ut}}^{\mbox{\tiny ub}}(R,\expVAR) > 0$ if and only if $\expVAR > \tilde{E}(R)$. 
	\end{enumerate}
\end{proposition}

Recall that for the typical code, i.e., any code with $\RE \approx E_{\mbox{\tiny trc}}(R)$, all TP type classes are populated and all TE type classes are empty (see Figure \ref{fig:POPULATION3}). Now, for any $\expVAR$ in the range $(E_{\mbox{\tiny trc}}(R), E_{\mbox{\tiny ex}}(R))$, all TE type classes are still empty, but now, also all TP type classes that are associated with the set $\calU(R,\expVAR)$ are also empty (see Figure \ref{fig:POPULATION4}). The dominant error event in these codebooks is caused by relatively distant pairs of codewords that have a joint composition $Q_{XX'}^{*}$, which is the maximizer of (\ref{UT_LB_EXPONENT}). We conclude that $E_{\mbox{\tiny trc}}(R)$ exhibits a phase transition in the $\expVAR$ axis. Below the $E_{\mbox{\tiny trc}}(R)$ curve, TE type classes become populated, and above it, TP type classes become empty.      

When $\expVAR$ reaches $E_{\mbox{\tiny ex}}(R)$, 
the set $\calU(R,E_{\mbox{\tiny ex}})$ is a subset of $\tilde{\calU}(R) = \{Q_{XX'}\in \calQ(Q_{X}):~ R< I_{Q}(X;X') \leq 2R\}$, and thus
\begin{align}
E_{\mbox{\tiny ut}}^{\mbox{\tiny lb}}(R,E_{\mbox{\tiny ex}}) 
&= \max_{\calU(R,E_{\mbox{\tiny ex}})} \{2R - I_{Q}(X;X')\} 
\leq \max_{\tilde{\calU}(R)} \{2R - I_{Q}(X;X')\} = R.
\end{align}
It means that the lower bound of Theorem \ref{THEOREM_UPPER_TAIL} is at least as high as the probability of any codebook in the ensemble, given by $\DEXE \exp\{-nH_{Q}(X)e^{nR}\}$, 
which implies the existence of codebooks with $\RE \approx E_{\mbox{\tiny ex}}(R)$. 
We have the following corollary, which is proved in Appendix F.
\begin{corollary} \label{COROLLARY_ONE}
	If $\expVAR < E_{\mbox{\tiny ex}}(R)$, then there exists at least one code with $\RE \geq \expVAR$.
\end{corollary}

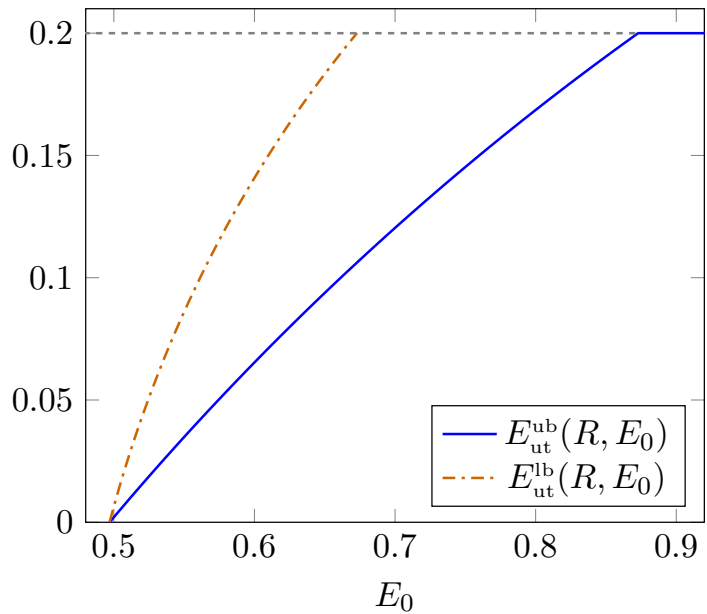
\begin{figure}[ht!]
	\centering
	\begin{tikzpicture}[scale=1.2]
	\begin{axis}[
	disabledatascaling,
	scaled y ticks=false,
	yticklabel style={/pgf/number format/fixed,
		/pgf/number format/precision=3},
	xlabel={$\expVAR$},
	xmin=0.48, xmax=0.92,
	ymin=0, ymax=0.21,
	legend pos=south east,
	]
	
	\addplot[smooth,color=blue,thick]
	table[row sep=crcr] 
	{
		0.497	1.35E-05	\\
		0.498	0.000697803	\\
		0.499	0.001380024	\\
		0.5	0.002061182	\\
		0.501	0.002742018	\\
		0.502	0.003422537	\\
		0.503	0.00410053	\\
		0.504	0.004777473	\\
		0.505	0.005453726	\\
		0.506	0.006130208	\\
		0.507	0.006804281	\\
		0.508	0.007477058	\\
		0.509	0.008148801	\\
		0.51	0.008821119	\\
		0.511	0.009491472	\\
		0.512	0.010160134	\\
		0.513	0.010827771	\\
		0.514	0.011495615	\\
		0.515	0.012162298	\\
		0.516	0.01282689	\\
		0.517	0.01349047	\\
		0.518	0.014153886	\\
		0.519	0.014816944	\\
		0.52	0.015477513	\\
		0.521	0.016137081	\\
		0.522	0.016796115	\\
		0.523	0.017455323	\\
		0.524	0.018112184	\\
		0.525	0.018767786	\\
		0.526	0.019422481	\\
		0.527	0.020077744	\\
		0.528	0.020731081	\\
		0.529	0.021382759	\\
		0.53	0.022033462	\\
		0.531	0.02268452	\\
		0.532	0.023334375	\\
		0.533	0.023982172	\\
		0.534	0.024629004	\\
		0.535	0.025275819	\\
		0.536	0.025922235	\\
		0.537	0.026566193	\\
		0.538	0.027209196	\\
		0.539	0.027851808	\\
		0.54	0.028494539	\\
		0.541	0.029134984	\\
		0.542	0.029774197	\\
		0.543	0.030412648	\\
		0.544	0.03105161	\\
		0.545	0.031688704	\\
		0.546	0.032324169	\\
		0.547	0.0329587	\\
		0.548	0.033593726	\\
		0.549	0.03422751	\\
		0.55	0.034859263	\\
		0.551	0.035490093	\\
		0.552	0.036121043	\\
		0.553	0.036751553	\\
		0.554	0.037379632	\\
		0.555	0.038006798	\\
		0.556	0.038633708	\\
		0.557	0.039260682	\\
		0.558	0.039885424	\\
		0.559	0.040508961	\\
		0.56	0.041131868	\\
		0.561	0.041755232	\\
		0.562	0.042376782	\\
		0.563	0.042996727	\\
		0.564	0.043615778	\\
		0.565	0.044235454	\\
		0.566	0.044853848	\\
		0.567	0.045470235	\\
		0.568	0.046085737	\\
		0.569	0.046701487	\\
		0.57	0.047316759	\\
		0.571	0.047929622	\\
		0.572	0.048541609	\\
		0.573	0.049153467	\\
		0.574	0.049765336	\\
		0.575	0.050375024	\\
		0.576	0.050983528	\\
		0.577	0.051591527	\\
		0.578	0.05219993	\\
		0.579	0.05280657	\\
		0.58	0.053411625	\\
		0.581	0.054015821	\\
		0.582	0.054620765	\\
		0.583	0.055224388	\\
		0.584	0.055826026	\\
		0.585	0.056426812	\\
		0.586	0.057027968	\\
		0.587	0.057628606	\\
		0.588	0.058226857	\\
		0.589	0.058824264	\\
		0.59	0.059421662	\\
		0.591	0.060019018	\\
		0.592	0.06061424	\\
		0.593	0.061208298	\\
		0.594	0.061801968	\\
		0.595	0.062395991	\\
		0.596	0.062988295	\\
		0.597	0.063579034	\\
		0.598	0.064169006	\\
		0.599	0.064759723	\\
		0.6	0.06534914	\\
		0.601	0.065936589	\\
		0.602	0.066523218	\\
		0.603	0.067110332	\\
		0.604	0.067696889	\\
		0.605	0.068281076	\\
		0.606	0.06886445	\\
		0.607	0.069447929	\\
		0.608	0.070031314	\\
		0.609	0.070612609	\\
		0.61	0.071192756	\\
		0.611	0.071772626	\\
		0.612	0.072352798	\\
		0.613	0.072931296	\\
		0.614	0.073508243	\\
		0.615	0.074084533	\\
		0.616	0.074661517	\\
		0.617	0.075237245	\\
		0.618	0.075811019	\\
		0.619	0.076384002	\\
		0.62	0.076957579	\\
		0.621	0.077530562	\\
		0.622	0.078101189	\\
		0.623	0.078671032	\\
		0.624	0.079241086	\\
		0.625	0.079810995	\\
		0.626	0.080378857	\\
		0.627	0.080945584	\\
		0.628	0.081512141	\\
		0.629	0.082078947	\\
		0.63	0.082644122	\\
		0.631	0.083207759	\\
		0.632	0.083770843	\\
		0.633	0.084334572	\\
		0.634	0.084897085	\\
		0.635	0.085457656	\\
		0.636	0.086017464	\\
		0.637	0.086577967	\\
		0.638	0.087137842	\\
		0.639	0.087695372	\\
		0.64	0.088252144	\\
		0.641	0.088809228	\\
		0.642	0.089366118	\\
		0.643	0.089921001	\\
		0.644	0.09047476	\\
		0.645	0.09102845	\\
		0.646	0.091582339	\\
		0.647	0.092134636	\\
		0.648	0.092685407	\\
		0.649	0.093235724	\\
		0.65	0.093786635	\\
		0.651	0.09433637	\\
		0.652	0.094884174	\\
		0.653	0.095431239	\\
		0.654	0.095979097	\\
		0.655	0.096526292	\\
		0.656	0.097071152	\\
		0.657	0.097615278	\\
		0.658	0.098159814	\\
		0.659	0.098704106	\\
		0.66	0.099246428	\\
		0.661	0.099787637	\\
		0.662	0.100328871	\\
		0.663	0.100870257	\\
		0.664	0.101410088	\\
		0.665	0.101948402	\\
		0.666	0.102486356	\\
		0.667	0.103024857	\\
		0.668	0.103562218	\\
		0.669	0.104097657	\\
		0.67	0.104632379	\\
		0.671	0.105167988	\\
		0.672	0.105702899	\\
		0.673	0.106235484	\\
		0.674	0.106767357	\\
		0.675	0.107299733	\\
		0.676	0.107831816	\\
		0.677	0.108361966	\\
		0.678	0.108891011	\\
		0.679	0.109420172	\\
		0.68	0.109949436	\\
		0.681	0.110477182	\\
		0.682	0.111003418	\\
		0.683	0.111529385	\\
		0.684	0.11205585	\\
		0.685	0.112581212	\\
		0.686	0.113104658	\\
		0.687	0.11362745	\\
		0.688	0.114151136	\\
		0.689	0.114674128	\\
		0.69	0.115194808	\\
		0.691	0.115714793	\\
		0.692	0.116235368	\\
		0.693	0.116755604	\\
		0.694	0.117273942	\\
		0.695	0.117791181	\\
		0.696	0.118308623	\\
		0.697	0.118826122	\\
		0.698	0.119342137	\\
		0.699	0.119856647	\\
		0.7	0.120370974	\\
		0.701	0.120885754	\\
		0.702	0.121399464	\\
		0.703	0.121911264	\\
		0.704	0.122422494	\\
		0.705	0.122934573	\\
		0.706	0.123445977	\\
		0.707	0.123955103	\\
		0.708	0.124463539	\\
		0.709	0.12497265	\\
		0.71	0.125481375	\\
		0.711	0.125988236	\\
		0.712	0.126494001	\\
		0.713	0.127000054	\\
		0.714	0.127506118	\\
		0.715	0.12801073	\\
		0.716	0.128513843	\\
		0.717	0.129016856	\\
		0.718	0.129520275	\\
		0.719	0.130022656	\\
		0.72	0.130523132	\\
		0.721	0.131023121	\\
		0.722	0.131523912	\\
		0.723	0.132024048	\\
		0.724	0.132521936	\\
		0.725	0.133019139	\\
		0.726	0.133517098	\\
		0.727	0.134014626	\\
		0.728	0.13451032	\\
		0.729	0.135004924	\\
		0.73	0.135499896	\\
		0.731	0.135994833	\\
		0.732	0.13648835	\\
		0.733	0.136980371	\\
		0.734	0.137472371	\\
		0.735	0.137964732	\\
		0.736	0.138456088	\\
		0.737	0.138945542	\\
		0.738	0.139434586	\\
		0.739	0.139924389	\\
		0.74	0.140413552	\\
		0.741	0.1409005	\\
		0.742	0.141386765	\\
		0.743	0.141873863	\\
		0.744	0.142360486	\\
		0.745	0.142845307	\\
		0.746	0.143329039	\\
		0.747	0.143813216	\\
		0.748	0.144297314	\\
		0.749	0.144780024	\\
		0.75	0.145261238	\\
		0.751	0.145742509	\\
		0.752	0.146224097	\\
		0.753	0.14670471	\\
		0.754	0.147183422	\\
		0.755	0.147661801	\\
		0.756	0.148140893	\\
		0.757	0.148619363	\\
		0.758	0.149095648	\\
		0.759	0.14957125	\\
		0.76	0.150047761	\\
		0.761	0.150523754	\\
		0.762	0.150997974	\\
		0.763	0.151471106	\\
		0.764	0.151944758	\\
		0.765	0.152418287	\\
		0.766	0.152890457	\\
		0.767	0.153361134	\\
		0.768	0.15383194	\\
		0.769	0.15430302	\\
		0.77	0.154773154	\\
		0.771	0.155241388	\\
		0.772	0.155709363	\\
		0.773	0.156178007	\\
		0.774	0.156646044	\\
		0.775	0.157111925	\\
		0.776	0.157577124	\\
		0.777	0.158043304	\\
		0.778	0.158508923	\\
		0.779	0.158972798	\\
		0.78	0.159435586	\\
		0.781	0.159898964	\\
		0.782	0.160362177	\\
		0.783	0.16082406	\\
		0.784	0.16128445	\\
		0.785	0.16174504	\\
		0.786	0.162205861	\\
		0.787	0.162665766	\\
		0.788	0.16312377	\\
		0.789	0.163581584	\\
		0.79	0.164040027	\\
		0.791	0.164497876	\\
		0.792	0.164953596	\\
		0.793	0.165408648	\\
		0.794	0.165864725	\\
		0.795	0.166320211	\\
		0.796	0.166773981	\\
		0.797	0.167226664	\\
		0.798	0.167680006	\\
		0.799	0.168133142	\\
		0.8	0.168584975	\\
		0.801	0.169035313	\\
		0.802	0.16948592	\\
		0.803	0.169936718	\\
		0.804	0.170386626	\\
		0.805	0.170834632	\\
		0.806	0.171282517	\\
		0.807	0.171730989	\\
		0.808	0.17217888	\\
		0.809	0.17262467	\\
		0.81	0.173069845	\\
		0.811	0.173516003	\\
		0.812	0.173961583	\\
		0.813	0.174405475	\\
		0.814	0.174848278	\\
		0.815	0.175291807	\\
		0.816	0.175735088	\\
		0.817	0.176177094	\\
		0.818	0.176617604	\\
		0.819	0.177058449	\\
		0.82	0.177499443	\\
		0.821	0.177939574	\\
		0.822	0.178377802	\\
		0.823	0.178815974	\\
		0.824	0.179254693	\\
		0.825	0.179692843	\\
		0.826	0.180128919	\\
		0.827	0.180564429	\\
		0.828	0.181000883	\\
		0.829	0.181436772	\\
		0.83	0.181870999	\\
		0.831	0.182304134	\\
		0.832	0.18273806	\\
		0.833	0.183171698	\\
		0.834	0.183604087	\\
		0.835	0.184034978	\\
		0.836	0.184466267	\\
		0.837	0.184897666	\\
		0.838	0.185328228	\\
		0.839	0.185756884	\\
		0.84	0.186185548	\\
		0.841	0.186614718	\\
		0.842	0.187043332	\\
		0.843	0.187469898	\\
		0.844	0.187895946	\\
		0.845	0.188322899	\\
		0.846	0.188749297	\\
		0.847	0.18917406	\\
		0.848	0.189597728	\\
		0.849	0.190022249	\\
		0.85	0.190446443	\\
		0.851	0.190869414	\\
		0.852	0.191290883	\\
		0.853	0.191712812	\\
		0.854	0.192134812	\\
		0.855	0.192556001	\\
		0.856	0.192975281	\\
		0.857	0.193394629	\\
		0.858	0.193814445	\\
		0.859	0.194233715	\\
		0.86	0.194650962	\\
		0.861	0.19506774	\\
		0.862	0.195485382	\\
		0.863	0.195902481	\\
		0.864	0.19631797	\\
		0.865	0.19673236	\\
		0.866	0.197147664	\\
		0.867	0.197562602	\\
		0.868	0.197976341	\\
		0.869	0.198388576	\\
		0.87	0.198801331	\\
		0.871	0.199214117	\\
		0.872	0.199626117	\\
		0.873	0.2	\\
		0.874	0.2	\\
		0.875	0.2	\\
		0.876	0.2	\\
		0.877	0.2	\\
		0.878	0.2	\\
		0.879	0.2	\\
		0.88	0.2	\\
		0.881	0.2	\\
		0.882	0.2	\\
		0.883	0.2	\\
		0.884	0.2	\\
		0.885	0.2	\\
		0.886	0.2	\\
		0.887	0.2	\\
		0.888	0.2	\\
		0.889	0.2	\\
		0.89	0.2	\\
		0.891	0.2	\\
		0.892	0.2	\\
		0.893	0.2	\\
		0.894	0.2	\\
		0.895	0.2	\\
		0.896	0.2	\\
		0.897	0.2	\\
		0.898	0.2	\\
		0.899	0.2	\\
		0.9	0.2	\\
		0.901	0.2	\\
		0.902	0.2	\\
		0.903	0.2	\\
		0.904	0.2	\\
		0.905	0.2	\\
		0.906	0.2	\\
		0.907	0.2	\\
		0.908	0.2	\\
		0.909	0.2	\\
		0.91	0.2	\\
		0.911	0.2	\\
		0.912	0.2	\\
		0.913	0.2	\\
		0.914	0.2	\\
		0.915	0.2	\\
		0.916	0.2	\\
		0.917	0.2	\\
		0.918	0.2	\\
		0.919	0.2	\\
		0.92	0.2	\\
		0.921	0.2	\\
		0.922	0.2	\\
		0.923	0.2	\\
		0.924	0.2	\\
		0.925	0.2	\\
		0.926	0.2	\\
		0.927	0.2	\\
		0.928	0.2	\\
		0.929	0.2	\\
		0.93	0.2	\\
		0.931	0.2	\\
		0.932	0.2	\\
		0.933	0.2	\\
		0.934	0.2	\\
		0.935	0.2	\\
		0.936	0.2	\\
		0.937	0.2	\\
		0.938	0.2	\\
		0.939	0.2	\\
		0.94	0.2	\\
		0.941	0.2	\\
		0.942	0.2	\\
		0.943	0.2	\\
		0.944	0.2	\\
		0.945	0.2	\\
		0.946	0.2	\\
		0.947	0.2	\\
		0.948	0.2	\\
		0.949	0.2	\\
		0.95	0.2	\\
		0.951	0.2	\\
		0.952	0.2	\\
		0.953	0.2	\\
		0.954	0.2	\\
		0.955	0.2	\\
		0.956	0.2	\\
		0.957	0.2	\\
		0.958	0.2	\\
		0.959	0.2	\\
		0.96	0.2	\\
		0.961	0.2	\\
		0.962	0.2	\\
		0.963	0.2	\\
		0.964	0.2	\\
		0.965	0.2	\\
		0.966	0.2	\\
		0.967	0.2	\\
		0.968	0.2	\\
		0.969	0.2	\\
		0.97	0.2	\\
		0.971	0.2	\\
		0.972	0.2	\\
		0.973	0.2	\\
		0.974	0.2	\\
		0.975	0.2	\\
		0.976	0.2	\\
		0.977	0.2	\\
		0.978	0.2	\\
		0.979	0.2	\\
		0.98	0.2	\\
		0.981	0.2	\\
		0.982	0.2	\\
		0.983	0.2	\\
		0.984	0.2	\\
		0.985	0.2	\\
		0.986	0.2	\\
		0.987	0.2	\\
		0.988	0.2	\\
		0.989	0.2	\\
		0.99	0.2	\\
		0.991	0.2	\\
		0.992	0.2	\\
		0.993	0.2	\\
		0.994	0.2	\\
		0.995	0.2	\\
		0.996	0.2	\\
		0.997	0.2	\\
		0.998	0.2	\\
		0.999	0.2	\\
		1	0.2	\\
	};
	\legend{}
	\addlegendentry{$E_{\mbox{\tiny ut}}^{\mbox{\tiny ub}}(R,\expVAR)$}	
	
	\addplot[smooth,color=black!20!orange,thick,dash pattern={on 4pt off 2pt on 1pt off 2pt}]
	table[row sep=crcr]
	{
		0.497	4.18E-05	\\
		0.498	0.002192015	\\
		0.499	0.00431301	\\
		0.5	0.006400509	\\
		0.501	0.008454936	\\
		0.502	0.01048132	\\
		0.503	0.012484595	\\
		0.504	0.01445588	\\
		0.505	0.016404667	\\
		0.506	0.018326678	\\
		0.507	0.020222221	\\
		0.508	0.022096091	\\
		0.509	0.023948526	\\
		0.51	0.025779757	\\
		0.511	0.027585573	\\
		0.512	0.029375091	\\
		0.513	0.031139669	\\
		0.514	0.032888343	\\
		0.515	0.034616903	\\
		0.516	0.036325551	\\
		0.517	0.038018824	\\
		0.518	0.03969255	\\
		0.519	0.041351222	\\
		0.52	0.0429907	\\
		0.521	0.044615436	\\
		0.522	0.046225566	\\
		0.523	0.047816988	\\
		0.524	0.049394099	\\
		0.525	0.050957029	\\
		0.526	0.052505903	\\
		0.527	0.054040846	\\
		0.528	0.055561979	\\
		0.529	0.057069424	\\
		0.53	0.058563298	\\
		0.531	0.060043718	\\
		0.532	0.061514901	\\
		0.533	0.062968741	\\
		0.534	0.06441354	\\
		0.535	0.065845306	\\
		0.536	0.067268194	\\
		0.537	0.068674202	\\
		0.538	0.070071517	\\
		0.539	0.071460203	\\
		0.54	0.072836328	\\
		0.541	0.074199993	\\
		0.542	0.075555267	\\
		0.543	0.076902209	\\
		0.544	0.078236937	\\
		0.545	0.079559545	\\
		0.546	0.080874047	\\
		0.547	0.0821805	\\
		0.548	0.083475065	\\
		0.549	0.084765597	\\
		0.55	0.086040508	\\
		0.551	0.087311493	\\
		0.552	0.088570879	\\
		0.553	0.089822589	\\
		0.554	0.091066673	\\
		0.555	0.092299372	\\
		0.556	0.09352457	\\
		0.557	0.094742317	\\
		0.558	0.095952661	\\
		0.559	0.097155652	\\
		0.56	0.098351338	\\
		0.561	0.099536022	\\
		0.562	0.100717249	\\
		0.563	0.101887591	\\
		0.564	0.103054546	\\
		0.565	0.104210728	\\
		0.566	0.105359904	\\
		0.567	0.106502117	\\
		0.568	0.107641081	\\
		0.569	0.108769491	\\
		0.57	0.109891069	\\
		0.571	0.111009496	\\
		0.572	0.112117528	\\
		0.573	0.113218855	\\
		0.574	0.114317125	\\
		0.575	0.115405155	\\
		0.576	0.11649019	\\
		0.577	0.117568664	\\
		0.578	0.118640615	\\
		0.579	0.119706083	\\
		0.58	0.120765108	\\
		0.581	0.121817728	\\
		0.582	0.122867511	\\
		0.583	0.123907427	\\
		0.584	0.124944563	\\
		0.585	0.125975426	\\
		0.586	0.127000054	\\
		0.587	0.128021967	\\
		0.588	0.129034226	\\
		0.589	0.130043823	\\
		0.59	0.131047313	\\
		0.591	0.132044729	\\
		0.592	0.133039545	\\
		0.593	0.134028341	\\
		0.594	0.135011151	\\
		0.595	0.13598801	\\
		0.596	0.136962354	\\
		0.597	0.137930797	\\
		0.598	0.138893373	\\
		0.599	0.139853491	\\
		0.6	0.140804424	\\
		0.601	0.141756307	\\
		0.602	0.14269907	\\
		0.603	0.143639455	\\
		0.604	0.144574135	\\
		0.605	0.145503143	\\
		0.606	0.146429826	\\
		0.607	0.147350884	\\
		0.608	0.148269646	\\
		0.609	0.149182829	\\
		0.61	0.150090462	\\
		0.611	0.150995852	\\
		0.612	0.151895737	\\
		0.613	0.152790147	\\
		0.614	0.153682366	\\
		0.615	0.154572398	\\
		0.616	0.155453776	\\
		0.617	0.156333011	\\
		0.618	0.157210108	\\
		0.619	0.158081861	\\
		0.62	0.158948297	\\
		0.621	0.159812644	\\
		0.622	0.160671716	\\
		0.623	0.161528724	\\
		0.624	0.1623805	\\
		0.625	0.163230237	\\
		0.626	0.164074781	\\
		0.627	0.164914161	\\
		0.628	0.165751547	\\
		0.629	0.166586946	\\
		0.63	0.167417231	\\
		0.631	0.168242431	\\
		0.632	0.169065687	\\
		0.633	0.169883896	\\
		0.634	0.170700185	\\
		0.635	0.171511465	\\
		0.636	0.172320848	\\
		0.637	0.17312834	\\
		0.638	0.173930872	\\
		0.639	0.174728471	\\
		0.64	0.175524219	\\
		0.641	0.176315071	\\
		0.642	0.177104095	\\
		0.643	0.177891295	\\
		0.644	0.178673647	\\
		0.645	0.179454197	\\
		0.646	0.180229934	\\
		0.647	0.181000883	\\
		0.648	0.181770071	\\
		0.649	0.182537499	\\
		0.65	0.183303174	\\
		0.651	0.184061136	\\
		0.652	0.184820353	\\
		0.653	0.185574873	\\
		0.654	0.186324719	\\
		0.655	0.187072871	\\
		0.656	0.187819331	\\
		0.657	0.188561162	\\
		0.658	0.189301322	\\
		0.659	0.190039813	\\
		0.66	0.19077372	\\
		0.661	0.191503064	\\
		0.662	0.192233684	\\
		0.663	0.192956861	\\
		0.664	0.193681321	\\
		0.665	0.194401271	\\
		0.666	0.195116733	\\
		0.667	0.195830606	\\
		0.668	0.196542893	\\
		0.669	0.197253596	\\
		0.67	0.197959863	\\
		0.671	0.198661716	\\
		0.672	0.199364863	\\
		0.673	0.200060778	\\
	};
	\addlegendentry{$E_{\mbox{\tiny ut}}^{\mbox{\tiny lb}}(R,\expVAR)$}
	
	\addplot[smooth,color=gray,thick,dash pattern={on 2pt off 2pt}]
	table[row sep=crcr] 
	{
		0	0.2	\\
		0.1	 0.2	\\
		0.2	 0.2	\\
		0.3	 0.2	\\
		0.4	 0.2	\\
		0.5	 0.2	\\
		0.6	 0.2	\\
		0.7	 0.2	\\
		0.8	 0.2	\\
		0.872	 0.2	\\
	};	
	
	\end{axis}
	
	\end{tikzpicture}
	\caption{Upper tail double--exponential rate functions for the $z$--channel with crossover probability $0.001$ and $R=0.2$.} \label{Z-Channel-Numeric-UT}
\end{figure}

Figure \ref{Z-Channel-Numeric-UT} illustrates the upper tail exponents (\ref{UT_UB_EXPONENT}) and (\ref{UT_LB_EXPONENT}) for the binary $z$--channel with crossover parameter 0.001, rate $R=0.2$, the symmetric input distribution, $Q_{X} = (\tfrac{1}{2},\tfrac{1}{2})$, and the ML decoder.
Due to the restriction in the lower bound of Theorem \ref{THEOREM_UPPER_TAIL}, note that $E_{\mbox{\tiny ut}}^{\mbox{\tiny lb}}(R,\expVAR)$ is applicable as long as $0 \leq E_{\mbox{\tiny ut}}^{\mbox{\tiny lb}}(R,\expVAR) \leq R$, while $E_{\mbox{\tiny ut}}^{\mbox{\tiny ub}}(R,\expVAR)$ is applicable for any $\expVAR$, but is truncated to $R$ for relatively high $\expVAR$. 
The lowest $\expVAR$ for which $E_{\mbox{\tiny ut}}^{\mbox{\tiny ub}}(R,\expVAR) = R$ is approximately $0.873$, which is strictly lower than the straight--line bound $E_{\mbox{\tiny sl}}(R) \approx 1.122$, but the truncation\footnote{We conjecture that this truncation to $R$ is artificial, and can be removed by deriving tighter LD bounds. More specifically, a tighter version of Fact 1 (Appendix A), which may lead to a tighter result in Lemma \ref{The_Lower_Tail} (Appendix B), which, in turn, may provide a tighter upper bound in Theorem \ref{THEOREM_UPPER_TAIL}} to $R$ prevents\footnote{Had the double--exponential rate of the upper bound strictly bigger than $R$, we were able to conclude the absentee of codebooks with error exponents above some threshold.} us from deducing a tighter upper bound to the reliability function.
In the entire range $(E_{\mbox{\tiny trc}}(R), E_{\mbox{\tiny ex}}(R))$, both $E_{\mbox{\tiny ut}}^{\mbox{\tiny lb}}(R,\expVAR)$ and $E_{\mbox{\tiny ut}}^{\mbox{\tiny ub}}(R,\expVAR)$ are strictly positive, such that the lower and the upper bounds on the probability of the upper tail are double--exponentially small. 

\section{Proof of Theorem \ref{THEOREM_LOWER_TAIL}}
\label{SEC_V}
\subsection{An Upper Bound on the Probability of the Lower Tail} \label{SEC_V1} 
Let $\calC_{n}$ be a constant composition code of rate $R$ and blocklength $n$ and let $\expVAR>0$ be given. Then,
\begin{align}
&\prob \left\{ -\frac{1}{n} \log P_{\mbox{\tiny e}} (\calC_{n}) \leq \expVAR \right\} \nonumber \\
&~~~~~~= \prob \left\{\frac{1}{M} \sum_{m=0}^{M-1} \sum_{m' \neq m} \sum_{\by \in \calY^{n}} W(\by|\bx_{m}) \cdot \frac{\exp\{n g(\hat{P}_{\bx_{m'}\by}) \}}{\sum_{\tilde{m}} \exp\{n g(\hat{P}_{\bx_{\tilde{m}}\by}) \}}  \geq e^{-n \cdot \expVAR} \right\} .
\end{align}
Let
\begin{align}
\label{Z_DEF}
Z_{m}(\by) = \sum_{\tilde{m} \neq m} \exp\{n g(\hat{P}_{\bx_{\tilde{m}}\by}) \},
\end{align}
fix $\epsilon>0$ arbitrarily small, and for every $\by \in \calY^{n}$, define the set
\begin{align}
\label{B_DEF}
\calB_{\epsilon}(m,\by) = \left\{\calC_{n}:~ Z_{m}(\by) \leq \exp\{n \alpha(R-\epsilon, \hat{P}_{\by})\}  \right\}.
\end{align}
Following the result of \cite[Appendix B]{MERHAV2017}, we know that, considering the ensemble of randomly selected constant composition codes of type $Q_{X}$,
\begin{align}
\label{B_DE_UB}
\prob \{\calB_{\epsilon}(m,\by)\} \leq \exp\{-e^{n\epsilon} + n\epsilon + 1\},
\end{align} 
for every $m \in \{0,1,\dotsc,M-1\}$ and $\by \in \calY^{n}$, and so, by the union bound,
\begin{align}
\label{B_UNION_DEF}
\prob \left\{\bigcup_{m=0}^{M-1} \bigcup_{\by \in \calY^{n}}\calB_{\epsilon}(m,\by)\right\}
\DEF  \prob \left\{\calB_{\epsilon}\right\} 
&\leq \sum_{m=0}^{M-1} \sum_{\by \in \calY^{n}} \prob \left\{\calB_{\epsilon}(m,\by)\right\} \\
&\leq \sum_{m=0}^{M-1} \sum_{\by \in \calY^{n}} \exp\{-e^{n\epsilon} + n\epsilon + 1\} \\
&= e^{nR} \cdot |\calY|^{n} \cdot \exp\{-e^{n\epsilon} + n\epsilon + 1\},
\end{align}
which still decays double--exponentially fast. Thus,
\begin{align}
&\prob \left\{ -\frac{1}{n} \log P_{\mbox{\tiny e}} (\calC_{n}) \leq \expVAR \right\} \nonumber \\
&= \prob \left\{\frac{1}{M} \sum_{m=0}^{M-1} \sum_{m' \neq m} \sum_{\by \in \calY^{n}} W(\by|\bx_{m}) \cdot \frac{\exp\{n g(\hat{P}_{\bx_{m'}\by}) \}}{\exp\{n g(\hat{P}_{\bx_{m}\by}) \} + Z_{m}(\by)}  \geq e^{-n \cdot \expVAR} \right\} \\
\label{EXP1}
&= \prob\left\{\calC_{n} \in \calB_{\epsilon}^{\mbox{\tiny c}}, \frac{1}{M} \sum_{m=0}^{M-1} \sum_{m' \neq m} \sum_{\by \in \calY^{n}} W(\by|\bx_{m}) \cdot \frac{\exp\{n g(\hat{P}_{\bx_{m'}\by}) \}}{\exp\{n g(\hat{P}_{\bx_{m}\by}) \} + Z_{m}(\by)}  \geq e^{-n \cdot \expVAR} \right\} \nonumber \\
&~~+ \prob\left\{\calC_{n} \in \calB_{\epsilon}, \frac{1}{M} \sum_{m=0}^{M-1} \sum_{m' \neq m} \sum_{\by \in \calY^{n}} W(\by|\bx_{m}) \cdot \frac{\exp\{n g(\hat{P}_{\bx_{m'}\by}) \}}{\exp\{n g(\hat{P}_{\bx_{m}\by}) \} + Z_{m}(\by)}  \geq e^{-n \cdot \expVAR} \right\} \\
\label{EXP2} 
&\leq \prob\left\{\calC_{n} \in \calB_{\epsilon}^{\mbox{\tiny c}}, \frac{1}{M} \sum_{m=0}^{M-1} \sum_{m' \neq m} \sum_{\by \in \calY^{n}} W(\by|\bx_{m}) \right. \nn \\
&\left. ~~~\times \min \left\{1, \frac{\exp\{n g(\hat{P}_{\bx_{m'}\by}) \}}{\exp\{n g(\hat{P}_{\bx_{m}\by}) \} + \exp\{n \alpha(R-\epsilon, \hat{P}_{\by})\}} \right\}  \geq e^{-n \cdot \expVAR} \right\} + \prob\{\calC_{n} \in \calB_{\epsilon}\} \\
\label{EXP3}
&\doteq \prob\left\{\calC_{n} \in \calB_{\epsilon}^{\mbox{\tiny c}}, \frac{1}{M} \sum_{m=0}^{M-1} \sum_{m' \neq m} \sum_{\by \in \calY^{n}} W(\by|\bx_{m}) \right. \nn \\
&\left. ~~~\times \exp\left\{-n \cdot [\max \{g(\hat{P}_{\bx_{m}\by}), \alpha(R-\epsilon, \hat{P}_{\by})\} - g(\hat{P}_{\bx_{m'}\by})]_{+} \right\}  \geq e^{-n \cdot \expVAR} \right\} +  \prob\{\calC_{n} \in \calB_{\epsilon}\} \\
\label{EXP4} 
&\doteq \prob\left\{\calC_{n} \in \calB_{\epsilon}^{\mbox{\tiny c}}, \frac{1}{M} \sum_{m=0}^{M-1} \sum_{m' \neq m} 
\exp\{-n \Gamma(\hat{P}_{\bx_{m}\bx_{m'}},R-\epsilon) \}  
\geq e^{-n \cdot \expVAR} \right\}  \\
\label{EXP5}
&\leq \prob\left\{\frac{1}{M} \sum_{m=0}^{M-1} \sum_{m' \neq m} \exp\{-n \Gamma(\hat{P}_{\bx_{m}\bx_{m'}},R-\epsilon) \}  
\geq e^{-n \cdot \expVAR} \right\},
\end{align}
where in \eqref{EXP2}, the inner terms in the first expression of (\ref{EXP1}) were upper--bounded according to (\ref{B_DEF}) as well as the trivial upper bound of one, and the indicators of the second summand were trivially upper--bounded by one.
In \eqref{EXP3}, we used the SME \eqref{SME}. 
In \eqref{EXP4}, the inner--most sum over $\by \in \calY^{n}$ was evaluated using the method of types, with the functional $\Gamma(Q_{XX'},R)$ defined in (\ref{Gamma_DEF}) (see \cite[Section 5]{MERHAV2017} for more details), and the fact that $\prob\{\calB_{\epsilon}\}$ is double--exponentially small was used. 
One of the difficulties in the statistical analysis of $N(Q_{XX'})$ \eqref{NQ_def} is that it is the sum of \textit{dependent}\footnote{This dependence can be demonstrated by the following extreme example. Let $Q_{X}$ be uniform over $\calX$ and let $Q_{XX'}(x,x')=1/|\calX|$ whenever $x=x'$ and $Q_{XX'}(x,x')=0$ otherwise. 
Then, without any prior knowledge, for every $m' \neq m$, $\prob \left\{\bX_{m} = \bX_{m'} \right\} = \prob \left\{(\bX_{m},\bX_{m'}) \in \calT(Q_{XX'}) \right\} \doteq \exp\{-n I_{Q}(X;X')\}$, where $I_{Q}(X;X')=\log |\calX|$.	
Now, conditioned on $\bX_{0}=\bX_{1}$ and $\bX_{1}=\bX_{2}$, it holds that $\bX_{0}=\bX_{2}$ with probability 1.} (though pairwise independent) binary random variables. 
This is different from the more commonly encountered type class enumerators (see, e.g., \cite{SBM}, \cite{MERHAV2014}, \cite{WM2017}), which are sums of \textit{independent} binary random variables.
Hence, existing results concerning the LD for type class enumerators of independent variables are not applicable, and thus, more refined tools from LD theory are required, like those of \cite{Suen}, that will allow us to handle dependency between terms\footnote{Also refer to \cite[Sec.\ IV--C]{Nazari_Graphs}, where bounds from \cite{Suen} were used to handle weak dependencies in joint types.}.
In spite of the statistical dependencies, it turns out, that the LD behavior of $N(Q_{XX'})$ and the ordinary type class enumerators are the same.
This can be seen in the following theorem, which is proved in Appendix B.
\begin{theorem} \label{2D_Large_Deviations}
	For any $s \in \mathbb{R}$,
	\begin{align}
	\prob \left\{ N(Q_{XX'}) \geq e^{n s} \right\} 
	&\doteq   e^{-n \cdot E(R,Q,s)} ,
	\end{align}
	where,
	\begin{align}
	E(R,Q,s) = \left\{   
	\begin{array}{l l}
	\left[I_{Q}(X;X')-2R\right]_{+}    & \quad \text{  $\left[2R - I_{Q}(X;X')\right]_{+} \geq s$  }   \\
	\infty    & \quad \text{  $\left[2R - I_{Q}(X;X')\right]_{+} < s$  }
	\end{array} \right. .
	\end{align}
\end{theorem}
Then, we rewrite \eqref{EXP5} in terms of the enumerators $N(Q_{XX'})$ and get
\begin{align}
&\prob \left\{ -\frac{1}{n} \log P_{\mbox{\tiny e}} (\calC_{n}) \leq \expVAR \right\} \nonumber \\
\label{WHYMAX0}
&\lexe \prob \left\{\sum_{Q_{XX'} \in \calQ(Q_{X})} N(Q_{XX'}) 
\exp\{-n \cdot \Gamma(Q_{XX'},R-\epsilon) \}  
\geq e^{n \cdot (R - \expVAR)} \right\} \\
\label{WHYMAX1}
&\doteq \prob \left\{\max_{Q_{XX'} \in \calQ(Q_{X})} N(Q_{XX'}) 
\exp\{-n \cdot \Gamma(Q_{XX'},R-\epsilon) \}  
\geq e^{n \cdot (R - \expVAR)} \right\} \\
&= \prob \left\{\bigcup_{Q_{XX'} \in \calQ(Q_{X})} N(Q_{XX'}) 
\exp\{-n \cdot \Gamma(Q_{XX'},R-\epsilon) \}  
\geq e^{n \cdot (R - \expVAR)} \right\} \\
&\doteq \sum_{Q_{XX'} \in \calQ(Q_{X})} \prob \left\{N(Q_{XX'}) \exp\{-n \cdot \Gamma(Q_{XX'},R-\epsilon) \}  
\geq e^{n \cdot (R - \expVAR)} \right\} \\
\label{WHYMAX2}
&\doteq \max_{Q_{XX'} \in \calQ(Q_{X})} \prob \left\{N(Q_{XX'}) \geq \exp\left\{n \cdot (\Psi(R-\epsilon,\expVAR,Q_{XX'})+\epsilon) \right\} \right\} .
\end{align}
where the steps to (\ref{WHYMAX1}) and (\ref{WHYMAX2}) are due to the SME of \eqref{SME}.
Thanks to Theorem \ref{2D_Large_Deviations}, the last expression decays exponentially with rate $E_{\mbox{\tiny lt}}^{\mbox{\tiny ub}}(R,\expVAR,\epsilon)$, which is given by 
\begin{align}
&E_{\mbox{\tiny lt}}^{\mbox{\tiny ub}}(R,\expVAR,\epsilon) \nn \\
&= \min_{Q_{XX'} \in \calQ(Q_{X})} \left\{   
\begin{array}{l l}
\left[I_{Q}(X;X')-2R\right]_{+}    & \quad \text{  $\left[2R - I_{Q}(X;X')\right]_{+} \geq \Psi(R-\epsilon,\expVAR,Q_{XX'})+\epsilon$  }   \\
\infty    & \quad \text{  $\left[2R - I_{Q}(X;X')\right]_{+} < \Psi(R-\epsilon,\expVAR,Q_{XX'})+\epsilon$  }
\end{array} \right.  \\
&= \min_{\{Q_{XX'} \in \calQ(Q_{X}):~ [2R - I_{Q}(X;X')]_{+} \geq \Psi(R-\epsilon,\expVAR,Q_{XX'})+\epsilon\}} \left[I_{Q}(X;X')-2R\right]_{+},
\end{align}
with the convention that the minimum over an empty set is defined as infinity. 
Due to the arbitrariness of $\epsilon > 0$, it follows that
\begin{align}
\prob \left\{ -\frac{1}{n} \log P_{\mbox{\tiny e}} (\calC_{n}) \leq \expVAR \right\} 
\lexe \exp\{-n \cdot E_{\mbox{\tiny lt}}^{\mbox{\tiny ub}}(R,\expVAR)\} ,
\end{align}
which proves the upper bound of Theorem \ref{THEOREM_LOWER_TAIL}.

\subsection{A Lower Bound on the Probability of the Lower Tail}
\label{SUBsection_LB_LT}
For a given $m$, $m' \neq m$, and $\by \in \calY^{n}$, define
\begin{align}
\label{Zmmtag_DEF}
Z_{mm'}(\by) = \sum_{\tilde{m}\in\{0,1,\ldots,M-1\} \setminus \{m,m'\}} \exp \{n g(\hat{P}_{\bx_{\tilde{m}}\by})\}.
\end{align}
Let $\sigma>0$ and define the set
\begin{align}
\hat{\calB}_{n}(\sigma,m,m',\by) = \left\{\calC_{n}:~ Z_{mm'}(\by) \geq \exp\{n \cdot (\beta(R,\hat{P}_{\by}) + \sigma)\}  \right\},
\end{align}
and its complement $\hat{\calG}_{n}(\sigma,m,m',\by)$,
where $\beta(R,Q_{Y})$ is defined as in \eqref{Beta_DEF}. Let
\begin{align}
\label{B_hat_DEF}
\hat{\calB}_{n}(\sigma)
= \bigcup_{m=0}^{M-1} \bigcup_{m' \neq m} \bigcup_{\by \in \calY^{n}} \hat{\calB}_{n}(\sigma,m,m',\by),
\end{align}
and
\begin{align}
\label{G_hat_DEF}
\hat{\calG}_{n}(\sigma) = \hat{\calB}_{n}^{\mbox{\tiny c}}(\sigma).
\end{align}
Let $\epsilon > 0$ be arbitrary and define
\begin{align}
\label{Lambda_Tilde_DEF}
\tilde{\Lambda}(Q_{XX'},R,\epsilon) &= \min_{Q_{Y|XX'}} \{ D(Q_{Y|X} \| W |Q_{X}) + I_{Q}(X';Y|X) \nonumber \\
&~~+ [\max\{g(Q_{XY}), \beta(R,Q_{Y})+\epsilon\} - g(Q_{X'Y})]_{+} \}.
\end{align}
We get the following
\begin{align}
&\prob \left\{ -\frac{1}{n} \log P_{\mbox{\tiny e}} (\calC_{n}) \leq \expVAR \right\} \nonumber \\
\label{aterminus1}
&= \prob \left\{\frac{1}{M} \sum_{m=0}^{M-1} \sum_{m' \neq m} \sum_{\by \in \calY^{n}} W(\by|\bx_{m}) \right. \nn \\ 
&\left. ~~~~\times \frac{\exp\{n g(\hat{P}_{\bx_{m'}\by}) \}}{\exp\{n g(\hat{P}_{\bx_{m}\by}) \} + \exp\{n g(\hat{P}_{\bx_{m'}\by}) \} + Z_{mm'}(\by)}  \geq e^{-n \cdot \expVAR} \right\} \\
\label{aterminus2}
&\geq \prob \left\{\calC_{n} \in \hat{\calG}_{n}(\epsilon), \frac{1}{M} \sum_{m=0}^{M-1} \sum_{m' \neq m} \sum_{\by \in \calY^{n}} W(\by|\bx_{m}) \right. \nn \\ 
&\left. ~~~~\times \frac{\exp\{n g(\hat{P}_{\bx_{m'}\by}) \}}{\exp\{n g(\hat{P}_{\bx_{m}\by}) \} + \exp\{n g(\hat{P}_{\bx_{m'}\by}) \} + Z_{mm'}(\by)}  \geq e^{-n \cdot \expVAR} \right\} \\
\label{aterminus3}
&\geq \prob \left\{\calC_{n} \in \hat{\calG}_{n}(\epsilon), \frac{1}{M} \sum_{m=0}^{M-1} \sum_{m' \neq m} \sum_{\by \in \calY^{n}} W(\by|\bx_{m}) \right. \nn \\
&\left. ~~~~\times \frac{\exp\{n g(\hat{P}_{\bx_{m'}\by}) \}}{\exp\{n g(\hat{P}_{\bx_{m}\by}) \} + \exp\{n g(\hat{P}_{\bx_{m'}\by}) \} + \exp\{n \cdot[\beta(R,\hat{P}_{\by}) + \epsilon]\} }  \geq e^{-n \cdot \expVAR} \right\} \\
\label{aterminus4}
&\doteq \prob \left\{\calC_{n} \in \hat{\calG}_{n}(\epsilon), \frac{1}{M} \sum_{m=0}^{M-1} \sum_{m' \neq m} \sum_{\by \in \calY^{n}} W(\by|\bx_{m}) \right. \nn \\
&\left.~~~~ \times  \exp\{n \cdot [\max\{ g(\hat{P}_{\bx_{m}\by}), \beta(R,\hat{P}_{\by}) + \epsilon\} - g(\hat{P}_{\bx_{m'}\by})]_{+} \}   \geq e^{-n \cdot \expVAR} \right\} \\
\label{aterminus5}
&\doteq \prob \left\{\calC_{n} \in \hat{\calG}_{n}(\epsilon), \frac{1}{M} \sum_{m=0}^{M-1} \sum_{m' \neq m} \exp \{-n \cdot \tilde{\Lambda}(\hat{P}_{\bx_{m}\bx_{m'}},R,\epsilon)\} \geq e^{-n \cdot \expVAR} \right\} \\
\label{ToContinue0}
&= \prob \left\{\calC_{n} \in \hat{\calG}_{n}(\epsilon),  \sum_{Q_{XX'} \in \calQ(Q_{X})} N(Q_{XX'}) \cdot
\exp \{-n \cdot \tilde{\Lambda}(Q_{XX'},R,\epsilon)\}
\geq e^{n \cdot (R -\expVAR)} \right\} ,
\end{align}
where \eqref{aterminus1} follows from the definitions of the probability of error and $Z_{mm'}(\by)$ in \eqref{PROBABILITYofErrorDEF} and \eqref{Zmmtag_DEF}, respectively. 
In \eqref{aterminus2}, we lower--bounded by intersecting with the event $\calC_{n} \in \hat{\calG}_{n}(\epsilon)$.
In \eqref{aterminus3}, the definition of the set $\hat{\calG}_{n}(\cdot)$ in \eqref{G_hat_DEF} was used, 
in \eqref{aterminus4}, the exponential equivalence $e^{nB}/(e^{nA}+e^{nB}+e^{nC}) \doteq \exp\{-n \cdot [\max\{A,C\}-B]_{+}\}$,
in \eqref{aterminus5}, the method of types and the definition of $\tilde{\Lambda}(Q_{XX'},R,\epsilon)$ in \eqref{Lambda_Tilde_DEF}, 
and in \eqref{ToContinue0}, the definition of the type class enumerators $N(Q_{XX'})$ in \eqref{NQ_def}.

Next, we simplify the expression of $\tilde{\Lambda}(Q_{XX'},R,\epsilon)$.
First, note that for any $\hQ_{XY}$ with marginals $Q_{X}$ and $Q_{Y}$
\begin{align}
\beta(R,Q_{Y}) 
&= \max_{\{Q_{\tilde{X}|Y}:~ Q_{\tilde{X}}=Q_{X}\}} \{g(Q_{\tilde{X}Y}) + [R - I_{Q}(\tilde{X};Y)]_{+}\} \\
&\geq \max_{\{Q_{\tilde{X}|Y}:~ Q_{\tilde{X}}=Q_{X}\}} g(Q_{\tilde{X}Y}) \\
&\geq g(\hQ_{XY}).
\end{align}
Then,
\begin{align}
&\tilde{\Lambda}(Q_{XX'},R,\epsilon) \nn \\
&= \min_{Q_{Y|XX'}} \{ D(Q_{Y|X} \| W |Q_{X}) + I_{Q}(X';Y|X) \nonumber \\
&~~~+ [\max\{g(Q_{XY}), \beta(R,Q_{Y})+\epsilon\} - g(Q_{X'Y})]_{+} \} \\
\label{TERM2EXPa1}
&= \min_{Q_{Y|XX'}} \{ D(Q_{Y|X} \| W |Q_{X}) + I_{Q}(X';Y|X) + [\beta(R,Q_{Y}) + \epsilon - g(Q_{X'Y})]_{+} \} \\
\label{TERM2EXPa2}
&= \min_{Q_{Y|XX'}} \{ D(Q_{Y|X} \| W |Q_{X}) + I_{Q}(X';Y|X) + \beta(R,Q_{Y}) - g(Q_{X'Y}) + \epsilon \} \\
\label{TERM2EXPa3}
&= \Lambda(Q_{XX'},R) + \epsilon,
\end{align}
where \eqref{TERM2EXPa1} is due to $\beta(R,Q_{Y}) \geq g(Q_{XY})$, \eqref{TERM2EXPa2} is because $\beta(R,Q_{Y}) \geq g(Q_{X'Y})$, and \eqref{TERM2EXPa3} follows the definition in \eqref{LAMBDA_DEF}.
Let us now define 
\begin{align}
\calG_{0} = \left\{\calC_{n}:~ \sum_{Q_{XX'} \in \calQ(Q_{X})} N(Q_{XX'}) \cdot \exp \{-n \cdot ( \Lambda(Q_{XX'},R) + \epsilon)\}
\geq e^{n \cdot (R -\expVAR)} \right\},
\end{align}
such that, continuing from \eqref{ToContinue0}:
\begin{align}
\prob \left\{ -\frac{1}{n} \log P_{\mbox{\tiny e}} (\calC_{n}) \leq \expVAR \right\} 
\label{TERM2Call1}
&\gexe \prob \left\{\hat{\calG}_{n}(\epsilon) \cap \calG_{0} \right\} \\
&= \prob \left\{ \bigcap_{m=0}^{M-1} \bigcap_{m' \neq m}\bigcap_{\by\in\calY^n}\hat{\calG}_{n}(\epsilon,m,m',\by) \middle| \calG_{0} \right\} \cdot \prob \left\{ \calG_{0} \right\}  \\
&= \left(1 - \prob \left\{ \bigcup_{m=0}^{M-1} \bigcup_{m' \neq m}\bigcup_{\by\in\calY^n}\hat{\calB}_{n}(\epsilon,m,m',\by) \middle| \calG_{0} \right\} \right) \cdot \prob \left\{ \calG_{0} \right\}  \\
&\geq \left(1 - \sum_{m=0}^{M-1} \sum_{m' \neq m}\sum_{\by\in\calY^n} \prob \left\{ \hat{\calB}_{n}(\epsilon,m,m',\by) \middle| \calG_{0}  \right\} \right) \cdot \prob \left\{ \calG_{0} \right\}  \\  
\label{ToCall9}
&= \prob \left\{ \calG_{0} \right\} - \sum_{m=0}^{M-1} \sum_{m' \neq m}\sum_{\by\in\calY^n} \prob \left\{ \hat{\calB}_{n}(\epsilon,m,m',\by) \cap \calG_{0} \right\} .
\end{align}

\subsubsection*{Assessing $\prob\{\calG_{0}\}$ in \eqref{ToCall9}}
Now,
\begin{align}
\prob\{\calG_{0}\}
&= \prob \left\{ \sum_{Q_{XX'} \in \calQ(Q_{X})} N(Q_{XX'}) \cdot
\exp \{-n \cdot (\Lambda(Q_{XX'},R) + \epsilon)\}
\geq e^{n \cdot (R -\expVAR)} \right\} \\
\label{ENDofPROOF1}
&\doteq \sum_{Q_{XX'} \in \calQ(Q_{X})} \prob \left\{ N(Q_{XX'})
\geq \exp \{n \cdot(\Lambda(Q_{XX'},R) + R - \expVAR + \epsilon)\} \right\} \\
\label{ENDofPROOF2}
&\doteq \max_{Q_{XX'} \in \calQ(Q_{X})} \prob \left\{N(Q_{XX'}) \geq \exp\left\{n \cdot (\Xi(R,\expVAR,Q_{XX'}) + \epsilon) \right\} \right\},
\end{align}
where \eqref{ENDofPROOF1} and \eqref{ENDofPROOF2} follow by the SME and are similar to the steps between \eqref{WHYMAX0}--\eqref{WHYMAX2}. 
Thanks to Theorem \ref{2D_Large_Deviations}, the last expression decays exponentially with rate $E_{\mbox{\tiny lt}}^{\mbox{\tiny lb}}(R,\expVAR,\epsilon)$, which is given by 
\begin{align}
&E_{\mbox{\tiny lt}}^{\mbox{\tiny lb}}(R,\expVAR,\epsilon) \nn \\
&= \min_{Q_{XX'} \in \calQ(Q_{X})} \left\{   
\begin{array}{l l}
\left[I_{Q}(X;X')-2R\right]_{+}    & \quad \text{  $\left[2R - I_{Q}(X;X')\right]_{+} \geq \Xi(R,\expVAR,Q_{XX'}) + \epsilon$  }   \\
\infty    & \quad \text{  $\left[2R - I_{Q}(X;X')\right]_{+} < \Xi(R,\expVAR,Q_{XX'}) + \epsilon$  }
\end{array} \right. \\
&= \min_{\{Q_{XX'} \in \calQ(Q_{X}):~ [2R - I_{Q}(X;X')]_{+} \geq \Xi(R,\expVAR,Q_{XX'}) + \epsilon\}} \left[I_{Q}(X;X')-2R\right]_{+} , 
\end{align}
and thus
\begin{align}
\label{ToCall13}
\prob\{\calG_{0}\} \doteq \exp\{-n \cdot E_{\mbox{\tiny lt}}^{\mbox{\tiny lb}}(R,\expVAR,\epsilon)\}.
\end{align}


\subsubsection*{Upper--bounding $\prob \{ \hat{\calB}_{n}(\epsilon,m,m',\by) \cap \calG_{0} \}$ in \eqref{ToCall9}}

Define the type class enumerator 
\begin{align}
\label{NQ_def2}
N_{\by}(Q_{XY}) = \sum_{m=0}^{M-1} \IND \left\{ (\bX_{m}, \by) \in \calT(Q_{XY}) \right\}. 
\end{align} 
Then, we have the following
\begin{align}
&\prob \{ \hat{\calB}_{n}(\epsilon,\hat{m},\ddot{m},\by) \cap \calG_{0} \} \nn \\
&=\prob \left\{ \sum_{\tilde{m}\in\{0,1,\ldots,M-1\} \setminus \{\hat{m},\ddot{m}\}} \exp \{n g(\hat{P}_{\bX_{\tilde{m}}\by})\} \geq \exp\{n \cdot(\beta(R,\hat{P}_{\by}) + \epsilon)\}, \nn \right. \\
&\left. ~~~~~~~~~~~~~
\sum_{m=0}^{M-1} \sum_{m' \neq m} \exp \{-n \cdot (\Lambda(\hat{P}_{\bX_{m}\bX_{m'}},R) + \epsilon)\} \geq e^{n \cdot (R - \expVAR)} \right\} \\
&\leq \prob \left\{ \sum_{\tilde{m}\in\{0,1,\ldots,M-1\}} \exp \{n g(\hat{P}_{\bX_{\tilde{m}}\by})\} \geq \exp\{n \cdot(\beta(R,\hat{P}_{\by}) + \epsilon)\}, \nn \right. \\
&\left. ~~~~~~~~~~~~~
\sum_{m=0}^{M-1} \sum_{m' \neq m} \exp \{-n \cdot (\Lambda(\hat{P}_{\bX_{m}\bX_{m'}},R) + \epsilon)\} \geq e^{n \cdot (R - \expVAR)} \right\} \\
&=\prob \left\{ \sum_{Q_{XY}} N_{\by}(Q_{XY}) \exp \{n g(Q_{XY})\} \geq \exp\{n \cdot(\beta(R,\hat{P}_{\by}) + \epsilon)\}, \nn \right. \\
&\left. ~~~~~~~~~~~~~
\sum_{Q_{XX'}} N(Q_{XX'}) \exp \{-n \cdot (\Lambda(Q_{XX'},R) + \epsilon)\} \geq e^{n \cdot (R - \expVAR)} \right\} \\
\label{abterminus0}
&\doteq \prob \left\{ \bigcup_{Q_{XY}} \left\{N_{\by}(Q_{XY}) \geq e^{n \cdot(\beta(R,\hat{P}_{\by}) - g(Q_{XY}) + \epsilon)} \right\}, 
\bigcup_{Q_{XX'}} \left\{ N(Q_{XX'}) \geq e^{n \cdot (\Xi(R,\expVAR,Q_{XX'}) + \epsilon)} \right\} \right\} \\
&\doteq \sum_{Q_{XY}} \sum_{Q_{XX'}} \prob \left\{N_{\by}(Q_{XY})^{l} \geq e^{n \cdot(\beta(R,\hat{P}_{\by}) - g(Q_{XY}) + \epsilon) \cdot l}, 
N(Q_{XX'})^{k} \geq e^{n \cdot (\Xi(R,\expVAR,Q_{XX'}) + \epsilon) \cdot k} \right\} \\
&\doteq \max_{Q_{XY}} \max_{Q_{XX'}} \prob \left\{N_{\by}(Q_{XY})^{l} \geq e^{n \cdot(\beta(R,\hat{P}_{\by}) - g(Q_{XY}) + \epsilon) \cdot l}, 
N(Q_{XX'})^{k} \geq e^{n \cdot (\Xi(R,\expVAR,Q_{XX'}) + \epsilon) \cdot k} \right\} \\
\label{abterminus1}
&\leq \max_{Q_{XY}} \max_{Q_{XX'}} \prob \left\{N_{\by}(Q_{XY})^{l} \cdot
N(Q_{XX'})^{k} \geq e^{n \cdot(\beta(R,\hat{P}_{\by}) - g(Q_{XY}) + \epsilon) \cdot l} \cdot e^{n \cdot (\Xi(R,\expVAR,Q_{XX'}) + \epsilon) \cdot k} \right\} \\
\label{abterminus2}
&\leq \max_{Q_{XY}} \max_{Q_{XX'}} \prob \left\{N_{\by}(Q_{XY})^{l} \cdot
N(Q_{XX'})^{k} \geq e^{n \cdot ([R - I_{Q}(X;Y)]_{+} + \epsilon) \cdot l} \cdot e^{n \cdot (\Xi(R,\expVAR,Q_{XX'}) + \epsilon) \cdot k} \right\} \\
\label{abterminus3}
&\leq \max_{Q_{XY}} \max_{Q_{XX'}}  
\frac{\mathbb{E} \left[ N_{\by}(Q_{XY})^{l} \cdot
N(Q_{XX'})^{k} \right]}{e^{n \cdot([R - I_{Q}(X;Y)]_{+} + \epsilon) \cdot l} \cdot e^{n \cdot (\Xi(R,\expVAR,Q_{XX'}) + \epsilon) \cdot k}} ,
\end{align}
where $k$ and $l$ are arbitrary positive integers, and where \eqref{abterminus0} follows from the definition of $\Xi(R,\expVAR,Q_{XX'})$ in \eqref{XI_DEF}. Step \eqref{abterminus1} is due to the fact that $\prob\{X\geq a, Y\geq b\} \leq \prob\{X\cdot Y \geq a \cdot b\}$, 
under the assumption that $a,b$ are positive.
In \eqref{abterminus2}, we use the definition of $\beta(R,Q_{Y})$ in \eqref{Beta_DEF}, which implies that $\beta(R,Q_{Y}) \geq g(Q_{XY}) + \left[R - I_{Q}(X;Y)\right]_{+}$ and \eqref{abterminus3} follows from Markov's inequality.
After optimizing over $l$ and $k$,
\begin{align}
\label{ToCall10}
\prob \{ \hat{\calB}_{n}(\epsilon,m,m',\by) \cap \calG_{0} \}
\lexe \max_{Q_{XY}} \max_{Q_{XX'}} \inf_{l \in \mathbb{N}} \inf_{k \in \mathbb{N}} 
\frac{\mathbb{E} \left[ N_{\by}(Q_{XY})^{l} \cdot
N(Q_{XX'})^{k} \right]}{e^{n \cdot ([R - I_{Q}(X;Y)]_{+} + \epsilon) \cdot l} \cdot e^{n \cdot (\Xi(R,\expVAR,Q_{XX'}) + \epsilon) \cdot k}} .
\end{align}
For $S \geq 0$, a joint distribution $Q_{UV}$, and an integer $j \in \mathbb{N}$, define the following quantity
\begin{align}
\label{DEF_F}
F(S,Q_{UV},j)
=\left\{   
\begin{array}{l l}
\exp \{n j \left(S - I_{Q}(U;V)\right) \}    & \quad \text{  $I_{Q}(U;V) < S$  }   \\
\exp \{n \left(S - I_{Q}(U;V)\right) \}    & \quad \text{  $I_{Q}(U;V) > S$  }
\end{array} \right. .
\end{align}
We use the following proposition, which is proved in Appendix G.
\begin{proposition} \label{Prop_Moments}
	Let $N(Q_{XX'})$ and $N_{\by}(Q_{XY})$ be as in \eqref{NQ_def} and \eqref{NQ_def2}, respectively. 
	Then, for any $k,l \in \mathbb{N}$,
	\begin{align}
	\mathbb{E} \left[ N_{\by}(Q_{XY})^{l}
	N(Q_{XX'})^{k} \right]
	&\lexe F(R,Q_{XY},l) \cdot F(2R,Q_{XX'},k).
	\end{align}
\end{proposition}
Next, substituting the result of Proposition \ref{Prop_Moments} back into \eqref{ToCall10} provides
\begin{align}
\prob \{ \hat{\calB}_{n}(\epsilon,m,m',\by) \cap \calG_{0} \}
\label{ToCall11}
&\lexe \max_{Q_{XY}}  \inf_{l \in \mathbb{N}} 
\frac{\exp\{n \cdot (l \cdot [R-I_{Q}(X;Y)]_{+} - [I_{Q}(X;Y)-R]_{+})\}}{\exp\{n \cdot ([R - I_{Q}(X;Y)]_{+} + \epsilon) \cdot l\}} \nn \\
&\times \max_{Q_{XX'}} \inf_{k \in \mathbb{N}} 
\frac{\exp\{n \cdot (k \cdot [2R-I_{Q}(X;X')]_{+} - [I_{Q}(X;X')-2R]_{+})\}}{\exp\{n \cdot (\Xi(R,\expVAR,Q_{XX'}) + \epsilon) \cdot k\}}.
\end{align} 
As for the left--hand term in \eqref{ToCall11}, we have that
\begin{align}
&-\frac{1}{n} \log\max_{Q_{XY}}  \inf_{l \in \mathbb{N}} 
\frac{\exp\{n \cdot (l \cdot [R-I_{Q}(X;Y)]_{+} - [I_{Q}(X;Y)-R]_{+})\}}{\exp\{n \cdot ([R - I_{Q}(X;Y)]_{+} + \epsilon) \cdot l\}} \nn \\
&=-\frac{1}{n} \log\max_{Q_{XY}}  \inf_{l \in \mathbb{N}} 
\exp\{-n \cdot \left([I_{Q}(X;Y)-R]_{+} + l \epsilon \right)\}\\
&=\min_{Q_{XY}}  \sup_{l \in \mathbb{N}} 
\left([I_{Q}(X;Y)-R]_{+} + l \epsilon \right)\\
&= \infty .
\end{align}
For the right--hand term in \eqref{ToCall11}, we get the following
\begin{align}
& -\frac{1}{n} \log \max_{Q_{XX'}} \inf_{k \in \mathbb{N}} 
\frac{\exp\{n \cdot (k \cdot [2R-I_{Q}(X;X')]_{+} - [I_{Q}(X;X')-2R]_{+})\}}{\exp\{n \cdot (\Xi(R,\expVAR,Q_{XX'}) + \epsilon) \cdot k\}} \nn \\
&= \min_{Q_{XX'}} \sup_{k \in \mathbb{N}} \left( k \cdot \left(\Xi(R,\expVAR,Q_{XX'}) + \epsilon  - [2R-I_{Q}(X;X')]_{+} \right) + [I_{Q}(X;X')-2R]_{+} \right) \\
&= \min_{\{Q_{XX'} \in \calQ(Q_{X}):~ [2R - I_{Q}(X;X')]_{+} \geq \Xi(R,\expVAR,Q_{XX'}) + \epsilon\}} \left[I_{Q}(X;X')-2R\right]_{+}  \\
&= E_{\mbox{\tiny lt}}^{\mbox{\tiny lb}}(R,\expVAR,\epsilon) .
\end{align}
Thus,
\begin{align}
\label{ToCall12}
\prob \{ \hat{\calB}_{n}(\epsilon,m,m',\by) \cap \calG_{0} \}
&\lexe e^{-n \infty} \cdot \exp\{-n \cdot E_{\mbox{\tiny lt}}^{\mbox{\tiny lb}}(R,\expVAR,\epsilon)\} .
\end{align}

\subsubsection*{Final Steps}
Finally, we continue from \eqref{ToCall9} and use the results of \eqref{ToCall13} and \eqref{ToCall12} to provide 
\begin{align}
&\prob \left\{ -\frac{1}{n} \log P_{\mbox{\tiny e}} (\calC_{n}) \leq \expVAR \right\} \nonumber \\ 
&\gexe \prob \left\{ \calG_{0} \right\} - \sum_{m=0}^{M-1} \sum_{m' \neq m}\sum_{\by\in\calY^n} \prob \left\{ \hat{\calB}_{n}(\epsilon,m,m',\by) \cap \calG_{0} \right\} \\
&\gexe \exp\{-n \cdot E_{\mbox{\tiny lt}}^{\mbox{\tiny lb}}(R,\expVAR,\epsilon)\} - \sum_{m=0}^{M-1} \sum_{m' \neq m}\sum_{\by\in\calY^n} e^{-n \infty} \cdot \exp\{-n \cdot E_{\mbox{\tiny lt}}^{\mbox{\tiny lb}}(R,\expVAR,\epsilon)\} \\
&\doteq \left(1 - e^{n2R} \cdot |\calY|^n \cdot e^{-n \infty} \right) \cdot \exp\{-n \cdot E_{\mbox{\tiny lt}}^{\mbox{\tiny lb}}(R,\expVAR,\epsilon)\} \\
&\doteq \exp\{-n \cdot E_{\mbox{\tiny lt}}^{\mbox{\tiny lb}}(R,\expVAR,\epsilon)\}.
\end{align}
Due to the arbitrariness of $\epsilon > 0$, it follows that
\begin{align}
\prob \left\{ -\frac{1}{n} \log P_{\mbox{\tiny e}} (\calC_{n}) \leq \expVAR \right\} 
\gexe \exp\{-n \cdot E_{\mbox{\tiny lt}}^{\mbox{\tiny lb}}(R,\expVAR)\} ,
\end{align}
which proves the lower bound of Theorem \ref{THEOREM_LOWER_TAIL}.

\section{Proof of the Upper Bound of Theorem \ref{THEOREM_UPPER_TAIL}}
\label{SEC_VI}
Let $Z_{mm'}(\by)$, $\hat{\calB}_{n}(\sigma)$, and $\hat{\calG}_{n}(\sigma)$ be defined as in \eqref{Zmmtag_DEF}, \eqref{B_hat_DEF}, and \eqref{G_hat_DEF}, respectively. 
One of the main ingredients in the proof of the upper bound on the probability of the lower tail in Subsection \ref{SEC_V1} is the fact that $Z_{m}(\by)$ is lower--bounded by $\exp\{n \alpha(R,\hat{P}_{\by})\}$ with a probability that approaches one double--exponentially fast. 
In order to prove an upper bound on the probability of the upper tail, we start by showing that $\exp\{n \beta(R,\hat{P}_{\by})\}$ serves as an upper bound on $Z_{mm'}(\by)$, simultaneously for every $m \in \{0,1,\dotsc,M-1\}$, $m' \in \{0,1,\dotsc,M-1\} \setminus \{m\}$, and $\by \in \calY^{n}$, with probability that tends to one double--exponentially fast. 
More specifically, we have the following result, which is proved in Appendix H.   
\begin{proposition}
	\label{Double_Exponential_Result}
	For every $\sigma > 0$,
	\begin{align}
	\prob \left\{\hat{\calB}_{n}(\sigma)\right\} 
	&\DLEXE \exp \left\{-e^{n \sigma} \right\}.
	\end{align}
\end{proposition}
We start with
\begin{align}
&\prob \left\{ -\frac{1}{n} \log P_{\mbox{\tiny e}} (\calC_{n}) \geq \expVAR \right\} \nn \\ 
&= \prob \left\{\calC_{n} \in \hat{\calG}_{n}(\sigma), -\frac{1}{n} \log P_{\mbox{\tiny e}} (\calC_{n}) \geq \expVAR \right\} 
+ \prob \left\{\calC_{n} \in \hat{\calB}_{n}(\sigma), -\frac{1}{n} \log P_{\mbox{\tiny e}} (\calC_{n}) \geq \expVAR \right\} \\
\label{TWO_SUMMANDS}
&\leq \prob \left\{\calC_{n} \in \hat{\calG}_{n}(\sigma), -\frac{1}{n} \log P_{\mbox{\tiny e}} (\calC_{n}) \geq \expVAR \right\} 
+ \prob \left\{\calC_{n} \in \hat{\calB}_{n}(\sigma) \right\} .
\end{align}
As for the first term,
\begin{align}
&\prob \left\{\calC_{n} \in \hat{\calG}_{n}(\sigma), -\frac{1}{n} \log P_{\mbox{\tiny e}} (\calC_{n}) \geq \expVAR \right\} \nonumber \\  
\label{abcterminus1}
&= \prob \left\{\calC_{n} \in \hat{\calG}_{n}(\sigma), \frac{1}{M} \sum_{m=0}^{M-1} \sum_{m' \neq m} \sum_{\by \in \calY^{n}} W(\by|\bx_{m}) \right. \nn \\
&\left. ~~~~\times \frac{\exp\{n g(\hat{P}_{\bx_{m'}\by}) \}}{\exp\{n g(\hat{P}_{\bx_{m}\by}) \} + \exp\{n g(\hat{P}_{\bx_{m'}\by}) \} + Z_{mm'}(\by)}  \leq e^{-n \cdot \expVAR} \right\} \\
\label{abcterminus2}
&\leq \prob \left\{\calC_{n} \in \hat{\calG}_{n}(\sigma), \frac{1}{M} \sum_{m=0}^{M-1} \sum_{m' \neq m} \sum_{\by \in \calY^{n}} W(\by|\bx_{m}) \right. \nn \\
&\left. ~~~~\times \frac{\exp\{n g(\hat{P}_{\bx_{m'}\by}) \}}{\exp\{n g(\hat{P}_{\bx_{m}\by}) \} + \exp\{n g(\hat{P}_{\bx_{m'}\by}) \} + \exp\{n \cdot[\beta(R,\hat{P}_{\by}) + \sigma]\} }  \leq e^{-n \cdot \expVAR} \right\} \\
\label{abcterminus3}
&\DEXE \prob \left\{\calC_{n} \in \hat{\calG}_{n}(\sigma), \frac{1}{M} \sum_{m=0}^{M-1} \sum_{m' \neq m} \sum_{\by \in \calY^{n}} W(\by|\bx_{m}) \right. \nn \\
&\left.~~~~ \times  \exp\{n \cdot [\max\{ g(\hat{P}_{\bx_{m}\by}), \beta(R,\hat{P}_{\by}) + \sigma\} - g(\hat{P}_{\bx_{m'}\by})]_{+} \}   \leq e^{-n \cdot \expVAR} \right\} \\
\label{abcterminus4}
&\DEXE \prob \left\{\calC_{n} \in \hat{\calG}_{n}(\sigma), \frac{1}{M} \sum_{m=0}^{M-1} \sum_{m' \neq m} \exp \{-n \cdot \tilde{\Lambda}(\hat{P}_{\bx_{m}\bx_{m'}},R,\sigma)\} \leq e^{-n \cdot \expVAR} \right\} \\
\label{abcterminus5}
&= \prob \left\{\calC_{n} \in \hat{\calG}_{n}(\sigma),  \sum_{Q_{XX'} \in \calQ(Q_{X})} N(Q_{XX'}) \cdot
\exp \{-n \cdot \tilde{\Lambda}(Q_{XX'},R,\sigma)\}
\leq e^{n \cdot (R -\expVAR)} \right\} \\
\label{abcterminus6}
&\leq \prob \left\{\sum_{Q_{XX'} \in \calQ(Q_{X})} N(Q_{XX'}) \cdot
\exp \{-n \cdot \tilde{\Lambda}(Q_{XX'},R,\sigma)\}
\leq e^{n \cdot (R -\expVAR)} \right\} ,
\end{align}
where \eqref{abcterminus1} follows from the definitions of the probability of error and $Z_{mm'}(\by)$ in \eqref{PROBABILITYofErrorDEF} and \eqref{Zmmtag_DEF}, respectively. 
In \eqref{abcterminus2}, the definition of the set $\hat{\calG}_{n}(\sigma)$ in \eqref{G_hat_DEF} was used, 
in \eqref{abcterminus3}, the exponential equivalence $e^{nB}/(e^{nA}+e^{nB}+e^{nC}) \doteq \exp\{-n \cdot [\max\{A,C\}-B]_{+}\}$,
in \eqref{abcterminus4}, the method of types and the definition of $\tilde{\Lambda}(Q_{XX'},R,\sigma)$ in \eqref{Lambda_Tilde_DEF},
in \eqref{abcterminus5}, the definition of the type class enumerators $N(Q_{XX'})$ in \eqref{NQ_def},
and in \eqref{abcterminus6}, the event $\calC_{n} \in \hat{\calG}_{n}(\sigma)$ was taken out.

Next,
\begin{align}
&\prob \left\{\calC_{n} \in \hat{\calG}_{n}(\sigma), -\frac{1}{n} \log P_{\mbox{\tiny e}} (\calC_{n}) \geq \expVAR \right\} \nonumber \\
&\DLEXE \prob \left\{ \sum_{Q_{XX'} \in \calQ(Q_{X})} N(Q_{XX'}) 
\cdot \exp \{-n \cdot \tilde{\Lambda}(Q_{XX'},R,\sigma)\} \leq e^{n \cdot (R - \expVAR)} \right\} \\
\label{SME_Equality1}
&\DEXE \prob \left\{ \max_{Q_{XX'} \in \calQ(Q_{X})} N(Q_{XX'}) 
\cdot \exp \{-n \cdot \tilde{\Lambda}(Q_{XX'},R,\sigma)\} \leq e^{n \cdot (R - \expVAR)} \right\} \\
\label{INTERSECTION2}
&= \prob \left\{\bigcap_{Q_{XX'} \in \calQ(Q_{X})} \left\{ N(Q_{XX'}) 
\leq e^{n \cdot (\tilde{\Lambda}(Q_{XX'},R,\sigma) + R - \expVAR)} \right\} \right\}, 
\end{align}
where \eqref{SME_Equality1} is due to the SME.

If $\expVAR$ is relatively small, then
for every $Q_{XX'} \in \calQ(Q_{X})$, either $I_{Q}(X;X') \geq 2R$ or $2R - I_{Q}(X;X') \leq \tilde{\Lambda}(Q_{XX'},R,\sigma) + R - \expVAR$, and we have an intersection of polynomially many events whose probabilities all tend to one.
Hence, for every $\sigma > 0$, we assume that $\expVAR$ is sufficiently large, so there must exist at least one $Q_{XX'} \in \calQ(Q_{X})$ for which $I_{Q}(X;X') \leq 2R$ and $\tilde{\Lambda}(Q_{XX'},R,\sigma) + R - \expVAR \leq 2R - I_{Q}(X;X')$, such that (\ref{INTERSECTION2}) decays double exponentially fast, according to Lemma \ref{The_Lower_Tail} in Appendix B.
We define the set
\begin{align}
\label{V_TILDE_DEF}
\tilde{\calV}(R,\expVAR,\sigma) 
\DEF \{&Q_{XX'} \in \calQ(Q_{X}):~ I_{Q}(X;X') \leq 2R, \nn \\ 
&\tilde{\Lambda}(Q_{XX'},R,\sigma) + I_{Q}(X;X') - R \leq \expVAR \}.
\end{align}
Then, 
\begin{align}
&\prob \left\{\calC_{n} \in \hat{\calG}_{n}(\sigma), -\frac{1}{n} \log P_{\mbox{\tiny e}} (\calC_{n}) \geq \expVAR \right\} \nonumber \\
\label{CH4_A}
&\DLEXE \prob \left\{\bigcap_{Q_{XX'} \in \calQ(Q_{X})} \left\{ N(Q_{XX'}) 
\leq e^{n \cdot (\tilde{\Lambda}(Q_{XX'},R,\sigma) + R - \expVAR)} \right\} \right\} \\
\label{CH4_B}
&\leq \prob \left\{\bigcap_{Q_{XX'} \in \tilde{\calV}(R,\expVAR,\sigma)} \left\{ N(Q_{XX'}) \leq e^{n \cdot (\tilde{\Lambda}(Q_{XX'},R,\sigma) + R - \expVAR)} \right\} \right\} .
\end{align}
Since $\tilde{\Lambda}(Q_{XX'},R,\sigma) + R - \expVAR \leq 2R - I_{Q}(X;X')$, we obtain 
\begin{align}
& \prob \left\{\bigcap_{Q_{XX'} \in \tilde{\calV}(R,\expVAR,\sigma)} \left\{ N(Q_{XX'}) \leq e^{n \cdot (\tilde{\Lambda}(Q_{XX'},R,\sigma) + R - \expVAR)} \right\} \right\} \\
&\leq \min_{Q_{XX'} \in \tilde{\calV}(R,\expVAR,\sigma)} \prob \left\{ N(Q_{XX'}) \leq e^{n \cdot (\tilde{\Lambda}(Q_{XX'},R,\sigma) + R - \expVAR)} \right\} \\
\label{TERM2EXP}
&\DLEXE \min_{Q_{XX'} \in \tilde{\calV}(R,\expVAR,\sigma)} \exp \left\{- \min \left(e^{n(2R - I_{Q}(X;X'))}, e^{nR} \right) \right\} \\
&= \min_{Q_{XX'} \in \tilde{\calV}(R,\expVAR,\sigma)} \exp \left\{- e^{n \cdot \min\{2R - I_{Q}(X;X'),R\}} \right\} \\
&= \exp \left\{- \exp\left\{n \cdot \max_{Q_{XX'} \in \tilde{\calV}(R,\expVAR,\sigma)} \min\{2R - I_{Q}(X;X'),R\} \right\} \right\},
\end{align}
where \eqref{TERM2EXP} follows from Lemma \ref{The_Lower_Tail} in Appendix B. 
Let us define
\begin{align}
E_{1}(R,\expVAR,\sigma) = \max_{Q_{XX'} \in \tilde{\calV}(R,\expVAR,\sigma)} \min\{2R - I_{Q}(X;X'),R\},
\end{align}
such that    
\begin{align}
\prob \left\{\calC_{n} \in \hat{\calG}_{n}(\sigma), -\frac{1}{n} \log P_{\mbox{\tiny e}} (\calC_{n}) \geq \expVAR \right\}  
\label{DE_MID_RESULT}
\DLEXE \exp \left\{- \exp\left\{n \cdot E_{1}(R,\expVAR,\sigma) \right\} \right\}.
\end{align}


\subsubsection*{Final Steps}
Finally, it follows from \eqref{DE_MID_RESULT} and Proposition \ref{Double_Exponential_Result} that
\begin{align}
\prob \left\{ -\frac{1}{n} \log P_{\mbox{\tiny e}} (\calC_{n}) \geq \expVAR \right\}  
&\leq \prob \left\{\calC_{n} \in \hat{\calG}_{n}(\sigma), -\frac{1}{n} \log P_{\mbox{\tiny e}} (\calC_{n}) \geq \expVAR \right\} 
+ \prob \left\{\calC_{n} \in \hat{\calB}_{n}(\sigma) \right\} \\
&\DLEXE \exp \left\{- e^{n \cdot E_{1}(R,\expVAR,\sigma)} \right\} + \exp \left\{-e^{n \sigma} \right\} \\
&\DEXE \exp \left\{- \exp\{n \cdot \min[ E_{1}(R,\expVAR,\sigma),\sigma] \} \right\}.
\end{align}
As a last step, we optimize over $\sigma>0$, which resulting in
\begin{align}
\prob \left\{ -\frac{1}{n} \log P_{\mbox{\tiny e}} (\calC_{n}) > \expVAR \right\}  
\DLEXE  \exp \left\{- \exp\left\{n \cdot \sup_{\sigma > 0} \min[ E_{1}(R,\expVAR,\sigma),\sigma] \right\} \right\}.
\end{align}

\subsubsection*{A Simplified Expression}

Note that $E_{1}(R,\expVAR,\sigma)$ is continuous and monotonically non--increasing in $\sigma$, hence we can solve for the optimal $\sigma > 0$ by finding the maximal $\sigma$ for which $\sigma \leq E_{1}(R,\expVAR,\sigma)$. 
Let us abbreviate $I_{Q}(X;X')$ by $I_{Q}$, and then
\begin{align}
&E_{1}(R,\expVAR,\sigma) \nn \\
&= \max_{Q_{XX'} \in \tilde{\calV}(R,\expVAR,\sigma)} \min\{2R - I_{Q},R\} \\
\label{TERM2EXPb1}
&= \max_{\{Q_{XX'} \in \calQ(Q_{X}):~ I_{Q} \leq 2R \}} \inf_{\mu \geq 0} \left\{ \min\{2R - I_{Q},R\} + \mu \cdot (\expVAR - \tilde{\Lambda}(Q_{XX'},R,\sigma) - I_{Q} + R) \right\}\\
\label{TERM2EXPb2}
&= \max_{\{Q_{XX'} \in \calQ(Q_{X}):~ I_{Q} \leq 2R \}} \inf_{\mu \geq 0} \left\{ \min\{2R - I_{Q},R\} + \mu \cdot (\expVAR - \Lambda(Q_{XX'},R) -\sigma - I_{Q} + R) \right\}\\
&= \max_{\{Q_{XX'} \in \calQ(Q_{X}):~ I_{Q} \leq 2R \}} \inf_{\mu \geq 0} \left\{ \min\{2R - I_{Q},R\} + \mu \cdot (\expVAR - \Lambda(Q_{XX'},R) - I_{Q} + R) - \mu \sigma \right\},
\end{align}
where \eqref{TERM2EXPb1} is due to \eqref{V_TILDE_DEF} and the fact that $\max_{\{Q:~ g(Q) \geq 0\}} f(Q) = \max_{Q} \inf_{\mu \geq 0} \{f(Q) + \mu \cdot g(Q)\}$
and \eqref{TERM2EXPb2} is true thanks to \eqref{TERM2EXPa3}.
Now, we would like to solve for
\begin{align}
\sigma 
&\leq \max_{\{Q_{XX'} \in \calQ(Q_{X}):~ I_{Q} \leq 2R \}} \inf_{\mu \geq 0} \left\{ \min\{2R - I_{Q},R\} + \mu \cdot (\expVAR - \Lambda(Q_{XX'},R) - I_{Q} + R) - \mu \sigma \right\}, 
\end{align}
which is equivalent to the statement 
\begin{align}
&\exists Q_{XX'} \in \calQ(Q_{X}) ~\text{s.t.}~ I_{Q} \leq 2R,  
~~\forall \mu \geq 0: \nn \\
&\sigma 
\leq  \min\{2R - I_{Q},R\} + \mu \cdot (\expVAR - \Lambda(Q_{XX'},R) - I_{Q} + R) - \mu \sigma , 
\end{align}
or,
\begin{align}
&\exists Q_{XX'} \in \calQ(Q_{X}) ~\text{s.t.}~ I_{Q} \leq 2R,  
~~\forall \mu \geq 0: \nn \\
& \sigma
\leq \frac{\min\{2R - I_{Q},R\} + \mu \cdot (\expVAR - \Lambda(Q_{XX'},R) - I_{Q} + R)}{1 + \mu} , 
\end{align}
or, equivalently,
\begin{align}
\sigma
&\leq \max_{\{Q_{XX'} \in \calQ(Q_{X}):~ I_{Q} \leq 2R \}} \inf_{\mu \geq 0} \left\{ \frac{\min\{2R - I_{Q},R\} + \mu \cdot (\expVAR - \Lambda(Q_{XX'},R) - I_{Q} + R)}{1 + \mu} \right\} .
\end{align}
For simplicity, let us denote
\begin{align}
A &= \min\{2R - I_{Q},R\}, \\
B &= \expVAR - \Lambda(Q_{XX'},R) - I_{Q} + R,
\end{align}
such that
\begin{align}
\sigma
&\leq \max_{\{Q_{XX'} \in \calQ(Q_{X}):~ I_{Q} \leq 2R \}} \inf_{\mu \geq 0} \left\{ \frac{A + \mu B}{1 + \mu} \right\} \\ 
&= \max_{\{Q_{XX'} \in \calQ(Q_{X}):~ I_{Q} \leq 2R \}}
\min \{A,B\} \\
&= \max \left\{   
\begin{array}{l l}
\max_{\{Q_{XX'} \in \calQ(Q_{X}):~ I_{Q} \leq 2R,~B \geq 0 \}}
\min \{A,B\}  \\
\max_{\{Q_{XX'} \in \calQ(Q_{X}):~ I_{Q} \leq 2R,~B < 0 \}}
\min \{A,B\}   
\end{array} \right. \\
\label{abcdterminus1}
&= \max \left\{   
\begin{array}{l l}
\max_{\{Q_{XX'} \in \calQ(Q_{X}):~ I_{Q} \leq 2R,~B \geq 0 \}}
\min \{A,B\}  \\
\max_{\{Q_{XX'} \in \calQ(Q_{X}):~ I_{Q} \leq 2R,~B < 0 \}} B   
\end{array} \right. \\
\label{abcdterminus2}
&= \max_{\{Q_{XX'} \in \calQ(Q_{X}):~ I_{Q} \leq 2R,~B \geq 0 \}} \min \{A,B\}  \\
\label{abcdterminus3}
&= \max_{Q_{XX'} \in \calV(R,\expVAR)}
 \min\{2R - I_{Q},\expVAR - \Lambda(Q_{XX'},R) - I_{Q} + R,R\}\\
\label{abcdterminus4}
&= E_{\mbox{\tiny ut}}^{\mbox{\tiny ub}}(R,\expVAR),
\end{align}
where \eqref{abcdterminus1} and \eqref{abcdterminus2} are due to the fact that $A \geq 0$, while \eqref{abcdterminus3} and \eqref{abcdterminus4} follow from the definitions in (\ref{V_DEF}) and (\ref{UT_UB_EXPONENT}), respectively. Thus,
\begin{align}
\prob \left\{ -\frac{1}{n} \log P_{\mbox{\tiny e}} (\calC_{n}) \geq \expVAR \right\}  
&\DLEXE  \exp \left\{- \exp\left\{n \cdot \sup_{\sigma > 0} \min[ E_{1}(R,\expVAR,\sigma),\sigma] \right\} \right\} \\
&=  \exp \left\{- \exp\left\{n \cdot \sup_{0< \sigma \leq E_{\mbox{\tiny ut}}^{\mbox{\tiny ub}}(R,\expVAR)} \sigma \right\} \right\} \\
&=  \exp \left\{- e^{n \cdot E_{\mbox{\tiny ut}}^{\mbox{\tiny ub}}(R,\expVAR)} \right\} ,
\end{align}
and the proof of the upper bound of Theorem \ref{THEOREM_UPPER_TAIL} is complete.

\section{Proof of the Lower Bound of Theorem \ref{THEOREM_UPPER_TAIL}}
\label{SEC_VII}

Let the sets $\calB_{\epsilon}(m,\by)$ and $\calB_{\epsilon}$ be as defined in \eqref{B_DEF} and \eqref{B_UNION_DEF}, respectively. 
Also define $\calG_{\epsilon}(m,\by) = \calB_{\epsilon}^{\mbox{\tiny c}}(m,\by)$ and $\calG_{\epsilon} = \calB_{\epsilon}^{\mbox{\tiny c}}$. 
Let $\expVAR>0$ be given. 
Then,
\begin{align}
&\prob \left\{ -\frac{1}{n} \log P_{\mbox{\tiny e}} (\calC_{n}) \geq \expVAR \right\} \nonumber \\
\label{abcdeterminus1}
&= \prob \left\{\frac{1}{M} \sum_{m=0}^{M-1} \sum_{m' \neq m} \sum_{\by \in \calY^{n}} W(\by|\bx_{m}) \cdot \frac{\exp\{n g(\hat{P}_{\bx_{m'}\by}) \}}{\exp\{n g(\hat{P}_{\bx_{m}\by}) \} + Z_{m}(\by)}  \leq e^{-n \cdot \expVAR} \right\} \\
&\geq \prob \left\{\frac{1}{M} \sum_{m=0}^{M-1} \sum_{m' \neq m} \sum_{\by \in \calY^{n}} W(\by|\bx_{m}) \cdot \frac{\exp\{n g(\hat{P}_{\bx_{m'}\by}) \}}{\exp\{n g(\hat{P}_{\bx_{m}\by}) \} + Z_{m}(\by)}  \leq e^{-n \cdot \expVAR} ,~ \calC_{n} \in \calG_{\epsilon} \right\}  \\
\label{ToCall3}
&\DGEXE \prob \left\{\frac{1}{M} \sum_{m=0}^{M-1} \sum_{m' \neq m} 
\exp\{-n \Gamma(\hat{P}_{\bx_{m}\bx_{m'}},R-\epsilon) \}  
\leq e^{-n \cdot \expVAR} ,~ \calC_{n} \in \calG_{\epsilon} \right\},
\end{align}
where \eqref{abcdeterminus1} follows from the definitions of the probability of error and $Z_{m}(\by)$ in \eqref{PROBABILITYofErrorDEF} and \eqref{Z_DEF}, respectively.
Step \eqref{ToCall3} follows from the same considerations as in eqs.\ (\ref{EXP1})--(\ref{EXP4}). 
Now, define the event
\begin{align}
\calE_{0} = \left\{\frac{1}{M} \sum_{m=0}^{M-1} \sum_{m' \neq m} \exp\{-n \Gamma(\hat{P}_{\bx_{m}\bx_{m'}},R-\epsilon) \}  
\leq e^{-n \cdot \expVAR}  \right\} ,
\end{align}
such that, continuing from \eqref{ToCall3},
\begin{align}
\prob \left\{\calC_{n} \in \calE_{0} ,~ \calC_{n} \in \calG_{\epsilon} \right\} 
&= \prob \left\{ \bigcap_{\bar{m}=0}^{M-1}\bigcap_{\by\in\calY^n}\calG_\epsilon(\bar{m},\by) \middle| \calE_{0}  \right\} \cdot \prob \left\{ \calE_{0}  \right\}  \\
&= \left(1 - \prob \left\{ \bigcup_{\bar{m}=0}^{M-1}\bigcup_{\by\in\calY^n} \calB_\epsilon(\bar{m},\by) \middle| \calE_{0}  \right\} \right) \cdot \prob \left\{ \calE_{0}  \right\}  \\
&\geq \left(1 - \sum_{\bar{m}=0}^{M-1}\sum_{\by\in\calY^n} \prob \left\{ \calB_\epsilon(\bar{m},\by) \middle| \calE_{0}  \right\} \right) \cdot \prob \left\{ \calE_{0}  \right\}  \\  
\label{ToCall4}
&= \prob \left\{ \calE_{0}  \right\} - \sum_{\bar{m}=0}^{M-1}\sum_{\by\in\calY^n} \prob \left\{ \calB_\epsilon(\bar{m},\by) \cap \calE_{0}  \right\} .
\end{align}

\subsubsection*{Lower--bounding $\prob\{\calE_{0}\}$ in \eqref{ToCall4}}

First of all, note that
\begin{align}
\prob \left\{ \calE_{0}  \right\}
&= \prob \left\{\frac{1}{M} \sum_{m=0}^{M-1} \sum_{m' \neq m} 
\exp\{-n \Gamma(\hat{P}_{\bx_{m}\bx_{m'}},R-\epsilon) \}  
\leq e^{-n \cdot \expVAR} \right\} \\
\label{Cauchy1}
&= \prob \left\{ \sum_{Q_{XX'} \in \calQ(Q_{X})} N(Q_{XX'}) 
\exp\{-n \Gamma(Q_{XX'},R-\epsilon) \} \leq e^{n \cdot (R - \expVAR)} \right\} \\
\label{Cauchy2}
&\DEXE \prob \left\{ \max_{Q_{XX'} \in \calQ(Q_{X})} N(Q_{XX'}) 
\exp\{-n \Gamma(Q_{XX'},R-\epsilon) \} \leq e^{n \cdot (R - \expVAR)} \right\} \\
\label{INTERSECTION}
&= \prob \left\{\bigcap_{Q_{XX'} \in \calQ(Q_{X})} \left\{ N(Q_{XX'}) 
\leq e^{n \cdot (\Gamma(Q_{XX'},R-\epsilon) + R - \expVAR)} \right\} \right\} ,
\end{align}
where in \eqref{Cauchy1}, the definition of $N(Q_{XX'})$ in \eqref{NQ_def} was used,
and \eqref{Cauchy2} is due to the SME in \eqref{SME}.

Now, if there exists at least one $Q_{XX'} \in \calQ(Q_{X})$ for which $I_{Q}(X;X') < 2R$ and $2R - I_{Q}(X;X') > \Gamma(Q_{XX'},R-\epsilon) + R - \expVAR$, then this $Q_{XX'}$ alone is responsible for a double exponential decay of the probability of the event $\{  N(Q_{XX'})  \leq  e^{n \cdot (\Gamma(Q_{XX'},R-\epsilon) + R - \expVAR)} \}$ (thanks to Lemma \ref{The_Lower_Tail} in Appendix B), 
such that the probability in (\ref{INTERSECTION}), 
which is of the intersection over all $Q_{XX'} \in \calQ(Q_{X})$, decays double exponentially fast.
On the other hand, if for every $Q_{XX'} \in \calQ(Q_{X})$, either $I_{Q}(X;X') \geq 2R$ or $2R - I_{Q}(X;X') \leq \Gamma(Q_{XX'},R-\epsilon) + R - \expVAR$, then we have an intersection of polynomially many events whose probabilities all tend to one.
Thus, this probability is exponentially equal to one if and only if for every $Q_{XX'} \in \calQ(Q_{X})$, either $I_{Q}(X;X') \geq 2R$ or $2R - I_{Q}(X;X') \leq \Gamma(Q_{XX'},R-\epsilon) + R - \expVAR$, or equivalently,
\begin{align}
\label{Condition0}
2R  \leq   \min_{Q_{XX'} \in \calQ(Q_{X})}   \left\{  I_{Q}(X;X')  +  [\Gamma(Q_{XX'},R-\epsilon) + R - \expVAR]_{+}   \right\} .         
\end{align}
Let us now find what is the maximum value of $\expVAR$ for which this inequality holds true. The condition is equivalent to
\begin{align}
\min_{Q_{XX'} \in \calQ(Q_{X})}  \max_{0 \leq a \leq 1}  \{ & I_{Q}(X;X') + a  \left(\Gamma(Q_{XX'},R-\epsilon) + R - \expVAR \right)  \}  \geq   2R ,
\end{align}
or
\begin{align}
&\forall {Q_{XX'} \in \calQ(Q_{X})} ~  \exists a \in [0,1]: ~~ I_{Q}(X;X')  + a  \left(\Gamma(Q_{XX'},R-\epsilon) + R - \expVAR \right) \geq   2R ,
\end{align}
or
\begin{align}
&\forall {Q_{XX'} \in \calQ(Q_{X})} ~  \exists a \in [0,1]: ~~ \Gamma(Q_{XX'},R-\epsilon) + R + \frac{1}{a} \left(I_{Q}(X;X') - 2R\right) \geq \expVAR ,
\end{align}
or, equivalently, 
\begin{align}
\expVAR &\leq   \min_{Q_{XX'} \in \calQ(Q_{X})}  \max_{0 \leq a \leq 1}   \left\{ \Gamma(Q_{XX'},R-\epsilon) + R + \frac{1}{a} \left(I_{Q}(X;X') - 2R\right) \right\}  \\
&= \min_{Q_{XX'} \in \calQ(Q_{X})}  \left[ \Gamma(Q_{XX'},R-\epsilon) + R +  \left\{ 
\begin{array}{l l}
I_{Q}(X;X') - 2R     & \quad \text{$2R \geq I_{Q}(X;X')$  }\\
\infty               & \quad \text{$2R    <  I_{Q}(X;X')$  } 
\end{array} \right.    \right]   \\
&= \min_{\{Q_{XX'} \in \calQ(Q_{X}): ~ I_{Q}(X;X') \leq 2R \}} \left\{ \Gamma(Q_{XX'},R-\epsilon) + I_{Q}(X;X') - R \right\}  \\
&\leq  E_{\mbox{\tiny trc}}(R).
\end{align}
Thus, we assume that $\expVAR > E_{\mbox{\tiny trc}}(R)$, which ensures that there exists at least one $Q_{XX'} \in \calQ(Q_{X})$ for which $I_{Q}(X;X') \leq 2R$ and $ \Gamma(Q_{XX'},R-\epsilon) + R - \expVAR \leq 2R - I_{Q}(X;X')$, such that the probability in (\ref{INTERSECTION}) decays double exponentially fast. Define
\begin{align}
\styleA_{1} 
&= \{Q_{XX'} \in \calQ(Q_{X}):~ I_{Q}(X;X') > 2R \} \\
\styleA_{2} 
&= \{Q_{XX'} \in \calQ(Q_{X}):~ I_{Q}(X;X') \leq 2R, ~ \Gamma(Q_{XX'},R-\epsilon) + I_{Q}(X;X') - R \leq \expVAR + \epsilon \} \\
\styleA_{3} 
&= \{Q_{XX'} \in \calQ(Q_{X}):~ I_{Q}(X;X') \leq 2R, ~ \Gamma(Q_{XX'},R-\epsilon) + I_{Q}(X;X') - R > \expVAR + \epsilon \} .
\end{align}
Defining the event
\begin{align}
\calF_{0} = \bigcap_{Q_{XX'} \in \styleA_{1} \cup \styleA_{2}} \left\{ N(Q_{XX'}) = 0 \right\},
\end{align}
then considering the probability in (\ref{INTERSECTION}), we have that
\begin{align}
&\prob \left\{\bigcap_{Q_{XX'} \in \calQ(Q_{X})} \left\{ N(Q_{XX'}) 
\leq e^{n \cdot (\Gamma(Q_{XX'},R-\epsilon) + R - \expVAR)} \right\} \right\} \\  
&=\prob \left\{\bigcap_{Q_{XX'} \in \styleA_{1} \cup \styleA_{2} \cup \styleA_{3}} \left\{ N(Q_{XX'}) 
\leq e^{n \cdot (\Gamma(Q_{XX'},R-\epsilon) + R - \expVAR)} \right\} \right\} \\
&\geq \prob \left\{\bigcap_{Q_{XX'} \in \styleA_{3}} \left\{ N(Q_{XX'}) 
\leq e^{n \cdot (\Gamma(Q_{XX'},R-\epsilon) + R - \expVAR)} \right\}, \bigcap_{Q_{XX'} \in \styleA_{1} \cup \styleA_{2}} \left\{ N(Q_{XX'}) = 0 \right\} \right\} \\
&= \prob \left\{\bigcap_{Q_{XX'} \in \styleA_{3}} \left\{ N(Q_{XX'}) 
\leq e^{n \cdot (\Gamma(Q_{XX'},R-\epsilon) + R - \expVAR)} \right\} \middle| \calF_{0} \right\} \cdot \prob \left\{  \calF_{0} \right\} \\
&= \left(1 - \prob \left\{\bigcup_{Q_{XX'} \in \styleA_{3}} \left\{ N(Q_{XX'}) 
\geq e^{n \cdot (\Gamma(Q_{XX'},R-\epsilon) + R - \expVAR)} \right\} \middle| \calF_{0} \right\} \right) \cdot \prob \left\{  \calF_{0} \right\} \\
\label{ToCall1}
&\geq \left(1 - \sum_{Q_{XX'} \in \styleA_{3}} \prob \left\{ N(Q_{XX'}) \geq e^{n \cdot (\Gamma(Q_{XX'},R-\epsilon) + R - \expVAR)} \middle| \calF_{0} \right\} \right) \cdot \prob \left\{  \calF_{0} \right\}.
\end{align}
Next, it follows from Markov's inequality that
\begin{align}
&\prob \left\{ N(Q_{XX'}) \geq e^{n \cdot (\Gamma(Q_{XX'},R-\epsilon) + R - \expVAR)} \middle| \calF_{0} \right\} \\
&~~~\leq \frac{\mathbb{E} \left[ N(Q_{XX'}) \middle| \calF_{0} \right]}{e^{n \cdot (\Gamma(Q_{XX'},R-\epsilon) + R - \expVAR)}} \\
&~~~= \frac{\mathbb{E} \left[ \sum_{m=0}^{M-1} \sum_{m' \neq m} \IND \left\{(\bX_{m},\bX_{m'}) \in \calT(Q_{XX'}) \right\} \middle| \calF_{0} \right]}{e^{n \cdot (\Gamma(Q_{XX'},R-\epsilon) + R - \expVAR)}} \\
&~~~= \frac{\sum_{m=0}^{M-1} \sum_{m' \neq m} \prob \left\{ (\bX_{m},\bX_{m'}) \in \calT(Q_{XX'}) \middle| \calF_{0} \right\}}{e^{n \cdot (\Gamma(Q_{XX'},R-\epsilon) + R - \expVAR)}} \\
&~~~\leq \frac{e^{n 2R} \cdot \prob \left\{ (\bX_{0},\bX_{1}) \in \calT(Q_{XX'}) \middle| \calF_{0} \right\}}{e^{n \cdot (\Gamma(Q_{XX'},R-\epsilon) + R - \expVAR)}} .
\end{align} 
We continue from \eqref{ToCall1} and get that
\begin{align}
&\prob \left\{\bigcap_{Q_{XX'} \in \calQ(Q_{X})} \left\{ N(Q_{XX'}) 
\leq e^{n \cdot (\Gamma(Q_{XX'},R-\epsilon) + R - \expVAR)} \right\} \right\} \\  
&\geq \left(1 - \sum_{Q_{XX'} \in \styleA_{3}} \frac{e^{n 2R} \cdot \prob \left\{ (\bX_{0},\bX_{1}) \in \calT(Q_{XX'}) \middle| \calF_{0} \right\}}{e^{n \cdot (\Gamma(Q_{XX'},R-\epsilon) + R - \expVAR)}} \right) \cdot \prob \left\{  \calF_{0} \right\} \\
\label{ToCall2}
&= \prob \left\{  \calF_{0} \right\} - \sum_{Q_{XX'} \in \styleA_{3}} \frac{e^{n 2R} \cdot \prob \left\{ (\bX_{0},\bX_{1}) \in \calT(Q_{XX'}) , \calF_{0} \right\}}{e^{n \cdot (\Gamma(Q_{XX'},R-\epsilon) + R - \expVAR)}} .
\end{align}
In order to upper--bound the probabilities in the summation in \eqref{ToCall2}, we define the following truncated enumerators 
\begin{align} \label{TRUN_NQ_def}
\tilde{N}(Q_{XX'}) \DEF \sum_{m=2}^{M-1} \sum_{m' \in \{2,3,\ldots,M-1\} \setminus \{m\}} \IND \left\{(\bx_{m},\bx_{m'}) \in \calT(Q_{XX'}) \right\},
\end{align}
and the event
\begin{align}
\calF_{1} = \bigcap_{Q_{XX'} \in \styleA_{2}} \left\{ \tilde{N}(Q_{XX'}) = 0 \right\}.
\end{align} 
Then,
\begin{align}
&\prob \left\{ (\bX_{0},\bX_{1}) \in \calT(Q_{XX'}) , \calF_{0} \right\} \\
&=\prob \left\{ (\bX_{0},\bX_{1}) \in \calT(Q_{XX'}) , \bigcap_{\hat{Q}_{XX'} \in \styleA_{1} \cup \styleA_{2}} \left\{ N(\hat{Q}_{XX'}) = 0 \right\} \right\} \\
&=\prob \left\{ (\bX_{0},\bX_{1}) \in \calT(Q_{XX'}) , \bigcap_{\hat{Q}_{XX'} \in \styleA_{1} \cup \styleA_{2}} \bigcap_{m=0}^{M-1} \bigcap_{m' \in \{0,1,\ldots,M-1\} \setminus \{m\}} \left\{(\bX_{m},\bX_{m'}) \notin \calT(\hat{Q}_{XX'}) \right\} \right\} \\
&\leq \prob \left\{ (\bX_{0},\bX_{1}) \in \calT(Q_{XX'}) , \bigcap_{\hat{Q}_{XX'} \in \styleA_{1} \cup \styleA_{2}} \bigcap_{m=2}^{M-1} \bigcap_{m' \in \{2,3,\ldots,M-1\} \setminus \{m\}} \left\{(\bX_{m},\bX_{m'}) \notin \calT(\hat{Q}_{XX'}) \right\} \right\} \\
&= \prob \left\{ (\bX_{0},\bX_{1}) \in \calT(Q_{XX'}) \right\} \nn \\
&~~~~~~~~~~~~~~~~~~~~\times \prob \left\{ \bigcap_{\hat{Q}_{XX'} \in \styleA_{1} \cup \styleA_{2}} \bigcap_{m=2}^{M-1} \bigcap_{m' \in \{2,3,\ldots,M-1\} \setminus \{m\}} \left\{(\bX_{m},\bX_{m'}) \notin \calT(\hat{Q}_{XX'}) \right\} \right\} \\
&= \prob \left\{ (\bX_{0},\bX_{1}) \in \calT(Q_{XX'}) \right\} \cdot \prob \left\{ \bigcap_{\hat{Q}_{XX'} \in \styleA_{1} \cup \styleA_{2}} \left\{ \tilde{N}(\hat{Q}_{XX'}) = 0 \right\} \right\} \\
&\leq \prob \left\{ (\bX_{0},\bX_{1}) \in \calT(Q_{XX'}) \right\} \cdot \prob \left\{ \bigcap_{\hat{Q}_{XX'} \in \styleA_{2}} \left\{ \tilde{N}(\hat{Q}_{XX'}) = 0 \right\} \right\} \\
&= \prob \left\{ (\bX_{0},\bX_{1}) \in \calT(Q_{XX'}) \right\} \cdot \prob \left\{ \calF_{1} \right\}. 
\end{align}
Substituting it back into \eqref{ToCall2}, now yields
\begin{align}
&\prob \left\{\bigcap_{Q_{XX'} \in \calQ(Q_{X})} \left\{ N(Q_{XX'}) 
\leq e^{n \cdot (\Gamma(Q_{XX'},R-\epsilon) + R - \expVAR)} \right\} \right\} \\  
&\geq \prob \left\{  \calF_{0} \right\} - \sum_{Q_{XX'} \in \styleA_{3}} \frac{e^{n 2R} \cdot \prob \left\{ (\bX_{0},\bX_{1}) \in \calT(Q_{XX'}) , \calF_{0} \right\}}{e^{n \cdot (\Gamma(Q_{XX'},R-\epsilon) + R - \expVAR)}} \\
&\geq \prob \left\{  \calF_{0} \right\} - \sum_{Q_{XX'} \in \styleA_{3}} \frac{e^{n 2R} \cdot \prob \left\{ (\bX_{0},\bX_{1}) \in \calT(Q_{XX'}) \right\} \cdot \prob \left\{ \calF_{1} \right\}}{e^{n \cdot (\Gamma(Q_{XX'},R-\epsilon) + R - \expVAR)}} \\
&= \prob \left\{  \calF_{0} \right\} - \prob \left\{ \calF_{1} \right\} \cdot \sum_{Q_{XX'} \in \styleA_{3}} \frac{e^{n 2R} \cdot \prob \left\{ (\bX_{0},\bX_{1}) \in \calT(Q_{XX'}) \right\}}{e^{n \cdot (\Gamma(Q_{XX'},R-\epsilon) + R - \expVAR)}}.
\end{align}
Generally, it follows that $\prob \left\{\calF_{0}\right\} \leq \prob \left\{\calF_{1}\right\}$. First, we lower--bound $\prob \left\{\calF_{0}\right\}$.
The following proposition is proved in Appendix I:
\begin{proposition}
	\label{Super_Enum_Lower_Bound}
	If $\expVAR < E_{\mbox{\tiny ex}}(R)$, then
	\begin{align}
	\prob \left\{ \calF_{0}  \right\} 
	\DGEXE \exp \left\{- \exp\left\{n \cdot \max_{Q_{XX'} \in  \styleA_{2}} \{2R - I_{Q}(X;X')\} \right\} \right\} .
	\end{align}
\end{proposition}
In addition, we can easily prove that under the condition of $\expVAR < E_{\mbox{\tiny ex}}(R)$, $\prob \left\{ \calF_{1} \right\}$ can be upper--bounded by the same expression that lower--bounds $\prob \left\{ \calF_{0} \right\}$. 
We have that 
\begin{align}
\prob \left\{ \calF_{1} \right\}
&= \prob \left\{ \bigcap_{Q_{XX'} \in \styleA_{2}} \left\{ \tilde{N}(Q_{XX'}) = 0 \right\} \right\} \\
&\leq \min_{Q_{XX'} \in \styleA_{2}} \prob \left\{ \tilde{N}(Q_{XX'}) = 0 \right\} \\
\label{TERM2EXPc1}
&\DLEXE \min_{Q_{XX'} \in \styleA_{2}} \exp \left\{- \min \left(e^{n(2R - I_{Q}(X;X'))}, e^{nR} \right) \right\} \\
\label{TERM2EXPc2}
&= \min_{Q_{XX'} \in \styleA_{2}} \exp \left\{- e^{n(2R - I_{Q}(X;X'))} \right\} \\
\label{ToCall7}
&= \exp \left\{- \exp\left\{n \cdot \max_{Q_{XX'} \in \styleA_{2}} \{2R - I_{Q}(X;X')\} \right\} \right\},
\end{align}
where \eqref{TERM2EXPc1} is due to Lemma \ref{The_Lower_Tail} in Appendix B and \eqref{TERM2EXPc2} follows from the fact that $\expVAR < E_{\mbox{\tiny ex}}(R)$ is equivalent to $\min_{Q_{XX'} \in \styleA_{2}} I_{Q}(X;X') > R$ (Appendix I). 
Hence,
\begin{align}
\prob \left\{\calF_{0}\right\}  
\DEXE \prob \left\{\calF_{1}\right\} 
\DEXE \exp \left\{- \exp\left\{n \cdot \max_{Q_{XX'} \in \styleA_{2}} \{2R - I_{Q}(X;X')\} \right\} \right\}.
\end{align}
Using the definition of the set $\styleA_{3}$ provides    
\begin{align}
&\prob \left\{ \calE_{0} \right\} \nn \\  
&\DEXE \prob \left\{\bigcap_{Q_{XX'} \in \calQ(Q_{X})} \left\{ N(Q_{XX'}) 
\leq e^{n \cdot (\Gamma(Q_{XX'},R-\epsilon) + R - \expVAR)} \right\} \right\} \\
&\geq \prob \left\{  \calF_{0} \right\} - \prob \left\{ \calF_{1} \right\} \cdot \sum_{Q_{XX'} \in \styleA_{3}} \frac{e^{n 2R} \cdot \prob \left\{ (\bX_{0},\bX_{1}) \in \calT(Q_{XX'}) \right\}}{e^{n \cdot (\Gamma(Q_{XX'},R-\epsilon) + R - \expVAR)}}  \\
\label{ToCall5}
&\DEXE \left(1- \sum_{Q_{XX'} \in \styleA_{3}} \frac{e^{n \cdot (2R-I_{Q}(X;X'))} }{e^{n \cdot (\Gamma(Q_{XX'},R-\epsilon) + R - \expVAR)}}\right) \cdot \exp \left\{- \exp\left\{n \cdot \max_{Q_{XX'} \in \styleA_{2}} \{2R - I_{Q}(X;X')\} \right\} \right\}   \\
&\DEXE \left( 1- e^{-n \epsilon}  \right) \cdot\exp \left\{- \exp\left\{n \cdot \max_{Q_{XX'} \in \styleA_{2}} \{2R - I_{Q}(X;X')\} \right\} \right\}   \\
&\DEXE \exp \left\{- \exp\left\{n \cdot \max_{Q_{XX'} \in \styleA_{2}} \{2R - I_{Q}(X;X')\} \right\} \right\}.
\end{align}

\subsubsection*{Upper--bounding $\prob \left\{ \calB_\epsilon(\bar{m},\by) \cap \calE_{0}  \right\}$ in \eqref{ToCall4}}

Recall that 
\begin{align}
\prob \left\{ \calB_\epsilon(\bar{m},\by) \cap \calE_{0}  \right\} 
&=\prob \left\{\sum_{\tilde{m} \in \{0,1,\ldots,M-1\} \setminus \{\bar{m}\}} \exp\{n g(\hat{P}_{\bX_{\tilde{m}}\by})\} \le e^{n \cdot \alpha(R-\epsilon,\hat{P}_{\by})}, 
\right. \nn \\ &~~~~~~~~~\left. 
\sum_{m=0}^{M-1} \sum_{m' \neq m} 
\exp\{-n \Gamma(\hat{P}_{\bX_{m}\bX_{m'}},R-\epsilon) \}  
\leq e^{n \cdot (R- \expVAR)}  \right\} .
\end{align}
In order to upper--bound this probability, we do the following. In the first event, instead of summing over $\{0,1, \dotsc, M-1\} \setminus \{\bar{m}\}$, we sum over $\{\lfloor M/2 \rfloor,\lfloor M/2 \rfloor + 1, \dotsc, M-1\} \setminus \{\bar{m}\}$,  
and in the second event, instead of summing over $\{(m,m'):~ m,m' \in \{0,1, \dotsc, M-1\},~ m \neq m'\}$, we sum over $\{(m,m'):~ m,m' \in \{0,1, \dotsc, \lfloor M/2 \rfloor-1\},~ m \neq m'\}$, hence, the two events become independent:
\begin{align}
&\prob \left\{ \calB_\epsilon(\bar{m},\by) \cap \calE_{0}  \right\} \nn \\
\label{ToCall6}
&\leq \prob \left\{\sum_{\tilde{m} \in \{\lfloor M/2 \rfloor,\ldots,M-1\} \setminus \{\bar{m}\}} \exp\{n g(\hat{P}_{\bX_{\tilde{m}}\by})\} \le e^{n \cdot \alpha(R-\epsilon,\hat{P}_{\by})} \right\} 
\nn \\ &~~~
\times \prob \left\{ \sum_{m=0}^{\lfloor M/2 \rfloor-1} \sum_{m' \in \{0,1,\ldots,\lfloor M/2 \rfloor-1\} \setminus \{m\}} 
\exp\{-n \Gamma(\hat{P}_{\bX_{m}\bX_{m'}},R-\epsilon) \}  
\leq e^{n \cdot (R- \expVAR)}  \right\} .
\end{align}
As for the first factor in \eqref{ToCall6}, note that its sum has exponentially many terms as $Z_{m}(\by)$, and hence is also upper--bounded as in \eqref{B_DE_UB}. The second factor in \eqref{ToCall6} can be upper--bounded using similar analysis as in the proof in Section \ref{SEC_VI}, which results an upper bound similar to \eqref{ToCall7}. Thus, 
\begin{align}
\prob \left\{ \calB_\epsilon(\bar{m},\by) \cap \calE_{0}  \right\} 
\label{ToCall8}
&\leq \exp\{-e^{n\epsilon}+n\epsilon+1\} \cdot
\exp \left\{- \exp\left\{n \cdot \max_{Q_{XX'} \in \styleA_{2}} \{2R - I_{Q}(X;X')\} \right\} \right\}.
\end{align}

\subsubsection*{Final Steps}
Finally, we continue from \eqref{ToCall4} and use the results of \eqref{ToCall5} and \eqref{ToCall8} to obtain 
\begin{align}
&\prob \left\{ -\frac{1}{n} \log P_{\mbox{\tiny e}} (\calC_{n}) \geq \expVAR \right\} \nonumber \\ 
&\DGEXE \prob \left\{ \calE_{0}  \right\} - \sum_{\bar{m}=0}^{M-1}\sum_{\by\in\calY^n} \prob \left\{ \calB_\epsilon(\bar{m},\by) \cap \calE_{0}  \right\} \\
&\DGEXE \exp \left\{- \exp\left\{n \cdot \max_{Q_{XX'} \in \styleA_{2}} \{2R - I_{Q}(X;X')\} \right\} \right\} \nn \\
&~~- \sum_{\bar{m}=0}^{M-1}\sum_{\by\in\calY^n} \exp\{-e^{n\epsilon}+n\epsilon+1\} \cdot
\exp \left\{- \exp\left\{n \cdot \max_{Q_{XX'} \in \styleA_{2}} \{2R - I_{Q}(X;X')\} \right\} \right\} \\
&= \left(1 - e^{nR} \cdot |\calY|^n \cdot \exp\{-e^{n\epsilon}+n\epsilon+1\} \right) \cdot \exp \left\{- \exp\left\{n \cdot \max_{Q_{XX'} \in \styleA_{2}} \{2R - I_{Q}(X;X')\} \right\} \right\} \\
&\DEXE \exp \left\{- \exp\left\{n \cdot \max_{Q_{XX'} \in \styleA_{2}} \{2R - I_{Q}(X;X')\} \right\} \right\} ,
\end{align}
which proves the lower bound of Theorem \ref{THEOREM_UPPER_TAIL}.

\section*{Appendix A}
\renewcommand{\theequation}{A.\arabic{equation}}
\setcounter{equation}{0}  
\subsection*{Preliminaries}

The main purpose of this appendix is to provide the general setting and the main results that are borrowed from \cite{Suen}.  

Let $\{U_{\bk}\}_{\bk \in \calK}$, where $\calK$ is a set of multidimensional indexes, be a family of Bernoulli random variables. 
Let $G$ be a dependency graph for $\{U_{\bk}\}_{\bk \in \calK}$, i.e., a graph with vertex set $\calK$ such that if $\calA$ and $\calB$ are two disjoint subsets of $\calK$, and $G$ contains no edge between $\calA$ and $\calB$, then the families $\{U_{\bk}\}_{\bk \in \calA}$ and $\{U_{\bk}\}_{\bk \in \calB}$ are independent.   
Let $S = \sum_{\bk \in \calK} U_{\bk}$ and $\Delta = \mathbb{E}[S]$. Moreover, we write $\bi \sim \bj$ if $(\bi,\bj)$ is an edge in the dependency graph $G$. Let
\begin{align}
\LambdaBOOM =  \max_{\bi \in \calK} \mathbb{E} [U_{\bi}],
\end{align} 
\begin{align}
\label{OMEGA_I_DEF}
\Omega_{\bi} =  \sum_{\bj \in \calK,\bj \sim \bi} \mathbb{E} [U_{\bj}],
\end{align} 
\begin{align}
\Omega =  \max_{\bi \in \calK} \sum_{\bj \in \calK,\bj \sim \bi} \mathbb{E} [U_{\bj}],
\end{align}
and
\begin{align}
\Theta = \frac{1}{2} \sum_{\bi \in \calK} \sum_{\bj \in \calK, \bj \sim \bi} \mathbb{E} [U_{\bi}U_{\bj}].
\end{align}
The following result will be used in the proof of Lemma \ref{The_Lower_Tail} in Appendix B:
\begin{fact} \label{FACT1}
	 With notations as above, \cite[Theorem 10]{Suen} states that for any $0\leq a \leq 1$, 
	\begin{align}
	\prob \{S \leq a \Delta\} \leq \exp \left\{- \min \left(
	(1-a)^{2} \frac{\Delta^{2}}{8\Theta+2\Delta},
	(1-a) \frac{\Delta}{6\Omega}
	\right) \right\} .
	\end{align}
\end{fact}
The following result will be used in the proof of Lemma \ref{The_Upper_Tail_Thin_Sharp} in Appendix B:
\begin{fact} \label{FACT2}
	With notations as above, \cite[Theorem 3]{Suen} states that, 
	\begin{align}
	\prob \{S = 0\} \leq \exp \left\{- \min \left(
	\frac{\Delta^{2}}{8\Theta},
	\frac{\Delta}{6\Omega},
	\frac{\Delta}{2}
	\right) \right\} .
	\end{align}
\end{fact}
Next, define $\varphi(x)$, $0 \leq x \leq e^{-1}$, to be the smallest root $t$ of the equation 
\begin{align}
\label{varphi_DEF}
t = e^{x t}.
\end{align}
It is well known that $\varphi(x)$ is well defined in $[0,e^{-1}]$, 
in particular, $\varphi(x) = 1 + x + O(x^{2})$. The following lower bound will be useful in the proof of Proposition \ref{Super_Enum_Lower_Bound} in Appendix I.
\begin{fact} \label{FACT3}
	With notations as above, suppose further that $\Omega + \LambdaBOOM \leq e^{-1}$. Then, with $\varphi$ defined by (\ref{varphi_DEF}), \cite[Theorem 9]{Suen} states that
	\begin{align}
	\prob \{S=0\} \geq \exp \{-\Delta \cdot \varphi(\Omega + \LambdaBOOM)\}.
	\end{align} 
\end{fact}

\section*{Appendix B}
\renewcommand{\theequation}{B.\arabic{equation}}
\setcounter{equation}{0}  
\subsection*{Proof of Theorem \ref{2D_Large_Deviations}}
Let us abbreviate $\IND(m,m') \DEF \IND \left\{(\bx_{m},\bx_{m'}) \in \calT(Q_{XX'}) \right\}$, such that the enumerator $N(Q_{XX'})$ can also be written by
\begin{align} \label{ENUM_DEF}
N(Q_{XX'}) = \sum_{(m,m') \in [M]_{*}^{2}} \IND(m,m'),
\end{align}
where the set $[M]_{*}^{2}$ is an abbreviation for the set $\{(m,m'):~ m,m' \in \{0,1, \dotsc, M-1\},~ m \neq m'\}$.

Before proving Theorem \ref{2D_Large_Deviations}, we start with the following series of partial results, that are going to be instrumental in proving Theorem \ref{2D_Large_Deviations}.
\begin{lemma} \label{Pairwise_Independence}
	For any two pairs $(i,j),(i,k) \in [M]_{*}^{2}$, $j \neq k$,
	\begin{align}
	\mathbb{E} [\IND(i,j)\IND(i,k)] \doteq \exp\{-2 n I_{Q}(X;X') \}.
	\end{align}
\end{lemma}
\textbf{Proof:} Since all codewords are independent, it follows by the method of types that
\begin{align}
&\mathbb{E} [\IND(i,j)\IND(i,k)] \nonumber \\
&=\prob \left\{(\bX_{i},\bX_{j}) \in \calT(Q_{XX'}), (\bX_{i},\bX_{k}) \in \calT(Q_{XX'}) \right\}  \\
&= \sum_{\bx \in \calT(Q_{X})} \prob \{\bX_{i} = \bx\} \cdot \prob \left\{(\bx,\bX_{j}) \in \calT(Q_{XX'}), (\bx,\bX_{k}) \in \calT(Q_{XX'}) \right\} \\
\label{TERM2EXPf}
&= \sum_{\bx \in \calT(Q_{X})} \prob \{\bX_{i} = \bx\} \cdot \prob \left\{(\bx,\bX_{j}) \in \calT(Q_{XX'}) \right\} \cdot \prob \left\{(\bx,\bX_{k}) \in \calT(Q_{XX'}) \right\}\\
&\doteq \sum_{\bx \in \calT(Q_{X})} \prob \{\bX_{i} = \bx\} \cdot \exp\{- n I_{Q}(X;X') \} \cdot \exp\{- n I_{Q}(X;X') \}\\
&= \exp\{-2 n I_{Q}(X;X') \},
\end{align} 
where \eqref{TERM2EXPf} is because $\bX_{j}$ and $\bX_{k}$ are statistically independent. Lemma \ref{Pairwise_Independence} is proved.

Now, we have the following Lemma, which proposes an upper bound on the probability of the lower tail in the case of TP type classes.

\begin{lemma} \label{The_Lower_Tail}
	Let $\epsilon>0$ be given. Then, for any $Q_{XX'}$ such that $I_{Q}(X;X') \leq 2R - \epsilon$,
	\begin{align}
	\prob  \left\{ N(Q_{XX'})  
	\leq e^{-n\epsilon } \cdot \mathbb{E}[N(Q_{XX'})]   \right\} 
	\DLEXE \exp \left\{- \min \left(e^{n(2R - I_{Q}(X;X'))}, e^{nR} \right) \right\}.
	\end{align}
\end{lemma}
\textbf{Proof:}
We use the result of Fact \ref{FACT1}, that appears in Appendix A.
In our case, we have $a = e^{-n\epsilon }$ and $\Delta \doteq e^{n(2R - I_{Q}(X;X'))}$, and it only remains to assess the quantities $\Theta$ and $\Omega$. One can easily check that the indicator random variables $\IND(i,j)$ and $\IND(k,l)$ are independent as long as $i \neq k$ and $j \neq l$. Thus, we define our dependency graph in a way that each vertex $(i,j)$ is connected to exactly $e^{nR} + e^{nR} - 2$ vertices of the form $(i,l)$, $l \neq j$ or $(k,j)$, $k \neq i$. If the vertices $(i,j)$ and $(k,l)$ are connected, we denote it by $(i,j) \sim (k,l)$. Using the result of Lemma \ref{Pairwise_Independence}, we get that
\begin{align}
\Theta &= \frac{1}{2} \sum_{(i,j) \in [M]_{*}^{2}} \sum_{(k,l) \in [M]_{*}^{2}, (k,l) \sim (i,j)} \mathbb{E} [\IND(i,j)\IND(k,l)] \\
&\doteq \frac{1}{2} e^{2nR} \cdot (e^{nR} + e^{nR} - 2) \cdot e^{-2nI_{Q}(X;X')} \\
&\doteq e^{n(3R - 2I_{Q}(X;X'))},
\end{align}
and 
\begin{align}
\Omega &=  \max_{(i,j) \in [M]_{*}^{2}} \sum_{(k,l) \in [M]_{*}^{2}, (k,l) \sim (i,j)} \mathbb{E} [\IND(k,l)] \\
&\doteq (e^{nR} + e^{nR} - 2) \cdot e^{-nI_{Q}(X;X')} \\
&\doteq e^{n(R - I_{Q}(X;X'))} .
\end{align}
Then,
\begin{align}
\label{RESULT1}
\frac{\Delta}{6\Omega}
\doteq \frac{e^{n(2R - I_{Q}(X;X'))}}
{e^{n(R - I_{Q}(X;X'))}} = e^{nR},
\end{align}
and,
\begin{align}
\label{RESULT2}
\frac{\Delta^{2}}{8\Theta + 2\Delta}
&\doteq \frac{e^{n(4R - 2I_{Q}(X;X'))}}
{e^{n(3R - 2I_{Q}(X;X'))} + e^{n(2R - I_{Q}(X;X'))}} \\
&= \frac{e^{n(2R - I_{Q}(X;X'))}}
{e^{n(R - I_{Q}(X;X'))} + 1 } \\
&\doteq \frac{e^{n(2R - I_{Q}(X;X'))}}
{e^{n[R - I_{Q}(X;X')]_{+}} }. 
\end{align}
Hence,
\begin{align}
\label{DE_RESULT}
\prob  \left\{ N(Q_{XX'})  
\leq e^{-n\epsilon } \cdot \mathbb{E}[N(Q_{XX'})]   \right\} 
&\DLEXE \exp \left\{- \min \left(
\frac{e^{n(2R - I_{Q}(X;X'))}}
{e^{n[R - I_{Q}(X;X')]_{+}}  },
e^{nR}
\right) \right\} \\
&= \exp \left\{- \min \left(
e^{n(2R - I_{Q}(X;X'))},
e^{nR}
\right) \right\}.
\end{align}
Now, if $I_{Q}(X;X') \leq R$, we get
\begin{align}
\prob  \left\{ N(Q_{XX'})  
\leq e^{-n\epsilon } \cdot \mathbb{E}[N(Q_{XX'})]   \right\} 
&\DLEXE \exp \left\{- e^{nR} \right\} ,
\end{align}
and otherwise, if $R < I_{Q}(X;X') \leq 2R-\epsilon$,
\begin{align}
\prob  \left\{ N(Q_{XX'})  
\leq e^{-n\epsilon } \cdot \mathbb{E}[N(Q_{XX'})]   \right\} 
&\DLEXE \exp \left\{- e^{n(2R - I_{Q}(X;X'))} \right\} \\
&\leq \exp \left\{- e^{n\epsilon}  \right\},
\end{align} 
which completes the proof of Lemma \ref{The_Lower_Tail}. 

Before moving on to the upper tail, we need the following lemma, proved in Appendix C.
\begin{lemma} \label{Integer_Moments}
	For any $k \in \mathbb{N}$,
	\begin{align}
	\mathbb{E} \left[ N(Q_{XX'})^{k}  \right] 
	&\lexe      \left\{   
	\begin{array}{l l}
	\exp \{n k \left(2R - I_{Q}(X;X')\right) \}    & \quad \text{  $I_{Q}(X;X') < 2R$  }   \\
	\exp \{n\left(2R - I_{Q}(X;X')\right) \}    & \quad \text{  $I_{Q}(X;X') > 2R$  }
	\end{array} \right..
	\end{align}
\end{lemma}
Concerning the upper tail, we have the following result.
\begin{lemma} \label{The_Upper_Tail}
	Let $\epsilon>0$ be given. Then, for any $Q_{XX'}$ such that $I_{Q}(X;X') \leq 2R$,
	\begin{align}
	\prob  \left\{ N(Q_{XX'}) \geq e^{n\epsilon } \cdot \mathbb{E}[N(Q_{XX'})] \right\} \lexe e^{-n \infty}. 
	\end{align}
\end{lemma}
\textbf{Proof:} For any $k \in \mathbb{N}$, Markov's inequality and Lemma \ref{Integer_Moments} implies that
\begin{align}
\prob  \left\{ N(Q_{XX'}) \geq e^{n\epsilon } \cdot \mathbb{E}[N(Q_{XX'})]   \right\} 
&\leq \inf_{k \in \mathbb{N}}  \frac{\mathbb{E}[N(Q_{XX'})^{k}]}{e^{nk\epsilon} \cdot (\mathbb{E}[N(Q_{XX'})])^{k}}  \\
&\lexe \inf_{k \in \mathbb{N}}  \frac{\exp \left\{n k \left(2R - I_{Q}(X;X')\right) \right\}}{e^{nk\epsilon} \cdot (\exp \left\{n \left(2R - I_{Q}(X;X')\right) \right\})^{k}}  \\
&= \inf_{k \in \mathbb{N}} \exp\{-nk\epsilon\} ,
\end{align}
thus,
\begin{align}
\liminf_{n \to \infty} - \frac{1}{n} \log \prob  \left\{ N(Q_{XX'}) \geq e^{n\epsilon } \cdot \mathbb{E}[N(Q_{XX'})]   \right\} 
\geq \sup_{k \in \mathbb{N}} k \epsilon = \infty,
\end{align} 
which proves Lemma \ref{The_Upper_Tail}.

Next, we treat the TE type classes.
\begin{lemma} \label{The_Upper_Tail_Thin}
	Let $\epsilon>0$ be given. Then, for any $Q_{XX'}$ such that $I_{Q}(X;X') \geq 2R$,
	\begin{align}
	\prob  \left\{ N(Q_{XX'}) \geq e^{n\epsilon } \right\} \lexe e^{-n \infty}.
	\end{align} 
\end{lemma}
\textbf{Proof:} For any $k \in \mathbb{N}$, Markov's inequality and Lemma \ref{Integer_Moments} implies that
\begin{align}
\prob  \left\{ N(Q_{XX'}) \geq e^{n\epsilon }  \right\} 
&\leq \inf_{k \in \mathbb{N}}  \frac{\mathbb{E}[N(Q_{XX'})^{k}]}{e^{nk\epsilon}}  \\
&\lexe \inf_{k \in \mathbb{N}}  \frac{\exp \left\{n \left(2R - I_{Q}(X;X')\right) \right\}}{e^{nk\epsilon}} \\
&= \inf_{k \in \mathbb{N}} \exp\left\{-n  \left(I_{Q}(X;X') - 2R + k\epsilon\right) \right\} ,
\end{align}
and hence,
\begin{align}
\liminf_{n \to \infty} - \frac{1}{n} \log \prob  \left\{ N(Q_{XX'}) \geq e^{n\epsilon }  \right\} \geq \sup_{k \in \mathbb{N}} \left\{I_{Q}(X;X') - 2R + k\epsilon\right\} = \infty,
\end{align} 
which completes the proof of Lemma \ref{The_Upper_Tail_Thin}. Furthermore, we have 
\begin{lemma} \label{The_Upper_Tail_Thin_Sharp}
	For any $Q_{XX'}$ such that $I_{Q}(X;X') \geq 2R$,
	\begin{align}
	\prob \left\{ N(Q_{XX'}) \geq 1 \right\} \doteq \exp \{n(2R-I_{Q}(X;X'))\}.
	\end{align}
\end{lemma}
\textbf{Proof:} An upper bound simply follows from Markov's inequality:
\begin{align}
\prob \left\{ N(Q_{XX'}) \geq 1 \right\} \leq \mathbb{E}[N(Q_{XX'})] \doteq \exp \{n(2R-I_{Q}(X;X'))\}.
\end{align}
For the lower bound, we use Fact \ref{FACT2} from Appendix A. 
Similarly to \eqref{RESULT1} and \eqref{RESULT2}, we have
\begin{align}
\frac{\Delta^{2}}{8\Theta}
&\doteq \frac{e^{n(4R - 2I_{Q}(X;X'))}}
{e^{n(3R - 2I_{Q}(X;X'))}} 
= e^{nR},
\end{align}
and,
\begin{align}
\frac{\Delta}{6\Omega}
&\doteq \frac{e^{n(2R - I_{Q}(X;X'))}}
{e^{n(R - I_{Q}(X;X'))}} = e^{nR}.
\end{align}
Now, since $I_{Q}(X;X') \geq 2R$,
\begin{align}
\prob  \left\{ N(Q_{XX'}) = 0 \right\} 
&\leq \exp \left\{- \min \left(
e^{nR}, e^{nR}, \frac{1}{2} \cdot e^{n(2R - I_{Q}(X;X'))}
\right) \right\} \\
&= \exp \left\{- \frac{1}{2} \cdot e^{n(2R - I_{Q}(X;X'))} \right\} \\
\label{TAYLOR_of_Exponential}
&\leq 1 - \frac{1}{2} \cdot e^{n(2R - I_{Q}(X;X'))} + \frac{1}{8} \cdot e^{n(4R - 2I_{Q}(X;X'))} ,
\end{align}
where \eqref{TAYLOR_of_Exponential} is due to the fact that for $t \geq 0$, $e^{-t} \leq 1-t + \frac{1}{2} t^{2}$, and so, 
\begin{align}
\prob \left\{ N(Q_{XX'}) \geq 1 \right\} 
&= 1 - \prob  \left\{ N(Q_{XX'}) = 0 \right\} \\ 
&\geq \frac{1}{2} \cdot \exp\{n(2R - I_{Q}(X;X'))\} - \frac{1}{8} \cdot \exp\{n(4R - 2I_{Q}(X;X'))\} \\
&\doteq \exp\{n(2R-I_{Q}(X;X'))\},
\end{align}
which is compatible with the above upper bound, proving Lemma \ref{The_Upper_Tail_Thin_Sharp}. 

\subsubsection*{Proof of Theorem \ref{2D_Large_Deviations}:}
We use the results of Lemmas \ref{The_Lower_Tail}, \ref{The_Upper_Tail}, \ref{The_Upper_Tail_Thin}, and \ref{The_Upper_Tail_Thin_Sharp}, and get the following exponential rate of decay for $\prob \left\{ N(Q_{XX'}) \geq e^{n s} \right\}$: 
\begin{align}
E(R,Q,s) 
&= \left\{   
\begin{array}{l l}
I_{Q}(X;X')-2R   & \quad \text{$I_{Q}(X;X') \geq 2R, s \leq 0$}   \\
\infty           & \quad \text{$I_{Q}(X;X') \geq 2R, s > 0$}   \\
0                & \quad \text{$I_{Q}(X;X') \leq 2R, s \leq 2R - I_{Q}(X;X')$}   \\
\infty           & \quad \text{$I_{Q}(X;X') \leq 2R, s > 2R - I_{Q}(X;X')$}   
\end{array} \right. \\
&= \left\{   
\begin{array}{l l}
\left[I_{Q}(X;X')-2R\right]_{+}   & \quad \text{$I_{Q}(X;X') \geq 2R, s \leq [2R - I_{Q}(X;X')]_{+}$}   \\
\infty           & \quad \text{$I_{Q}(X;X') \geq 2R, s > [2R - I_{Q}(X;X')]_{+}$}   \\
\left[I_{Q}(X;X')-2R\right]_{+}  & \quad \text{$I_{Q}(X;X') \leq 2R, s \leq [2R - I_{Q}(X;X')]_{+}$}   \\
\infty           & \quad \text{$I_{Q}(X;X') \leq 2R, s > [2R - I_{Q}(X;X')]_{+}$}   
\end{array} \right. \\
&= \left\{   
\begin{array}{l l}
\left[I_{Q}(X;X')-2R\right]_{+}   & \quad \text{$[2R - I_{Q}(X;X')]_{+} \geq s$}   \\
\infty           & \quad \text{$[2R - I_{Q}(X;X')]_{+} < s$}    
\end{array} \right. ,
\end{align}
which proves Theorem \ref{2D_Large_Deviations}.

\section*{Appendix C}
\renewcommand{\theequation}{C.\arabic{equation}}
\setcounter{equation}{0}  
\subsection*{Proof of Lemma \ref{Integer_Moments}}

For a set of indices $\calJ$ let us denote $\calJ_{*}^{2} = \{(j,j') \in \calJ^{2}:~ j \neq j'\}$. Recall that $\IND(m,m') = \IND\{(\bX_{m},\bX_{m'}) \in \calT(Q_{XX'})\}$ and $N(Q_{XX'}) = \sum_{(m,m') \in [M]_{*}^{2}} \IND(m,m')$. We show by induction that
\begin{align}
\mathbb{E} \left[ N(Q_{XX'})^{k}  \right] 
&\lexe      \left\{   
\begin{array}{l l}
e^{n k (2R - I)}    & \quad \text{  $I < 2R$  }   \\
e^{-n (I - 2R) }    & \quad \text{  $I > 2R$  }
\end{array} \right. ,
\end{align}
where $I$ is a shorthand notation for $I_{Q}(X;X')$.
This clearly holds for $k=1$ by linearity of expectation. We assume it holds up to $k-1$ and show this for $k$.

\underline{Proof for $k$:} Assume that $\{(m_{i},m_{i}')\}_{i=1}^{k-1}$ are given, where $(m_{i},m_{i}') \in [M]_{*}^{2}$ for all $i \in [k-1]$. 
Let $\calM_{k-1} = \bigcup_{i=1}^{k-1} \{\{m_{i}\} \cup \{m_{i}'\}\}$ be the set of indices of the $k-1$ pairs of codeword indices $\{(m_{i},m_{i}')\}_{i=1}^{k-1}$.
We condition on all these codewords, and then compute expectation w.r.t.\ all other codewords. For any fixed $k$, the number of codewords in the first $k-1$ indicators is negligible to the number of all other codewords. Specifically, $|\calM_{k-1}| \leq 2(k-1)$ holds. Now, 
\begin{align}
\label{STEP0}
\sum_{(m_{k},m_{k}') \in [M]_{*}^{2}} \IND(m_{k},m_{k}')
&= \sum_{(m_{k},m_{k}') \in ([M] \setminus \calM_{k-1})_{*}^{2}} \IND(m_{k},m_{k}') \nn \\
&~~~ + \sum_{m_{k} \in \calM_{k-1}} \sum_{m_{k}' \in [M] \setminus \calM_{k-1}} \left(\IND(m_{k},m_{k}') +\IND(m_{k}',m_{k}) \right) \nn \\
&~~~+ \sum_{(m_{k},m_{k}') \in (\calM_{k-1})_{*}^{2}} \IND(m_{k},m_{k}') .
\end{align}
By \eqref{STEP0}, linearity of expectation, the independence of codewords assumption, and the trivial fact that $\IND(m_{k},m_{k}') \leq 1$,
\begin{align}
\mathbb{E} \left[ \sum_{(m_{k},m_{k}') \in [M]_{*}^{2}} \IND(m_{k},m_{k}') \middle| \{\bX_{l}\}_{l \in \calM_{k-1}} \right] 
&\lexe e^{n(2R-I)} + 4(k-1)e^{n(R-I)} + 4(k-1)^{2} \\
\label{STEP1}
&\doteq \max \{e^{n(2R-I)}, 1\}.
\end{align}
Now, 
\begin{align}
\mathbb{E} \left[ N(Q_{XX'})^{k}  \right] 
& = \sum_{\left\{\substack{(m_{i},m_{i}')\in[M]_{*}^{2}, \\ 1 \leq i \leq k}\right\}} \mathbb{E} \left[ \prod_{i=1}^{k} \IND(m_{i},m_{i}') \right] \\
\label{STEP2}
& = \sum_{\left\{\substack{(m_{i},m_{i}')\in[M]_{*}^{2}, \\ 1 \leq i \leq k-1}\right\}} \mathbb{E} \left[ \prod_{i=1}^{k-1} \IND(m_{i},m_{i}') 
\cdot \left( \sum_{(m_{k},m_{k}') \in [M]_{*}^{2}} \IND(m_{k},m_{k}') \right) \right] .
\end{align}
The expectation in \eqref{STEP2} is given by
\begin{align}
&\mathbb{E} \left[ \prod_{i=1}^{k-1} \IND(m_{i},m_{i}') 
\cdot \left( \sum_{(m_{k},m_{k}') \in [M]_{*}^{2}} \IND(m_{k},m_{k}') \right) \right] \nn \\
&=\mathbb{E} \left[ \mathbb{E} \left[ \prod_{i=1}^{k-1} \IND(m_{i},m_{i}') 
\cdot \left( \sum_{(m_{k},m_{k}') \in [M]_{*}^{2}} \IND(m_{k},m_{k}') \right) \middle| \{\bX_{l}\}_{l \in \calM_{k-1}} \right] \right] \\
\label{TUTTY1}
&=\mathbb{E} \left[ \prod_{i=1}^{k-1} \IND(m_{i},m_{i}') 
\cdot \mathbb{E} \left[ \left( \sum_{(m_{k},m_{k}') \in [M]_{*}^{2}} \IND(m_{k},m_{k}') \right) \middle| \{\bX_{l}\}_{l \in \calM_{k-1}} \right] \right] \\
\label{TUTTY2}
&\lexe \max \{e^{n(2R-I)}, 1\} \cdot \mathbb{E} \left[ \prod_{i=1}^{k-1} \IND(m_{i},m_{i}') \right],
\end{align}
where \eqref{TUTTY1} is due to the fact that upon conditioning on $\{\bX_{l}\}_{l \in \calM_{k-1}}$, $\prod_{i=1}^{k-1} \IND(m_{i},m_{i}')$ is fixed, and \eqref{TUTTY2} follows from \eqref{STEP1}.
Substituting it back into \eqref{STEP2} and using the induction assumption provides 
\begin{align}
\mathbb{E} \left[ N(Q_{XX'})^{k}  \right]
&\lexe \max \{e^{n(2R-I)}, 1\} \sum_{\left\{\substack{(m_{i},m_{i}')\in[M]_{*}^{2}, \\ 1 \leq i \leq k-1}\right\}} \mathbb{E} \left[ \prod_{i=1}^{k-1} \IND(m_{i},m_{i}') \right] \\
&= \max \{e^{n(2R-I)}, 1\} \cdot \mathbb{E} \left[ \left(N(Q_{XX'})\right)^{k-1}  \right] \\
&\lexe \max \{e^{n(2R-I)}, 1\} \cdot 
\left\{   
\begin{array}{l l}
e^{n (k-1) (2R - I)}    & \quad \text{  $I < 2R$  }   \\
e^{-n (I - 2R) }    & \quad \text{  $I > 2R$  }
\end{array} \right.  \\
&= \left\{   
\begin{array}{l l}
e^{n k (2R - I)}    & \quad \text{  $I < 2R$  }   \\
e^{-n (I - 2R) }    & \quad \text{  $I > 2R$  }
\end{array} \right. .
\end{align}
Thus, Lemma \ref{Integer_Moments} is proved.

\section*{Appendix D}
\renewcommand{\theequation}{D.\arabic{equation}}
\setcounter{equation}{0}  
\subsection*{Proof of Proposition \ref{LT_Properties}}

The monotonicity is straightforward, and follows the fact that $\calL(R,\expVAR)$ and $\calM(R,\expVAR)$, defined in \eqref{L_DEF} and \eqref{M_DEF}, respectively, become larger when $\expVAR$ grows. 
In order to show the fourth item, observe that when $\expVAR < \expVAR^{\mbox{\tiny min}}$, the set $\calL(R,\expVAR)$ is empty.
As for the second item, we seek a condition on $\expVAR$ such that $E_{\mbox{\tiny lt}}^{\mbox{\tiny ub}}(R,\expVAR)>0$:
\begin{align}
\min_{Q_{XX'} \in \calL(R,\expVAR)} [I_{Q}(X;X')-2R]_{+} > 0.
\end{align}
Explicitly, 
\begin{align}
\min_{\{Q_{XX'} \in \calQ(Q_{X}):~  [2R-I_{Q}(X;X')]_{+}  \geq \Gamma(Q_{XX'},R) + R - \expVAR \}} [I_{Q}(X;X')-2R]_{+} > 0,
\end{align}
and by using the identity $\min_{\{Q:~ g(Q) \leq 0\}} f(Q) = \min_{Q} \sup_{s \geq 0} \{f(Q) + s \cdot g(Q)\}$,
it can also be written as
\begin{align}
\min_{Q_{XX'} \in \calQ(Q_{X})} \sup_{s \geq 0} 
\left\{ s \cdot (\Gamma(Q_{XX'},R) + R - \expVAR - [2R-I_{Q}(X;X')]_{+})  + [I_{Q}(X;X')-2R]_{+} \right\} > 0, \nonumber
\end{align}
which means that for every $Q_{XX'} \in \calQ(Q_{X})$ there exists some $s \geq 0$, such that
\begin{align}
s \cdot (\Gamma(Q_{XX'},R) + R - \expVAR - [2R-I_{Q}(X;X')]_{+}) + [I_{Q}(X;X')-2R]_{+} > 0,
\end{align}
or equivalently,
\begin{align}
\expVAR < \Gamma(Q_{XX'},R) + R - [2R-I_{Q}(X;X')]_{+} + \frac{[I_{Q}(X;X')-2R]_{+}}{s}.
\end{align}
Thus, 
\begin{align}
\expVAR 
\label{TERM2EXPf1}
&< \min_{Q_{XX'} \in \calQ(Q_{X})} \sup_{s \geq 0} \left\{ \Gamma(Q_{XX'},R) + R - [2R-I_{Q}(X;X')]_{+} + \frac{[I_{Q}(X;X')-2R]_{+}}{s} \right\} \\
\label{TERM2EXPf2}
&= \min_{Q_{XX'} \in \calQ(Q_{X})} \left[ \Gamma(Q_{XX'},R) + R - [2R-I_{Q}(X;X')]_{+}  + \left\{ 
\begin{array}{l l}
0               & \quad \text{$I_{Q}(X;X') \leq 2R$  }\\
\infty          & \quad \text{$I_{Q}(X;X') > 2R$  } 
\end{array} \right. \right] \\
&= \min_{\{Q_{XX'} \in \calQ(Q_{X}):~ I_{Q}(X;X') \leq 2R\}} 
\left\{ \Gamma(Q_{XX'},R) + R - [2R-I_{Q}(X;X')]_{+} \right\} \\
&= \min_{\{Q_{XX'} \in \calQ(Q_{X}):~ I_{Q}(X;X') \leq 2R\}} 
\left\{ \Gamma(Q_{XX'},R) + I_{Q}(X;X') - R \right\} \\
&= E_{\mbox{\tiny trc}}(R),
\end{align} 
where the $\infty$ in \eqref{TERM2EXPf2} is because the maximizing $s \geq 0$ in \eqref{TERM2EXPf1} when $I_{Q}(X;X')>2R$ is $s^{*}=0$.
The proof of the third item is very similar to the proof of the second item and hence omitted.

\section*{Appendix E}
\renewcommand{\theequation}{E.\arabic{equation}}
\setcounter{equation}{0}  
\subsection*{Proof of Proposition \ref{UT_Properties}}

The monotonicity is immediate, since both $\calV(R,\expVAR)$ and $\calU(R,\expVAR)$, defined in \eqref{V_DEF} and \eqref{U_DEF}, respectively, become larger when $\expVAR$ grows. 
In order to show the second item, we seek a condition on $\expVAR$ such that $E_{\mbox{\tiny ut}}^{\mbox{\tiny lb}}(R,\expVAR)>0$:
\begin{align}
\max_{Q_{XX'} \in \calU(R,\expVAR)} \{2R - I_{Q}(X;X')\} > 0.
\end{align}
Explicitly, 
\begin{align}
\max_{\{Q_{XX'} \in \calQ(Q_{X}):~ I_{Q}(X;X') \leq 2R,~  \Gamma(Q_{XX'},R) + I_{Q}(X;X') - R \leq \expVAR \}} \{2R-I_{Q}(X;X')\} > 0,
\end{align}
and thanks to the fact that $\max_{\{Q:~ g(Q) \geq 0\}} f(Q) = \max_{Q} \inf_{\mu \geq 0} \{f(Q) + \mu \cdot g(Q)\}$,
it can also be written as
\begin{align}
\max_{\{Q_{XX'} \in \calQ(Q_{X}):~ I_{Q}(X;X') \leq 2R \}} \inf_{\mu \geq 0} \{&2R - I_{Q}(X;X') \nn \\
&+ \mu \cdot (\expVAR - \Gamma(Q_{XX'},R) - I_{Q}(X;X') + R)\} > 0,
\end{align}
or, equivalently,
\begin{align}
&\exists Q_{XX'} \in \calQ(Q_{X}) ~\text{s.t.}~ I_{Q}(X;X') \leq 2R,  
~~\forall \mu \geq 0: \nn \\
& \mu \cdot \expVAR > I_{Q}(X;X') - 2R + \mu \cdot (\Gamma(Q_{XX'},R) + I_{Q}(X;X') - R), 
\end{align}
or,
\begin{align}
\expVAR 
\label{TERM2EXPg1}
&> \min_{\{Q_{XX'} \in \calQ(Q_{X}):~ I_{Q}(X;X') \leq 2R \}} \sup_{\mu \geq 0} \left\{\frac{I_{Q}(X;X') - 2R}{\mu} +  \Gamma(Q_{XX'},R) + I_{Q}(X;X') - R \right\} \\
\label{TERM2EXPg2}
&= \min_{\{Q_{XX'} \in \calQ(Q_{X}):~ I_{Q}(X;X') \leq 2R \}} \left\{\Gamma(Q_{XX'},R) + I_{Q}(X;X') - R \right\} \\
&= E_{\mbox{\tiny trc}}(R),
\end{align}
where \eqref{TERM2EXPg2} is because the maximizing $\mu \geq 0$ in \eqref{TERM2EXPg1} is $\mu^{*}=\infty$, since $I_{Q}(X;X')\leq2R$.
The proof of the third item is very similar to the proof of the second item and hence omitted.

\section*{Appendix F}
\renewcommand{\theequation}{F.\arabic{equation}}
\setcounter{equation}{0}  
\subsection*{Proof of Corollary \ref{COROLLARY_ONE}}

The probability of any codebook in the ensemble is given asymptotically by $\exp\{-nH_{Q}(X)e^{nR}\}$, hence, in order to assure that a code exists, we demand that 
\begin{align}
\label{Code_Existence}
\prob \left\{ -\frac{1}{n} \log P_{\mbox{\tiny e}} (\calC_{n}) \geq \expVAR \right\} 
> \exp \{-nH_{Q}(X)e^{nR}\}.
\end{align}
Now, the lower bound of Theorem \ref{THEOREM_UPPER_TAIL} reads
\begin{align}
\prob \left\{ -\frac{1}{n} \log P_{\mbox{\tiny e}} (\calC_{n}) \geq \expVAR \right\}  
\DGEXE \exp \left\{- \exp\left\{n \cdot \max_{Q_{XX'} \in \calU(R,\expVAR)} \{2R - I_{Q}(X;X')\} \right\} \right\} ,
\end{align}
thus (\ref{Code_Existence}) will obviously be satisfied if 
\begin{align}
\max_{Q_{XX'} \in \calU(R,\expVAR)} \{2R - I_{Q}(X;X')\} < R,
\end{align}
or, equivalently, 
\begin{align}
\label{CONDITIONF4}
\min_{Q_{XX'} \in \calU(R,\expVAR)} I_{Q}(X;X') > R,
\end{align}
which is exactly (\ref{CONDITION1}).
Then, following some algebraic work, that can be found in \eqref{ALgebraicWorkA}--\eqref{ALgebraicWorkB}, we found that \eqref{CONDITIONF4} is equivalent to $\expVAR < E_{\mbox{\tiny ex}}(R)$.

\section*{Appendix G}
\renewcommand{\theequation}{G.\arabic{equation}}
\setcounter{equation}{0}  
\subsection*{Proof of Proposition \ref{Prop_Moments}}

For a set of indices ${\cal J}$ let us denote ${\cal J}_{*}^{2} = \{(j,j')\in{\cal J}^{2}\colon j\neq j'\}$.
Recall that ${\cal I}(m,m') = {\cal I}\{(\boldsymbol{X}_{m},\boldsymbol{X}_{m'})\in{\cal T}(Q_{XX'})\}$
and $N(Q_{XX'}) = \sum_{(m,m')\in[M]_{*}^{2}}{\cal I}(m,m')$.
Let us abbreviate $\IND(m) = \IND \left\{(\bX_{m},\by) \in \calT(Q_{XY}) \right\}$, such that 
\begin{align} \label{ENUM_DEF_With_y}
N_{\by}(Q_{XY}) = \sum_{m \in [M]} \IND(m).
\end{align}
Recall the definition of $F(S,Q_{UV},j)$ in \eqref{DEF_F}. We show by induction that 
\begin{align}
\E\left[N_{\boldsymbol{y}}(Q_{XY})^{l}N(Q_{XX'})^{k}\right]
\lexe F(R,Q_{XY},l) \cdot F(2R,Q_{XX'},k). \label{eq: enumerator result}
\end{align}

\underline{Checking for \mbox{$k=l=1$}:} Note that due to the symmetry of the random draw over the type class:
\begin{align}
\E\left[{\cal I}(m,m'){\cal I}(m)\right] & =\E\left[{\cal I}(m)\E\left[{\cal I}(m,m')\mid\boldsymbol{X}_{m}\right]\right]\\
& =\E\left[{\cal I}(m)\right]\cdot\E\left[{\cal I}(m,m')\right]
\end{align}
and similarly, $\E\left[{\cal I}(m,m'){\cal I}(m')\right]=\E\left[{\cal I}(m')\right]\cdot\E\left[{\cal I}(m,m')\right]$.
Thus, for $k=l=1$:
\begin{align}
& \E\left[N_{\boldsymbol{y}}(Q_{XY})N(Q_{XX'})\right] \nn \\
& =\sum_{(m,m')\in[M]_{*}^{2}}\sum_{r\in[M]}\E\left[{\cal I}(m,m'){\cal I}(r)\right]\\
& =\sum_{(m,m')\in[M]_{*}^{2}}\left(\sum_{r\in[M]\backslash\{m,m'\}}\E\left[{\cal I}(m,m')\right]\E\left[{\cal I}(r)\right]+\E\left[{\cal I}(m,m'){\cal I}(m)\right]+\E\left[{\cal I}(m,m'){\cal I}(m')\right]\right)\\
& =\sum_{(m,m')\in[M]_{*}^{2}}\sum_{r\in[M]}\E\left[{\cal I}(m,m')\right]\E\left[{\cal I}(r)\right]\\
&\doteq e^{n(2R-I_{Q}(X;X'))}\cdot e^{n(R-I_{Q}(X;Y))}.
\end{align}

\underline{Induction assumption:} Assume that (\ref{eq: enumerator result})
holds up for some $(k-1,l-1)$. We show by two inductive steps that this holds for $(k,l-1)$ and $(k-1,l)$ and thus for any $(k,l)$. 

\underline{Proof for \mbox{$(k,l-1)$}:} Assume that $\{(m_{i},m_{i}')\}_{i=1}^{k-1}$
and $\{r_{j}\}_{j=1}^{l-1}$ are given, where $(m_{i},m_{i}')\in[M]_{*}^{2}$ for all $i\in[k-1]$, and $r_{j}\in[M]$ for all $j\in[l-1]$. 
Let ${\cal M}_{k-1,l-1}=\bigcup_{i=1}^{k-1}\{\{m_{i}\} \cup  \{m_{i}'\}\}\cup\bigcup_{j=1}^{l-1}\{r_{j}\}$
be the set of indices of the $k-1$ pairs of codeword indices $\{(m_{i},m_{i}')\}_{i=1}^{k-1}$ and of the $l-1$ codeword indices $\{r_{j}\}_{j=1}^{l-1}$. 
Clearly $|{\cal M}_{k-1,l-1}|\leq2(k-1)+l-1 \DEF c_{k-1,l-1}$
holds. Now,
\begin{multline}
\sum_{(m_{k},m_{k}')\in[M]_{*}^{2}}{\cal I}(m_{k},m_{k}')=\sum_{(m_{k},m_{k}')\in([M]\backslash{\cal M}_{k-1,l-1})_{*}^{2}}{\cal I}(m_{k},m_{k}')\\
+\sum_{m_{k}\in{\cal M}_{k-1,l-1}}\sum_{m'_{k}\in[M] \setminus \calM_{k-1,l-1}} \left( {\cal I}(m_{k},m'_{k})+{\cal I}(m'_{k},m_{k})\right) +\sum_{(m_{k},m'_{k})\in (\calM_{k-1,l-1})_{*}^{2}}{\cal I}(m_{k},m'_{k}).\label{eq: decomposition}
\end{multline}
By (\ref{eq: decomposition}), linearity of expectation, the independence
of codewords assumption, and the fact that ${\cal I}(m_{k},m'_{k})\leq1$, 
\begin{align}
&\E\left[\sum_{(m_{k},m_{k}')\in[M]_{*}^{2}}{\cal I}(m_{k},m_{k}') \middle| \{\boldsymbol{X}_{s}\}_{s\in{\cal M}_{k-1,l-1}}\right] \nn \\ 
&~~~~~~~~ \dot{\leq} e^{n(2R-I_{Q}(X;X'))} + 2c_{k-1,l-1}e^{n(R-I_{Q}(X;X'))}+c_{k-1,l-1}^{2} \\
&~~~~~~~~ \dot{=} \max\{e^{n(2R-I_{Q}(X;X'))},1\}.\label{STEPA0}
\end{align}
Next, 
\begin{align}
& \E\left[N_{\boldsymbol{y}}(Q_{XY})^{l-1}N(Q_{XX'})^{k}\right]\nonumber \\
& =\sum_{\left\{\substack{(m_{i},m_{i}')\in[M]_{*}^{2}, \\ 1 \leq i \leq k}\right\}} \sum_{\left\{\substack{r_{j}\in[M], \\ 1 \leq j \leq l-1}\right\}} \E\left[\prod_{i=1}^{k}{\cal I}(m_{i},m_{i}')\prod_{j=1}^{l-1}{\cal I}(r_{j})\right]\\
\label{STEPA1}
& =\sum_{\left\{\substack{(m_{i},m_{i}')\in[M]_{*}^{2}, \\ 1 \leq i \leq k-1}\right\}} \sum_{\left\{\substack{r_{j}\in[M], \\ 1 \leq j \leq l-1}\right\}} \E\left[\prod_{i=1}^{k-1}{\cal I}(m_{i},m_{i}')\cdot\prod_{j=1}^{l-1}{\cal I}(r_{j})\left(\sum_{(m_{k},m_{k}')\in[M]_{*}^{2}}{\cal I}(m_{k},m_{k}')\right)\right] .
\end{align}
The expectation in \eqref{STEPA1} is given by
\begin{align}
&\E\left[\prod_{i=1}^{k-1}{\cal I}(m_{i},m_{i}')\cdot\prod_{j=1}^{l-1}{\cal I}(r_{j})\left(\sum_{(m_{k},m_{k}')\in[M]_{*}^{2}}{\cal I}(m_{k},m_{k}')\right)\right] \nn \\
&= \E\left[\E\left[\prod_{i=1}^{k-1}{\cal I}(m_{i},m_{i}')\cdot\prod_{j=1}^{l-1}{\cal I}(r_{j})\left(\sum_{(m_{k},m_{k}')\in[M]_{*}^{2}}{\cal I}(m_{k},m_{k}')\right) \middle| \{\boldsymbol{X}_{s}\}_{s\in{\cal M}_{k-1,l-1}}\right]\right]  \\
\label{TUTTY3}
&= \E\left[\prod_{i=1}^{k-1}{\cal I}(m_{i},m_{i}')\cdot\prod_{j=1}^{l-1}{\cal I}(r_{j}) \cdot \E\left[\left(\sum_{(m_{k},m_{k}')\in[M]_{*}^{2}}{\cal I}(m_{k},m_{k}')\right) \middle| \{\boldsymbol{X}_{s}\}_{s\in{\cal M}_{k-1,l-1}}\right]\right]  \\
\label{TUTTY4}
&\lexe \max\{e^{n(2R-I_{Q}(X;X'))},1\} \cdot \E\left[\prod_{i=1}^{k-1}{\cal I}(m_{i},m_{i}')\cdot\prod_{j=1}^{l-1}{\cal I}(r_{j}) \right],
\end{align}
where \eqref{TUTTY3} is thanks to the conditioning on $\{\boldsymbol{X}_{s}\}_{s\in{\cal M}_{k-1,l-1}}$, and \eqref{TUTTY4} is due to \eqref{STEPA0}. 
Substituting it back into \eqref{STEPA1} and using the induction assumption provides
\begin{align}
& \E\left[N_{\boldsymbol{y}}(Q_{XY})^{l-1}N(Q_{XX'})^{k}\right]\nonumber \\
&\lexe \max\{e^{n(2R-I_{Q}(X;X'))},1\} \cdot \sum_{\left\{\substack{(m_{i},m_{i}')\in[M]_{*}^{2}, \\ 1 \leq i \leq k-1}\right\}} \sum_{\left\{\substack{r_{j}\in[M], \\ 1 \leq j \leq l-1}\right\}} \E\left[\prod_{i=1}^{k-1}{\cal I}(m_{i},m_{i}')\cdot\prod_{j=1}^{l-1}{\cal I}(r_{j}) \right] \\
&=\max\{e^{n(2R-I_{Q}(X;X'))},1\}\cdot\E\left[N_{\boldsymbol{y}}(Q_{XY})^{l-1}N(Q_{XX'})^{k-1}\right]\\
&\lexe \max\{e^{n(2R-I_{Q}(X;X'))},1\} \cdot F(R,Q_{XY},l-1) \cdot F(2R,Q_{XX'},k-1)\\
&= F(R,Q_{XY},l-1) \cdot F(2R,Q_{XX'},k) ,
\end{align}
which completes the proof of the first inductive step. The proof of the second inductive step follows exactly the same lines and hence omitted. 
The proof of Proposition \ref{Prop_Moments} is complete.

\section*{Appendix H}
\renewcommand{\theequation}{H.\arabic{equation}}
\setcounter{equation}{0}  
\subsection*{Proof of Proposition \ref{Double_Exponential_Result}}

By the union bound,
\begin{align}
\prob \left\{\hat{\calB}_{n}(\sigma)\right\} 
&= \prob \left\{\bigcup_{m=0}^{M-1} \bigcup_{m' \neq m} \bigcup_{\by \in \calY^{n}}\hat{\calB}_{n}(\sigma,m,m',\by)\right\} \\   
\label{TERM2ContinueFrom}
&\leq \sum_{m=0}^{M-1} \sum_{m' \neq m} \sum_{\by \in \calY^{n}} \prob \left\{\hat{\calB}_{n}(\sigma,m,m',\by)\right\} .
\end{align}
Now,
\begin{align}
&\prob \left\{\hat{\calB}_{n}(\sigma,m,m',\by)\right\} \nn \\
\label{TERM2EXP2}
&=\prob \left\{\sum_{\tilde{m}\in\{0,1,\ldots,M-1\} \setminus \{m,m'\}} \exp \{n g(\hat{P}_{\bX_{\tilde{m}}\by})\} \geq \exp\{n \cdot (\beta(R,Q_{Y}) + \sigma)\} \right\}  \\
\label{TERM2EXP3}
&= \prob \left\{\sum_{Q_{XY}} N(Q_{XY}) e^{n g(Q_{XY})} \geq \exp\{n \cdot (\beta(R,Q_{Y}) + \sigma)\} \right\} \\
\label{TERM2EXP4}
&\doteq \sum_{Q_{XY}} \prob \left\{N(Q_{XY}) \geq \exp\{n(\beta(R,Q_{Y}) + \sigma - g(Q_{XY}))\} \right\} \\
\label{Two_Summands}
&= \sum_{\{Q_{XY}:~ I_{Q}(X;Y) \leq R\}} \prob \left\{N(Q_{XY}) \geq \exp\{n(\beta(R,Q_{Y}) + \sigma - g(Q_{XY}))\} \right\} \nn \\
&~+ \sum_{\{Q_{XY}:~ I_{Q}(X;Y) > R\}} \prob \left\{N(Q_{XY}) \geq \exp\{n(\beta(R,Q_{Y}) + \sigma - g(Q_{XY}))\} \right\}, 
\end{align}
where \eqref{TERM2EXP2} is due to the definition of $Z_{mm'}(\by)$ in \eqref{Zmmtag_DEF}, 
in \eqref{TERM2EXP3} we introduced the type class enumerator $N(Q_{XY})$, which is the number of codewords in $\calC_{n}$, other than $\bx_{m}$ and $\bx_{m'}$, that have a joint composition $Q_{XY}$ together with $\by$,
and where \eqref{TERM2EXP4} is due to the SME.
The first summand of \eqref{Two_Summands} is upper--bounded by
\begin{align}
&\prob \left\{N(Q_{XY}) \geq \exp\left\{n\left(\beta(R,Q_{Y}) + \sigma - g(Q_{XY})\right)\right\} \right\} \nn \\
&=\prob \left\{N(Q_{XY}) \geq \exp\left\{n\left(\sigma + \beta(R,Q_{Y}) - g(Q_{XY}) - \left[R - I_{Q}(X;Y)\right]_{+} + \left[R - I_{Q}(X;Y)\right]_{+} \right)\right\} \right\} \nn \\
\label{TERM2HELP}
&\leq \prob \left\{N(Q_{XY}) \geq \exp\left\{n\left(\sigma + \left[R - I_{Q}(X;Y)\right]_{+} \right)\right\} \right\} \\
\label{Gauss2}
&= \prob \left\{N(Q_{XY}) \geq e^{n (\sigma+R-I_{Q}(X;Y))}  \right\} \\
\label{Gauss3}
&\leq \exp \left\{-e^{nR} D(e^{-n[R-(\sigma+R-I_{Q}(X;Y))]}\|e^{-n I_{Q}(X;Y)})\right\} \\
&= \exp \left\{-e^{nR} D(e^{-n(I_{Q}(X;Y)-\sigma)}\|e^{-n I_{Q}(X;Y)})\right\} \\
\label{Gauss4}
&< \exp \left\{-e^{nR} \cdot e^{-n(I_{Q}(X;Y)-\sigma)} \cdot \left(\ln \frac{e^{-n(I_{Q}(X;Y)-\sigma)}}{e^{-n I_{Q}(X;Y)}} - 1 \right) \right\} \\
&= \exp \left\{-e^{n(R-I_{Q}(X;Y)+\sigma)} \cdot \left( n\sigma - 1 \right) \right\} \\
\label{Gauss5}
&\leq \exp \left\{-e^{n \sigma} \right\}.
\end{align}
In \eqref{TERM2HELP}, we use the definition of $\beta(R,Q_{Y})$ in \eqref{Beta_DEF}, which implies that $\beta(R,Q_{Y}) \geq g(Q_{XY}) + \left[R - I_{Q}(X;Y)\right]_{+}$, and for \eqref{Gauss2}, recall that $R \geq I_{Q}(X;Y)$.   
Step \eqref{Gauss3} is according to Chernoff's bound \cite[Appendix]{SBM}, \cite[Appendix B]{MERHAV2017}, \eqref{Gauss4} is due to the following lower bound to the binary divergence \cite[Sec.\ 6.3, p.\ 167]{MERHAV09}   
\begin{align}
\label{Binary_DIV_LB}
D(a\|b) > a \left(\ln \frac{a}{b}-1\right),
\end{align}
and \eqref{Gauss5} is true since $R \geq I_{Q}(X;Y)$.
Similarly, for the second summand of \eqref{Two_Summands}, we have
\begin{align}
&\prob \left\{N(Q_{XY}) \geq \exp\left\{n\left(\beta(R,Q_{Y}) + \sigma - g(Q_{XY})\right)\right\} \right\} \nn \\
\label{TUTZ1}
&\leq \prob \left\{N(Q_{XY}) \geq \exp\left\{n\left(\sigma + \left[R - I_{Q}(X;Y)\right]_{+} \right)\right\} \right\} \\
\label{TUTZ2}
&= \prob \left\{N(Q_{XY}) \geq e^{n \sigma} \right\} \\
\label{TUTZ3}
&\leq \exp \left\{-e^{nR} D(e^{-n(R-\sigma)}\|e^{-n I_{Q}(X;Y)})\right\} \\
\label{TUTZ4}
&< \exp \left\{-e^{nR} \cdot e^{-n(R-\sigma)} \cdot \left(\ln \frac{e^{-n(R-\sigma)}}{e^{-n I_{Q}(X;Y)}} - 1 \right) \right\} \\
&= \exp \left\{-e^{n\sigma} \cdot \left[ n(I_{Q}(X;Y) - R + \sigma) - 1 \right] \right\} \\
\label{TUTZ5}
&\leq \exp \left\{-e^{n \sigma} \right\},
\end{align}
where \eqref{TUTZ1} is true for the same reason as \eqref{TERM2HELP}, \eqref{TUTZ2} is because $I_{Q}(X;Y)>R$, \eqref{TUTZ3} is again due to Chernoff's bound, \eqref{TUTZ4} is true thanks to \eqref{Binary_DIV_LB}, and \eqref{TUTZ5} is due to $I_{Q}(X;Y) - R + \sigma > 0$. 
Hence, we conclude that for every $\sigma > 0$
\begin{align}
\prob \{\hat{\calB}_{n}(\sigma,m,m',\by)\}
&=\prob \{Z_{mm'}(\by) \geq \exp\{n \cdot (\beta(R,Q_{Y}) + \sigma)\} \} \\
&\DLEXE \exp \left\{-e^{n \sigma} \right\},
\end{align}
and so, continuing from \eqref{TERM2ContinueFrom}, this means that
\begin{align}
\prob \left\{\hat{\calB}_{n}(\sigma)\right\} 
&\DLEXE \sum_{m=0}^{M-1} \sum_{m' \neq m} \sum_{\by \in \calY^{n}} \exp \left\{-e^{n \sigma} \right\} \\
&\DEXE \exp \left\{-e^{n \sigma} \right\},
\end{align}
which completes the proof of the proposition.

\section*{Appendix I}
\renewcommand{\theequation}{I.\arabic{equation}}
\setcounter{equation}{0}  
\subsection*{Proof of Proposition \ref{Super_Enum_Lower_Bound}}

First, note that  
\begin{align}
\calF_{0}
= \left\{\sum_{Q_{XX'} \in \styleA_{1} \cup \styleA_{2}} N(Q_{XX'}) = 0  \right\}.
\end{align} 
Let us define
\begin{align}
N(\styleA_{1} \cup \styleA_{2}) 
&\DEF \sum_{Q_{XX'} \in \styleA_{1} \cup \styleA_{2}} N(Q_{XX'}), 
\end{align}
and the binary random variables 
\begin{align}
\IND(m,m',Q_{XX'}) \DEF \IND \left\{(\bX_{m},\bX_{m'}) \in \calT(Q_{XX'}) \right\},
\end{align}
such that,
\begin{align}
N(\styleA_{1} \cup \styleA_{2}) 
= \sum_{Q_{XX'} \in \styleA_{1} \cup \styleA_{2}} 
\sum_{m=0}^{M-1} \sum_{m' \neq m} \IND(m,m',Q_{XX'}).
\end{align}

In order to use Fact \ref{FACT3} that appears in Appendix A, 
let us first define an appropriate dependency graph.
One can easily check that the indicator random variables $\IND(i,j,Q)$ and $\IND(k,l,\tilde{Q})$ are independent as long as $i \neq k$, $j \neq l$, and $Q \neq \tilde{Q}$. Thus, we define our dependency graph in a way that each vertex $(i,j,Q)$ is connected to exactly $e^{nR} - 1$ vertices of the form $(k,j,Q)$, $k \neq i$,
to $e^{nR} - 1$ vertices of the form $(i,l,Q)$, $l \neq j$, and to exactly $|\styleA_{1} \cup \styleA_{2}| - 1$ vertices of the form $(i,j,\tilde{Q})$, $\tilde{Q} \neq Q$. Let us now examine the quantities $\Delta$, $\Omega$, and $\LambdaBOOM$. First,
\begin{align}
\Delta 
&= \mathbb{E} [N(\styleA_{1} \cup \styleA_{2})] \\
&= \sum_{Q_{XX'} \in \styleA_{1} \cup \styleA_{2}} \mathbb{E} [N(Q_{XX'})] \\
&\doteq \sum_{Q_{XX'} \in \styleA_{1} \cup \styleA_{2}} e^{n \cdot (2R - I_{Q}(X;X'))} \\
&\doteq \max_{Q_{XX'} \in \styleA_{1} \cup \styleA_{2}} e^{n \cdot (2R - I_{Q}(X;X'))} \\
&= \exp\left\{n \cdot \max_{Q_{XX'} \in \styleA_{1} \cup \styleA_{2}} \{2R - I_{Q}(X;X')\} \right\} \\
&= \exp\left\{n \cdot \max_{Q_{XX'} \in \styleA_{2}} \{2R - I_{Q}(X;X')\} \right\} ,
\end{align}
where the last equality follows from the definitions of $\styleA_{1}$ and $\styleA_{2}$ and the assumption that $\styleA_{2}$ is nonempty. 
Regarding the quantity $\Omega_{i,j,Q}$ of (\ref{OMEGA_I_DEF}), notice that it actually depends only on $Q$. 
Thus, for some $Q \in \styleA_{1} \cup \styleA_{2}$, 
\begin{align}
\Omega_{Q} 
&\doteq (e^{nR} + e^{nR} - 2) \cdot e^{-n I_{Q}(X;X')} + \sum_{\tilde{Q} \in \styleA_{1} \cup \styleA_{2} \setminus \{Q\}} e^{-n I_{\tilde{Q}}(X;X')} \\
&\doteq e^{n (R-I_{Q}(X;X'))} + \sum_{\tilde{Q} \in \styleA_{1} \cup \styleA_{2}} e^{-n I_{\tilde{Q}}(X;X')} \\
&\doteq e^{n (R-I_{Q}(X;X'))} + \max_{\tilde{Q} \in \styleA_{1} \cup \styleA_{2}} e^{-n I_{\tilde{Q}}(X;X')},
\end{align}
and hence
\begin{align}
\Omega 
= \max_{Q \in \styleA_{1} \cup \styleA_{2}} \Omega_{Q}
\doteq \max_{Q \in \styleA_{1} \cup \styleA_{2}} e^{n (R-I_{Q}(X;X'))} .
\end{align}
Furthermore,
\begin{align}
\LambdaBOOM \doteq  \max_{Q \in \styleA_{1} \cup \styleA_{2}} e^{-n I_{Q}(X;X')},
\end{align}
such that
\begin{align}
\Omega + \LambdaBOOM 
&\doteq \max_{Q \in \styleA_{1} \cup \styleA_{2}} e^{n (R-I_{Q}(X;X'))} \\
&= \max_{Q \in \styleA_{2}} e^{n (R-I_{Q}(X;X'))} .
\end{align}
Now, we would like to have $\Omega + \LambdaBOOM \in [0,e^{-1}]$. Specifically, if $\Omega + \LambdaBOOM \to 0$ as $n \to \infty$, then $\varphi(\Omega + \LambdaBOOM) \doteq 1$.
In order to have $\Omega + \LambdaBOOM \to 0$, we need that 
\begin{align}
\max_{Q \in \styleA_{2}} \{R-I_{Q}(X;X')\} < 0,
\end{align}
or
\begin{align}
\label{CONDITION1}
\min_{Q_{XX'} \in \styleA_{2}} I_{Q}(X;X') > R.
\end{align}
Let us abbreviate $I_{Q}(X;X')$ by $I_{Q}$. 
In order to find the highest $\expVAR$ for which (\ref{CONDITION1}) holds, let us derive $\min_{Q_{XX'} \in \styleA_{2}} I_{Q}(X;X')$ as follows:
\begin{align}
&\min_{Q_{XX'} \in \styleA_{2}} I_{Q} \nonumber \\
\label{ALgebraicWorkA}
&=\min_{\{Q_{XX'} \in \calQ(Q_{X}):~ I_{Q} \leq 2R,~
	\Gamma(Q,R-\epsilon) + I_{Q} - R \leq \expVAR \}} I_{Q} \\
\label{TERM2EXPd}
&=\min_{Q_{XX'} \in \calQ(Q_{X})} \sup_{\sigma \geq 0} \sup_{\mu \geq 0}
\left\{I_{Q} + \sigma \cdot (I_{Q} - 2R) + \mu \cdot (\Gamma(Q,R-\epsilon) + I_{Q} - R - \expVAR)\right\}, 
\end{align}
where in \eqref{TERM2EXPd} we used twice the fact that
$\min_{\{Q:~ g(Q) \leq 0\}} f(Q) = \min_{Q} \sup_{\sigma \geq 0} \{f(Q) + \sigma \cdot g(Q)\}$.
For \eqref{TERM2EXPd} to be strictly larger than $R$, it is equivalent to require that for all $Q_{XX'} \in \styleA_{2}$ there exist $\sigma \geq 0$ and $\mu \geq 0$ such that  
\begin{align}
I_{Q} + \sigma \cdot (I_{Q} - 2R) + \mu \cdot (\Gamma(Q,R-\epsilon) + I_{Q} - R - \expVAR) > R,
\end{align}
or, equivalently, 
\begin{align}
\expVAR < \frac{I_{Q} - R + \sigma \cdot (I_{Q} - 2R)}{\mu} +  \Gamma(Q,R-\epsilon) + I_{Q} - R. 
\end{align}
Thus,
\begin{align}
\expVAR 
\label{TERM2EXPe1}
&< \min_{Q_{XX'} \in \calQ(Q_{X})} \sup_{\mu \geq 0} \sup_{\sigma \geq 0} \left\{ \Gamma(Q,R-\epsilon) + I_{Q} - R + \frac{I_{Q} - R + \sigma \cdot (I_{Q} - 2R)}{\mu} \right\} \\ 
\label{TERM2EXPe2}
&= \min_{Q_{XX'} \in \calQ(Q_{X})} \sup_{\mu \geq 0}  \left[ \Gamma(Q,R-\epsilon) + I_{Q} - R + \left\{ 
\begin{array}{l l}
\frac{I_{Q}-R}{\mu}     & \quad \text{$I_{Q} \leq 2R$  }\\
\infty               & \quad \text{$I_{Q}>2R$  } 
\end{array} \right. \right] \\
\label{TERM2EXPe3}
&= \min_{\{Q_{XX'} \in \calQ(Q_{X}):~ I_{Q} \leq 2R\}} \sup_{\mu \geq 0}  \left\{ \Gamma(Q,R-\epsilon) + I_{Q} - R + \frac{I_{Q}-R}{\mu} \right\} \\
\label{TERM2EXPe4}
&= \min_{\{Q_{XX'} \in \calQ(Q_{X}):~ I_{Q} \leq 2R\}} \left[ \Gamma(Q,R-\epsilon) + I_{Q} - R + \left\{ 
\begin{array}{l l}
0               & \quad \text{$I_{Q} \leq R$  }\\
\infty          & \quad \text{$I_{Q}>R$  } 
\end{array} \right. \right] \\
&= \min_{\{Q_{XX'} \in \calQ(Q_{X}):~ I_{Q} \leq 2R,~ I_{Q} \leq R\}} \left\{ \Gamma(Q,R-\epsilon) + I_{Q} - R  \right\} \\
&= \min_{\{Q_{XX'} \in \calQ(Q_{X}):~ I_{Q} \leq R\}} \left\{ \Gamma(Q,R-\epsilon) + I_{Q} - R  \right\} \\
\label{ALgebraicWorkB}
&\equiv  E_{\mbox{\tiny ex}}(R,\epsilon),
\end{align}
where the $\infty$ in \eqref{TERM2EXPe2} is because the maximizing $\sigma \geq 0$ in \eqref{TERM2EXPe1} when $I_{Q}>2R$ is $\sigma^{*}=\infty$.
The $\infty$ in \eqref{TERM2EXPe4} is due to the fact that when $I_{Q}>R$, the maximizing $\mu \geq 0$ in \eqref{TERM2EXPe3} is $\mu^{*}=0$.
Note that the exponent function $E_{\mbox{\tiny ex}}(R,\epsilon)$ converges to $E_{\mbox{\tiny ex}}(R)$ when $\epsilon \downarrow 0$.
Finally, we use these results in Fact \ref{FACT3} and get the desired lower bound on $\prob\{\calF_{0}\}$.

%

\end{document}